%% file: resistiveMM.tex
\pdfoutput=1
\documentclass[preprint,12pt]{elsarticle}

\usepackage{amssymb}
\usepackage{amsmath}
\usepackage{float}
\usepackage{xcolor}

\newcommand{\up}[1]{\textcolor{black}{#1}}
\newcommand{\secRev}[1]{\textcolor{black}{#1}}

\usepackage{siunitx}
\usepackage{tikz}

\usepackage{lineno}

\usepackage{subcaption}
\usepackage{pdfpages}

\graphicspath{{figures/}}

\journal{Nuclear Instruments and Methods}

\begin{document}

\begin{frontmatter}




\title{Performances of a resistive Micromegas module for the Time Projection Chambers of the T2K Near Detector upgrade}

\newcommand{\LPNHE}{LPNHE Paris, Sorbonne Universit\'e, Universit\'e Paris Diderot, CNRS/IN2P3, Paris 75252, France}

\input{author.tex}

\begin{abstract}
An upgrade of the Near Detector of the T2K long baseline neutrino oscillation experiment, ND280, has been proposed. This upgrade will include two new Time Projection Chambers, each equipped with 16 resistive Micromegas modules for gas amplification. A first prototype of resistive Micromegas has been designed, built, installed in the HARP field cage, and exposed to a beam of charged particles at CERN. 
The data have been used to characterize the performances of the resistive Micromegas module. A spatial resolution of 300 $\mu m$ and a deposited energy resolution of 9\% were observed for horizontal electrons crossing the TPCs at 30~cm from the anode. Such performances fully satisfy the requirements for the upgrade of the ND280 TPC.

\end{abstract}

\begin{keyword}



\end{keyword}

\end{frontmatter}


\section{Introduction}
\input{intro.tex}
\label{sec:introduction}


\section{Resistive Micromegas}
\label{sec:Micromegas}
\input{mm.tex}

\section{Experimental setup}
\label{sec:setup}
\input{setup.tex}

\section{Track reconstruction}
\label{sec:reco}
\input{reco.tex}

\section{Gas quality}
\label{sec:gas}
\input{gas.tex}

\section{Micromegas gain}
\label{sec:gain}
\input{gain.tex}

\section{Characterization of charge spreading}
\label{sec:spread}
\input{spread.tex}

\section{Deposited energy resolution}
\label{sec:dedx}
\input{dedx.tex}

\section{Spatial resolution}
\label{sec:spatial}
\input{spatial.tex}

\section{Conclusions}
\label{sec:conclusion}
\input{conclusions.tex}

\section*{Acknowledgements}
We thank CERN for delivering the beam and the CERN Neutrino Platform (CENF) for their support during the data taking. We are also indebted to Rui de Oliveira and Olivier Pizzirusso for building the resistive Micromegas module and to Roberto Guida for his help with the gas system.

We acknowledge the support of CEA and CNRS/IN2P3, France; DFG, Germany; INFN,
Italy; National Science Centre (NCN) and Ministry of
Science and Higher Education, Poland; MINECO and ERDF funds, Spain. In addition,
participation of individual researchers and institutions
has been further supported by H2020 Grant No. RISE-GA644294-JENNIFER 2020.




\section*{References}

\bibliographystyle{elsarticle-num}
\bibliography{bibliography}

\end{document}

%% file: author.tex
\author[saclay]{D.~Atti\'e}
\author[ifj]{M.~Batkiewicz-Kwasniak}
\author[ifae]{J.~Boix}
\author[saclay]{S.~Bolognesi}
\author[cern]{S.~Bordoni}
\author[saclay]{D.~Calvet}
\author[bari]{M.G.~Catanesi}
\author[legnaro]{M.~Cicerchia}
\author[padova]{G.~Cogo}
\author[saclay]{P.~Colas}
\author[padova]{G.~Collazuol}
\author[ifj]{A.~Dabrowska}
\author[saclay]{A.~Delbart}
\author[lpnhe]{J.~Dumarchez}
\author[saclay]{S.~Emery-Schrenk}
\author[lpnhe]{C.~Giganti}
\author[legnaro]{F.~Gramegna}
\author[lpnhe]{M.~Guigue}
\author[aachen]{P.~Hamacher-Baumann}
\author[padova]{F.~Iacob}
\author[ifae]{C.~Jes\'{u}s-Valls\fnref{fnref1}}
\author[cern]{U.~Kose}
\author[wut]{R.~Kurjata}
\author[bari]{N.~Lacalamita}
\author[saclay,padova]{M.~Lamoureux}
\author[padova]{A.~Longhin}
\author[ifae]{T.~Lux}
\author[bari]{L.~Magaletti}
\author[legnaro]{T.~Marchi}
\author[padova]{M.~Mezzetto}
\author[ifj]{J.~Michalowski}
\author[ifae]{J.~Mundet}
\author[saclay]{L.~Munteanu}
\author[lpnhe]{Q.~V.~Nguyen}
\author[padova]{M.~Pari}
\author[lpnhe]{J.-M.~Parraud}
\author[bari]{C.~Pastore}
\author[padova]{A.~Pepato}
\author[lpnhe]{B.~Popov}
\author[ifj]{H.~Przybiliski}
\author[bari]{E.~Radicioni}
\author[saclay]{M.~Riallot}
\author[napoli]{C.~Riccio}
\author[aachen]{S.~Roth}
\author[napoli]{A.~C.~Ruggeri}
\author[wut]{A.~Rychter}
\author[geneva]{F.~S\'anchez}
\author[aachen]{J.~Steinmann}
\author[saclay,inr]{S.~Suvorov\fnref{fnref2}}
\author[ifj]{J.~Swierblewski}
\author[ifae]{D.~Vargas}
\author[saclay]{G.~Vasseur}
\author[ifj]{T.~Wachala}
\author[ifj]{A.~Zalewska}
\author[wut]{M.~Ziembicki}
\author[saclay]{M.~Zito}

\address[saclay]{IRFU, CEA Saclay, Gif-sur-Yvette, France}
\address[ifj]{H. Niewodniczanski Institute of Nuclear Physics PAN, Cracow, Poland}
\address[ifae]{Institut de Fisica d'Altes Energies (IFAE), The Barcelona Institute of Science and Technology, Bellaterra Spain}
\address[cern]{CERN, Geneva, Switzerland}
\address[bari]{INFN sezione di Bari, Universit\`a di Bari  e Politecnico di Bari, Italy}
\address[legnaro]{INFN: Laboratori Nazionali di Legnaro (LNL), Padova , Italy}
\address[padova]{INFN Sezione di Padova and Universit\`a di Padova, Dipartimento di Fisica, Padova, Italy}
\address[lpnhe]{LPNHE Paris, Sorbonne Universit\'e, Universit\'e Paris Diderot, CNRS/IN2P3, Paris 75252, France}
\address[aachen]{RWTH Aachen University, III.~Physikalisches Institut, Aachen, Germany}
\address[wut]{Warsaw University of technology, Warsaw, Poland}
\address[napoli]{INFN sezione di Napoli and Universit\`a  Federico II, Dipartimento di Fisica, Napoli, Italy}
\address[geneva]{University of Geneva, Section de Physique, DPNC, Geneva, Switzerland}
\address[inr]{Institute for Nuclear Research of the Russian Academy of Sciences, Moscow, Russia}

\fntext[fnref1]{cjesus@ifae.es}
\fntext[fnref2]{suvorov@inr.ru}

%% file: intro.tex

T2K (Tokai To Kamioka) is a long baseline neutrino oscillation experiment in Japan~\cite{Abe:2011ks} taking data since 2010. An intense beam of muon neutrinos or antineutrinos is produced at the J-PARC accelerator complex (Tokai) and measured twice, 
at a distance of 280 m from the production target with the near detector complex (ND280) and at a distance of 295 km with the water 
Cherenkov detector, Super Kamiokande (SK).

The combined measurement of the neutrino spectra before and after oscillations  allowed to observe for the first time the appearance of electron neutrinos in the muon neutrino beam~\cite{Abe:2011sj,Abe:2013hdq} and recently provided first hints, at  2~$\sigma$ Confidence Level, of charge-parity (CP) violation in the lepton sector by comparing $\nu_{\mu} \rightarrow \nu_e$ and $\overline{\nu}_{\mu} \rightarrow \overline{\nu}_e$ oscillation probabilities~\cite{Abe:2018wpn}.

The ND280 detector used for these analyses consists of a $\pi^0$ detector (P0D)~\cite{Assylbekov:2011sh} which is based on water and scintillator materials with the main purpose to measure neutral  current events and a tracker complex with two fine grained detectors (FGD)~\cite{Amaudruz:2012agx} and three
time projection chambers (TPC)~\cite{Abgrall:2010hi} designed to measure charge, momenta and particle identification for leptons and hadrons produced in neutrino interactions. The P0D and the 
tracker are surrounded by an electromagnetic calorimeter (ECAL)~\cite{Allan:2013ofa} which is enclosed
in the former UA1/NOMAD dipole magnet. In addition, Side Muon Range Detectors (SMRD)~\cite{Aoki:2012mf} are embedded in the iron yokes of the magnet.

To improve the measurements of oscillation parameters, and in particular observe CP violation with more than 3$\sigma$ CL, a continuation of the physics run beyond 2021, named T2K-II, has been proposed by the collaboration. 

The T2K-II proposal consists of an upgrade of the beam power from 485 kW to 1.3~MW~\cite{Friend:2017oav,Abe:2019fux} and an upgrade of the current Near Detector complex~\cite{Abe:2019whr}. 

\begin{figure}[!ht]
    \begin{minipage}{0.49\linewidth}
    \centering
    \includegraphics[width=\linewidth]{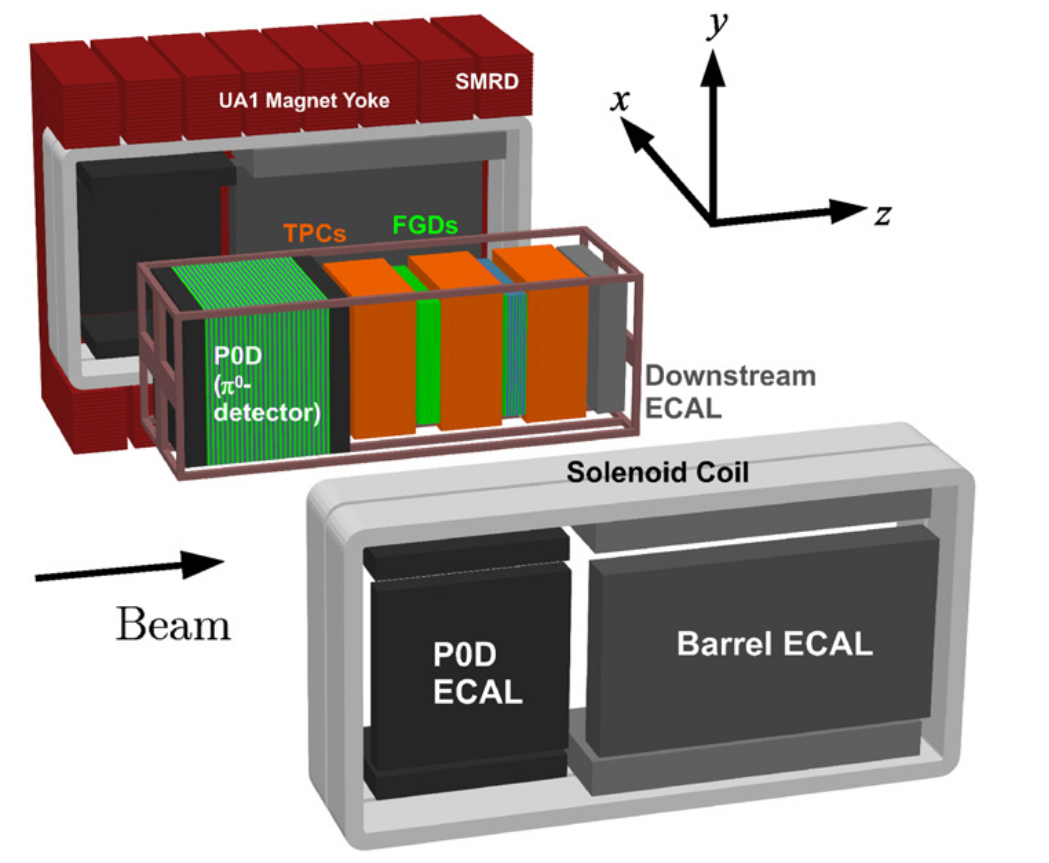}
    \end{minipage}
    \hfill
    \begin{minipage}{0.49\linewidth}
    \vspace{-0.12in}
    \centering
    \includegraphics[width=\linewidth]{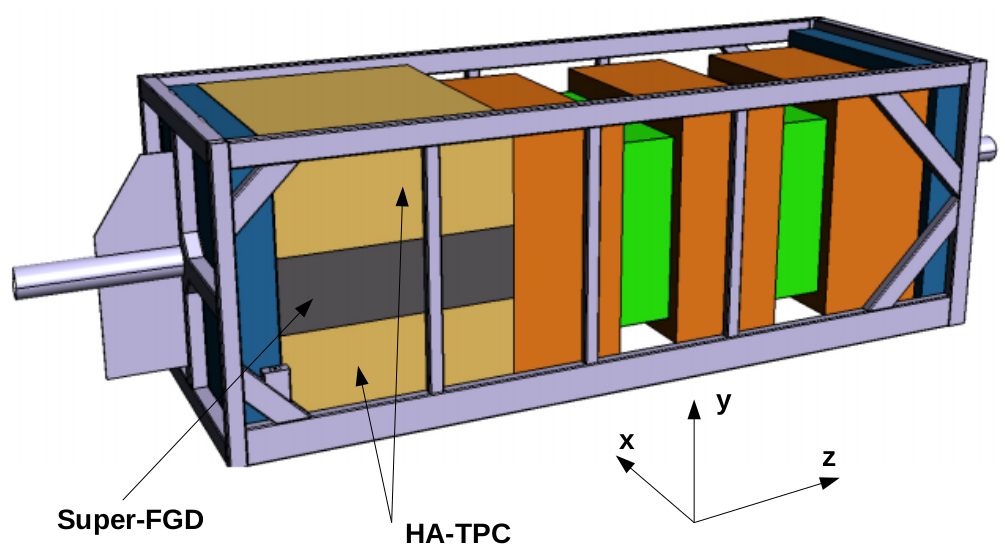}
    \end{minipage}
    \caption{\up{The schematic view of the current Near Detector complex (left) and the proposed upgraded configuration (right) with two additional horizontal TPCs. The ECal, solenoid coil, magnet yoke and SMRD will be kept at place but are not shown on the right scheme.}}
    \label{fig:ndUp}
\end{figure}

For the ND280 upgrade it is foreseen to remove the P0D and to replace it with a new tracker system consisting of a Super Fine Grained Detector (SuperFGD), two horizontal TPCs and a Time-of-Flight (TOF). The upgraded detector will have twice the target mass of the current FGDs, will improve the detection efficiency for leptons emitted at large angles with respect to the neutrino beam direction and will allow for a better reconstruction of the hadronic part of the interaction, thanks to the superior granularity of the SuperFGD. \up{The schematic views of the current and the proposed upgraded Near Detector complex are shown in Figure~\ref{fig:ndUp}.}

For the horizontal TPC a new readout system, based on the Resistive Micromegas (MM) technology is foreseen.
This novel technology will be used for  the first time in a full size experiment and is expected to provide better  performances in respect of spark resistance and either point resolution or reduction
of electronics channels. \up{Some experiments \secRev{(e.g. ATLAS~\cite{Kawamoto2013})} are also planning to use resistive Micromegas. But the ND280 will be the first detector to operate with the charge spreading in two dimensions across the resistive foil.} However, for the usage in \up{T2K}, the ionization energy loss (dE/dx) capabilities 
of the resistive Micromegas for different particles with momenta below 1 GeV/c have to be verified. 

In order to test these capabilities, a resistive MM prototype was designed, constructed and exposed to various charged particles in a testbeam at CERN and the performances are described in this paper.  
In Sect.~\ref{sec:Micromegas} we describe the resistive Micromegas module that was used for the test beam reported in Sect.~\ref{sec:setup}. The track reconstruction algorithms are shown in Sect.~\ref{sec:reco} and in Sect.~\ref{sec:gas} we discuss the gas quality during the test beam. 

In Sect.~\ref{sec:gain} and~\ref{sec:spread} we show the uniformity of the gain of the Micromegas and we  characterize the charge spreading induced by the resistive plane. Finally Sect.~\ref{sec:dedx} and Sect.~\ref{sec:spatial} are devoted to the observed performances in terms of deposited energy and spatial resolution.

%% file: mm.tex


The well known \textit{bulk} technology for building Micromegas detectors~\cite{Giomataris:2004aa} is used for the existing TPCs taking data as part of the ND280 detector since 2009. It provides an easy and robust manufacturing, with very limited dead area, the mesh being fixed on the whole detector surface between two arrays of cylindrical pillars. The result is a permanent dust-tight assembly, for dust particles down to 10 $\mu\rm{m}$ size. 

\begin{figure}[ht!]
    \center{\includegraphics[width=0.8\linewidth]{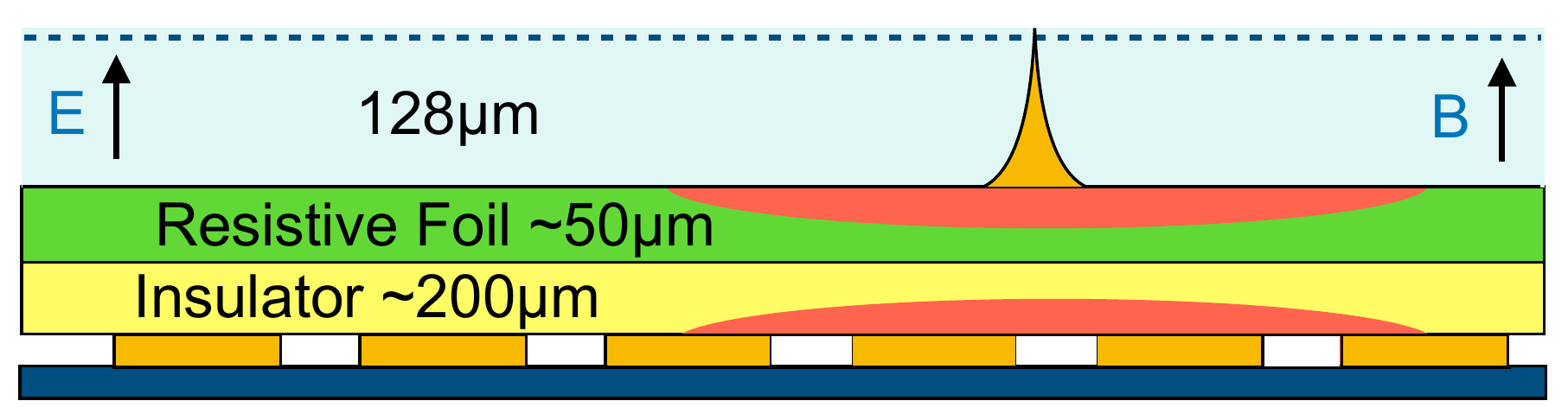}}
    \caption{\up{Schematic cross-section of the resistive MM.}}
    \label{fig:MMconcept}
\end{figure}

An improvement with respect to this technology, the Resistive Anode, has been developed for a Linear Collider (LC) TPC, where the  position resolution requires charge sharing between pads~\cite{Dixit:2003qg}, see Fig.~\ref{fig:MMconcept}. 
In the case of T2K, the resolution in reconstructing neutrino energy is limited by the Fermi motion of nuclei inside the nucleus, thus reducing the benefits of an improved spatial resolution. Nevertheless, resistive Micromegas for T2K will allow to keep the original spatial resolution capabilities by using  fewer and larger pads, thus reducing the number of electronics channels.

The resistive anode provides mainly two advantages: by spreading the charge between neighbouring pads it improves greatly the resolution with respect to the pitch/$\sqrt{12}$ provided by a mere hodoscope, and it suppresses the formation of sparks and limits their intensity. A further and novel improvement of this technique is a new High Voltage powering scheme, where the mesh is set to ground while the anode is set to a positive amplification voltage. The insulation of the resistive anode from the pads, hence from the electronics, allows a safe operation by a capacitive coupling readout, thus allowing to get rid of the cumbersome anti-spark protection circuitry necessary in the case of the standard bulk readout. Another advantage of the anode encapsulation is that the detection plane is fully equipotential, as the grounded mesh is at the potential of the detector frame and supporting mechanics. This allows a better uniformity of the electric field, especially near the module edges, and minimizes  track distortions. Moreover, it provides more flexibility in the operation, allowing the High Voltage on a module to be different from the neighbours without degrading the drift field uniformity. Even if a module has to be disconnected this does not affect the drift field. The goal of this test beam was to demonstrate that the encapsulated scheme with grounded mesh was functional. In parallel, a multi-module test of this technique with the LC TPC design demonstrated the suppression of the distortions~\cite{Colas:2010zz,Attie:2011zz}.

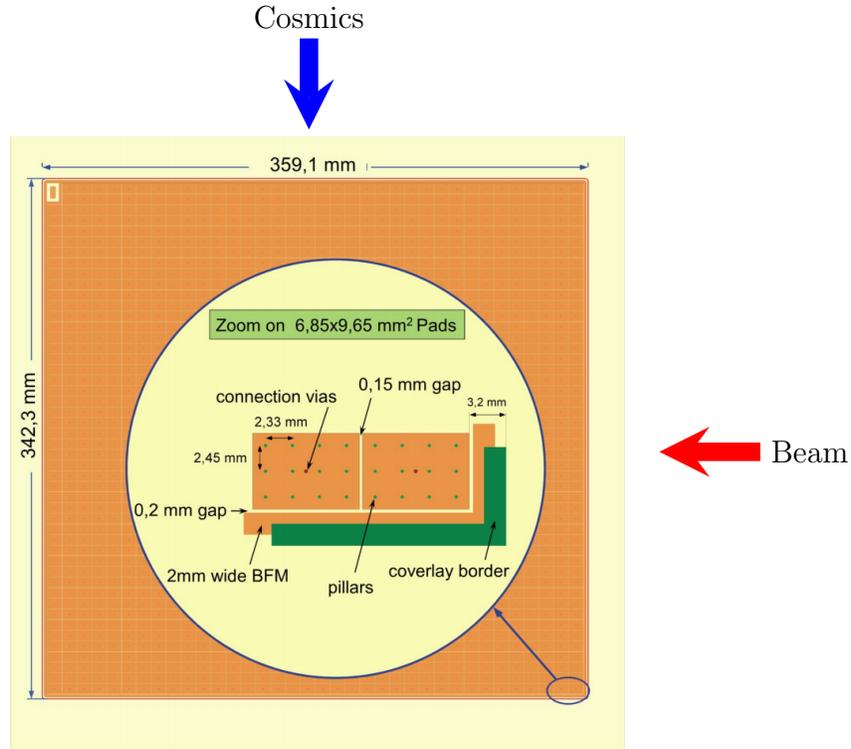
\begin{figure}[!ht]
\centering
\begin{tikzpicture}
\node [anchor=west] (beam) at (10,4) {Beam};
\node [anchor=south] (cosm) at (4,9.5) {Cosmics};
\begin{scope}
    \node[anchor=south west,inner sep=0] (image) at (0.0,0) { \includegraphics[width=0.6\textwidth]{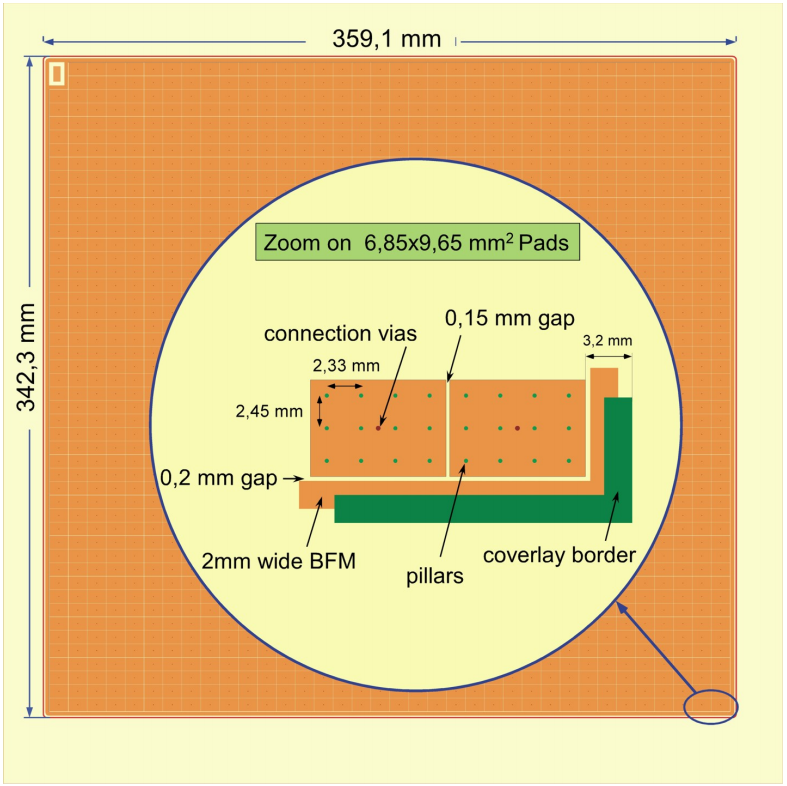}};
    \draw [-stealth, line width=7pt, red] (beam) -- ++(-2.0,0.0);
    \draw [-stealth, line width=7pt, blue] (cosm) -- ++(0.0,-1.5);
\end{scope}
\end{tikzpicture}
\caption{\up{Layout of the pads in the resistive Micromegas. The beam is coming from the right and cosmic rays come from the top. The horizontal line of the pads will be called \textit{row}, the vertical line will be called \textit{column}} }
\label{fig:MMLayout}
\end{figure}

The Micromegas used for the test beam was developed on the basis of the already existing Micromegas PCB used in the present ND280 TPC. This module has a sensitive area of $36 \times 34$~$cm^2$ and is covered by pads of $0.98 \times 0.70$~$cm^2$\up{, Fig \ref{fig:MMLayout} illustrates its layout with respect to the particle beam}. 
The thickness of the PCB is 2.2 mm and comprises three layers
of FR4 with blind vias in the inner layer. The
top conductive layer forming the anode pad plane is made of 25 $\mu m$ thick copper deposited
on FR4. The other conductive layers are used for the routing network, grounding and
pad-readout connectors.
The pad surface was covered by a 200 $\mu m$ insulating layer acting as the capacitance, and
then a 50 $\mu m$ kapton (Apical) with a thin Diamond-Like-Carbon (DLC) layer. The resistivity was 2.5 M\si{\ohm}/$\square$. On top of this surface, a bulk Micromegas was built, with a 128 $\mu m$ amplification gap. \up{ A cross-section scheme for the MM is shown in Fig \ref{fig:MMconcept}.}
The electronics used was the same that had been developed for the T2K TPCs and is based on the AFTER chip~\cite{Baron:2008zza}. 

%% file: setup.tex


\subsection{The setup}
During the test beam, the resistive Micromegas module was installed in the HARP field cage. A detailed description of the HARP field cage can be found in~\cite{prior2003harp}.
A cylindrical volume of 2 m long and 0.8 m diameter \up{hosts} the drift volume. The drift distance is 1.5 m. The field cage is made of Stesalit with a double interleaved strip pattern to avoid electric field inhomogeneities and high field gradients. A foil of individual aluminized Mylar strips has been glued inside the cylinder, and an aluminium foil has been glued onto the outside surface. 

The cathode is at one extremity of the field cage and at about 50 cm from the edge of the external cylinder. On its rear holes it hosts calibration sources. During the TPC operation a voltage of 25~kV \up{was} applied to the cathode generating an electric field in the drift volume of 167 V/cm. On the extremity opposite to the cathode a circular flange \up{closes} the cylinder where the Micromegas MM0 is installed.

The TPC has been operated using a premixed gas with $Ar:CF_4:iC_4H_{10}$ (95:3:2) 
which is the same mixture as used for the existing ND280 TPCs. A simple gas system with only one line has been set up to operate the detector. Before starting the data taking the TPC has been extensively flushed with Nitrogen gas. Later it has been flushed with the gas mixture for several hours (3 or 4 times the volume of the TPC) to remove impurities. During normal operation the gas flow was kept of about 25 litres/\up{hour}. 
\up{Temperature measurements on the exhaust line were taken to monitor environmental conditions which affect gas density and thus the performance of the detector in, for instance, the electron drift velocity.}

\subsection{The trigger}
\begin{figure}[ht]
    \centering
    \begin{minipage}[h]{0.49\linewidth}
        \includegraphics[width=\linewidth]{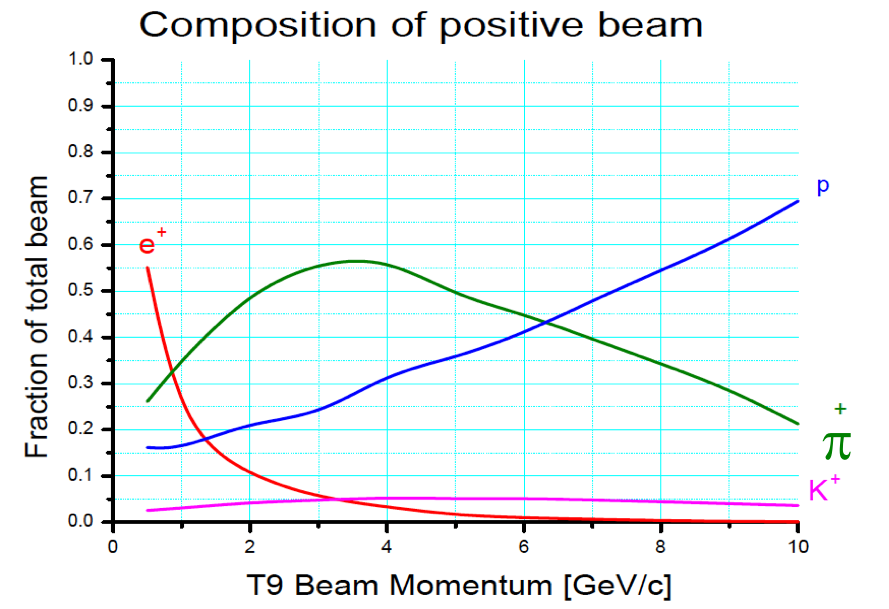}
    \end{minipage}
    \begin{minipage}[h]{0.49\linewidth}
        \includegraphics[width=\linewidth]{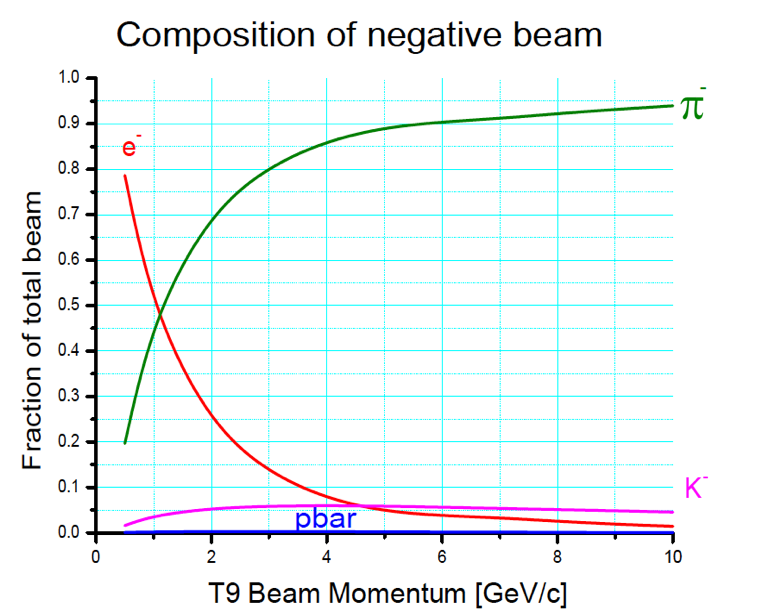}
    \end{minipage}
    \caption{Breakdown of the beam composition as a function of the momentum for hadron enriched beams at the T9 area at CERN. 
    }
    \label{fig:BeamSpectrum}
\end{figure}

Test beam data were taken using the T9 beamline with copper target to have an ``hadron enriched beam configuration''. The breakdown of the beam composition as a function of the momentum is shown in Figure~\ref{fig:BeamSpectrum}. Positively charged tracks were used for the analyses presented in this paper.

The beam composition at low energies is largely dominated by electrons and positrons. \up{From the Figure~\ref{fig:dedx_spectrum_all} one could see that the pion trigger accepts many of them.} The use of copper target allows to reduce the \up{electron} contamination by about a factor of 8.

\begin{figure}[ht]
    \centering
    \includegraphics[width=\linewidth]{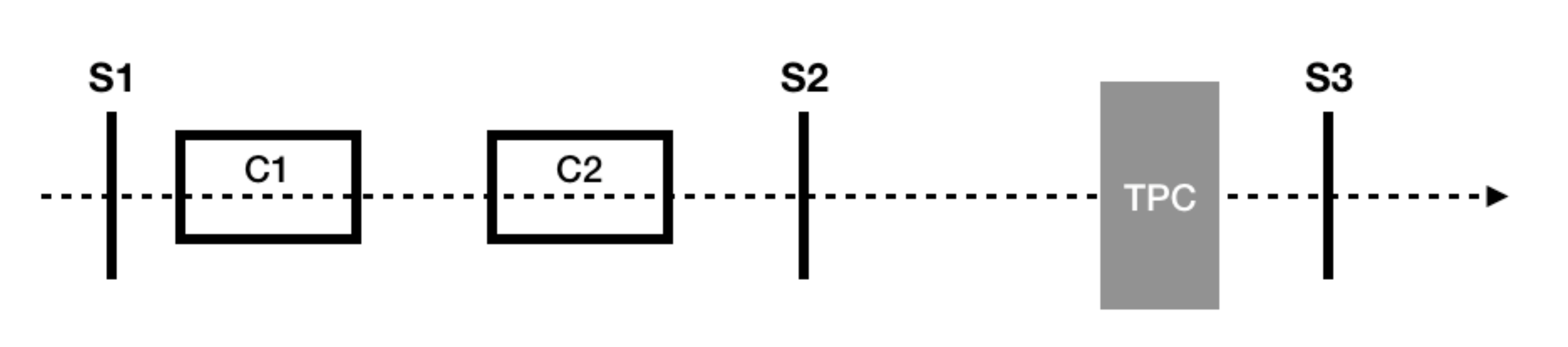}
    \caption{Sketch of the detectors used to tag the particles from the beam and set up trigger selections. The beam is coming from the left}
    \label{fig:triggerScheme}
\end{figure}

Charged particles have been tagged using three plastic scintillator detectors called respectively S1, S2, S3 and two Cherenkov detectors called C1 and C2. Figure~\ref{fig:triggerScheme} shows a cartoon of the locations for those detectors along the beamline. The selection for the different particle types was done by the combination of the NIM signals coming from those detectors. A summary of the different options is presented in Table~\ref{tab:trigger_sum}. 

To trigger on cosmic rays going through the TPC two scintillator plastic panels have been
positioned on the top and the bottom of the TPC. The surface is coated with reflective paint and two grooves are made to host fibres. The readout is done at one side with Hamamatsu MPPC while the other end-side of the fibres is made reflecting. Plastic bars are installed inside aluminum boxes to get good light-tightness conditions.

\begin{table}[ht]
    \centering
    \begin{tabular}{ll} \hline \hline
        Particle            &  Selection\\ \hline
        Positrons           & Scintillators + Cherenkov \\ 
        Protons (+Kaons)    & S1(delayed) * S2 (delay proton TOF between S1 and S2) \\
        Pions (+ muons)     & Scintillators $\overline{\mbox{protons}}$ * $\overline{\mbox{positrons}}$ \\
        Cosmic ray          & From the scintillators panels (only out of spill) \\ \hline \hline
    \end{tabular}
    \caption{Summary of the different signal combinations used to tag different beam particles.}
    \label{tab:trigger_sum}
\end{table}

\subsection{Collected data}
\label{sec:DataSamples}
In the standard condition, the Micromegas was operated with a voltage of 340 V. The settings chosen for the AFTER chip were a sampling time of 80~ns, a shaping time of 600~ns \up{ and the charge to saturate the ADC was 120 fC. Considering that $^{55}$Fe charge signature is larger than that of any of the tracks in this study and looking at \secRev{Fig~\ref{fig:gain_HV}}, where even at 380V the  gain kept increasing exponentially, we can be sure that this value was large enough to avoid saturation at all tested voltages}. 

The results described in this paper were obtained with proton, pion and electron triggers with momenta of 0.8~GeV/c and at three different drift distances (10, 30 and 80~cm). 

In addition we performed a scan of the MM voltage that was varied from 330 to 380~V and we took data with a 1 GeV/c pion trigger with different voltages applied to the detector (360 and 380~V). Finally also the shaping time was varied to 100 ns and 200 ns.  

During all the beam data taking period we were continuously collecting data with the cosmic trigger. In addition a radioactive source of $^{55}Fe$ \up{was} positioned at the cathode to have a reference point for energy calibration. 


%% file: reco.tex

As described in the previous section, different types of data were collected at the same time, including cosmics, beam, and a $^{55}Fe$  source. A simple 3D track reconstruction algorithm was developed in order to remove noise and distinguish between cosmics and beam tracks.
\begin{figure}[!ht]
    \begin{minipage}[b]{0.49\linewidth}
        \center{\includegraphics[width=\linewidth]{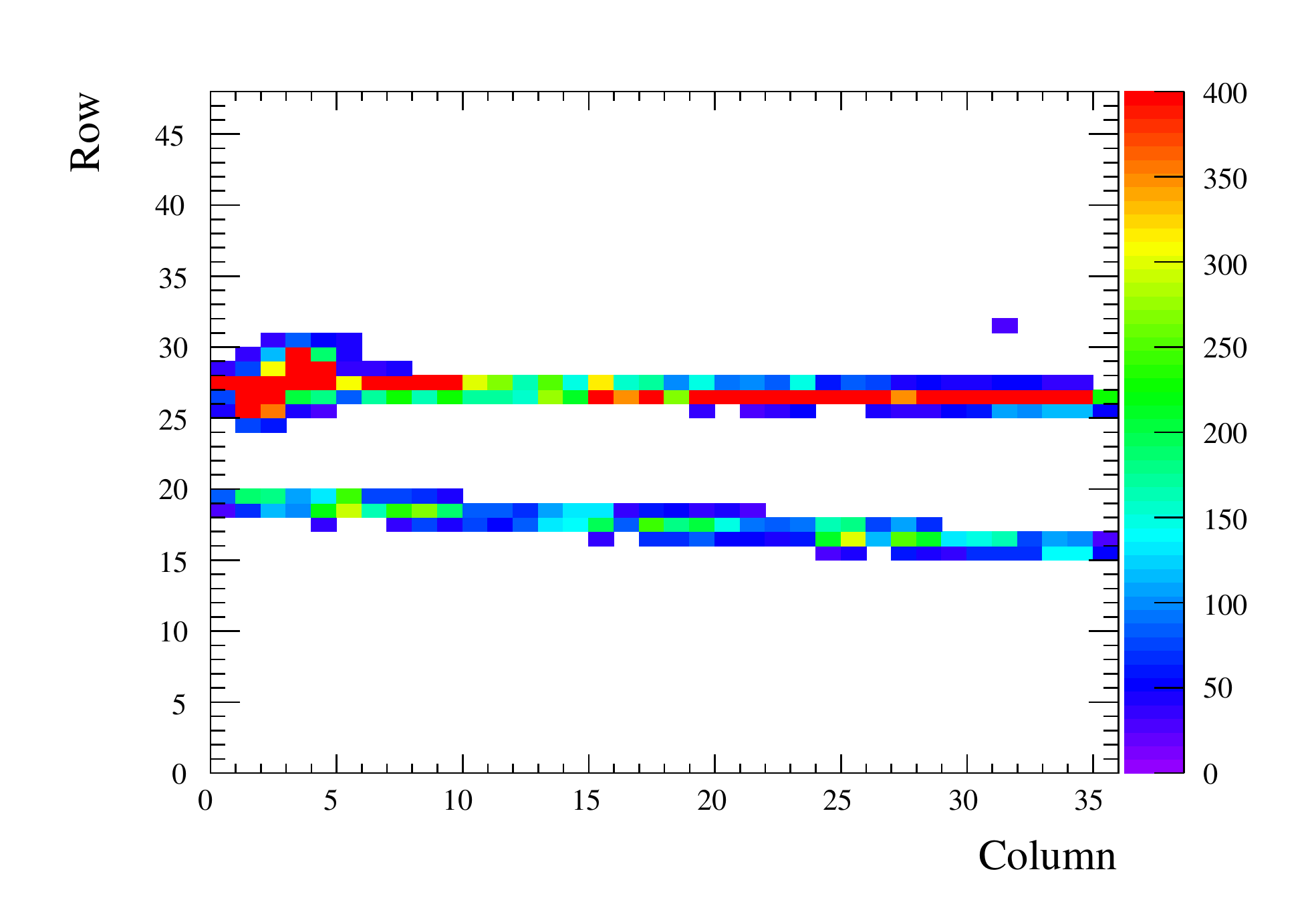} \\ a) Raw event}
    \end{minipage}
    \hfill
    \begin{minipage}[b]{0.49\linewidth}
        \center{\includegraphics[width=\linewidth]{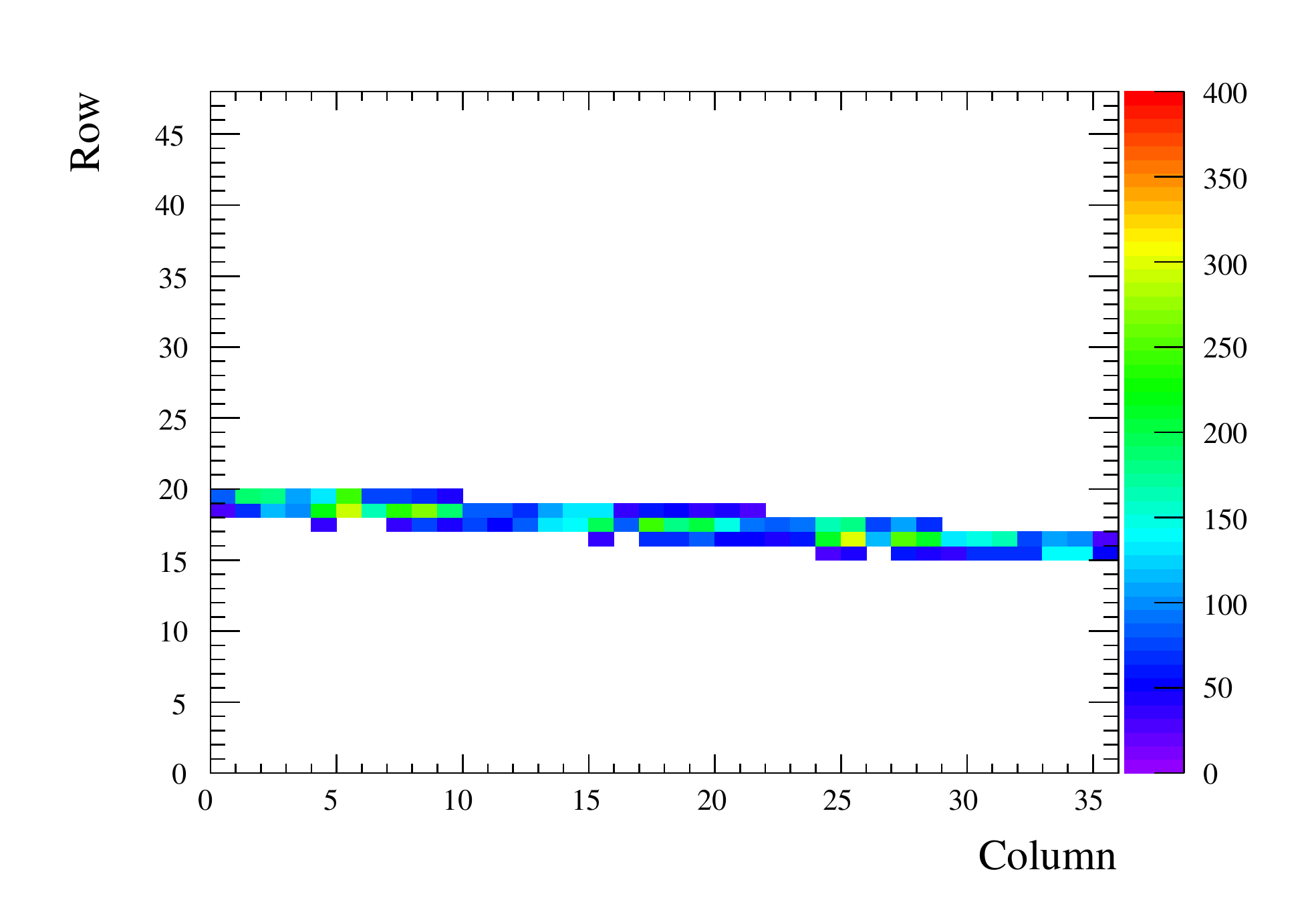} \\ b) One of the selected tracks}
    \end{minipage}
    \caption{The example of the track selection algorithm performance, that is based on the track start and end position matching.}
    \label{fig:sel3D}
\end{figure}
The reconstruction \up{looks for} the pads at the beginning and at the end of the module. \up{If it is possible to connect clusters at both sides of the MM with a straight line of hits, then this is supposed to be a track}. Such \up{simple} algorithm showed enough purity for beam and cosmic track separation and also for removal of noise and hits due to the $^{55}Fe$ source. An example of reconstructed tracks is shown in figure~\ref{fig:sel3D}.

\begin{figure}[!ht]
    \center{\includegraphics[width=0.9\linewidth]{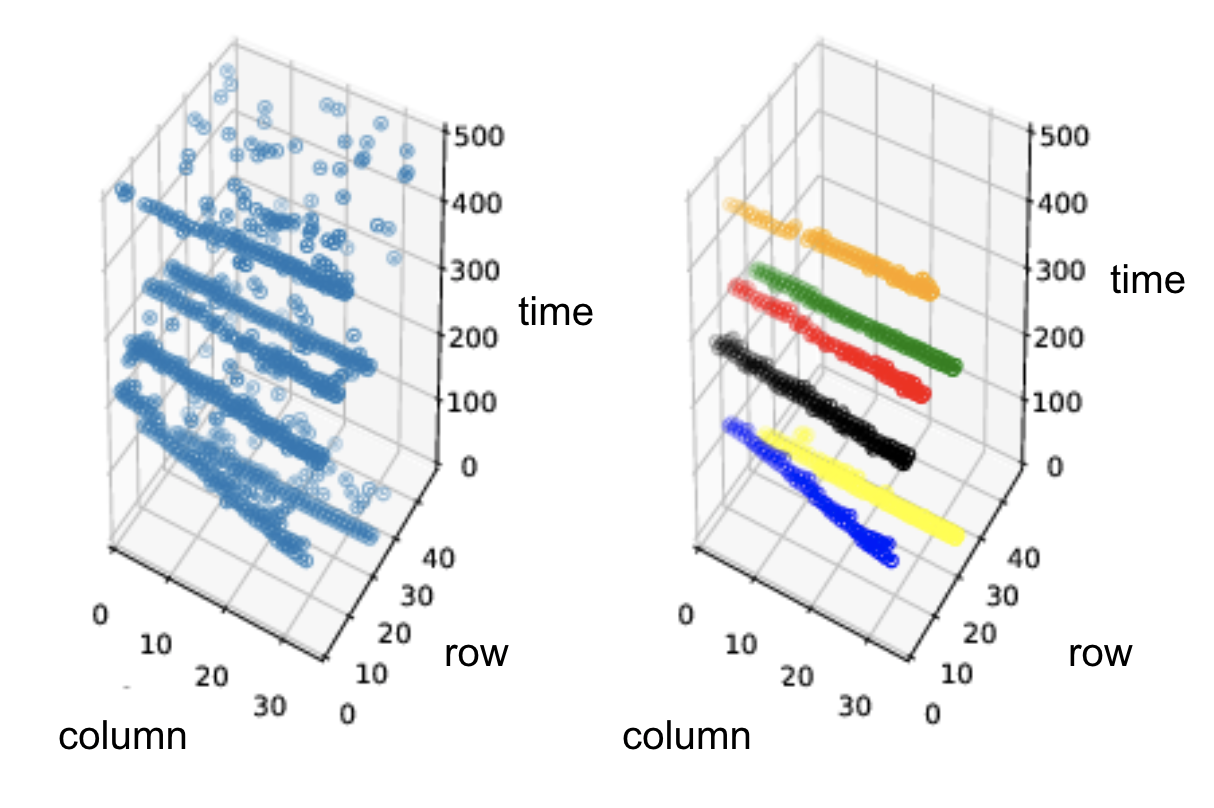}}
    \caption{Example of the DBSCAN 3D matching algorithm performance. The time is expressed in units of 80 ns.}
    \label{fig:selDBSCAN}
\end{figure}

We also considered the selection based on the 3D matching algorithm DBSCAN~\cite{Ester96adensity-based} (Figure~\ref{fig:selDBSCAN}). Both selections provide similar results in terms of number of reconstructed tracks and performances for point and energy resolution. 

%% file: gas.tex

The gas quality has been analyzed after the data taking by measuring \up{the electron attenuation length and the drift velocity.}

\subsection{Attenuation length}
\label{sec:att}

\begin{figure} [ht!]
  \centering
\begin{subfigure}[b]{.49\textwidth}
  \centering
  \includegraphics[width=1\linewidth]{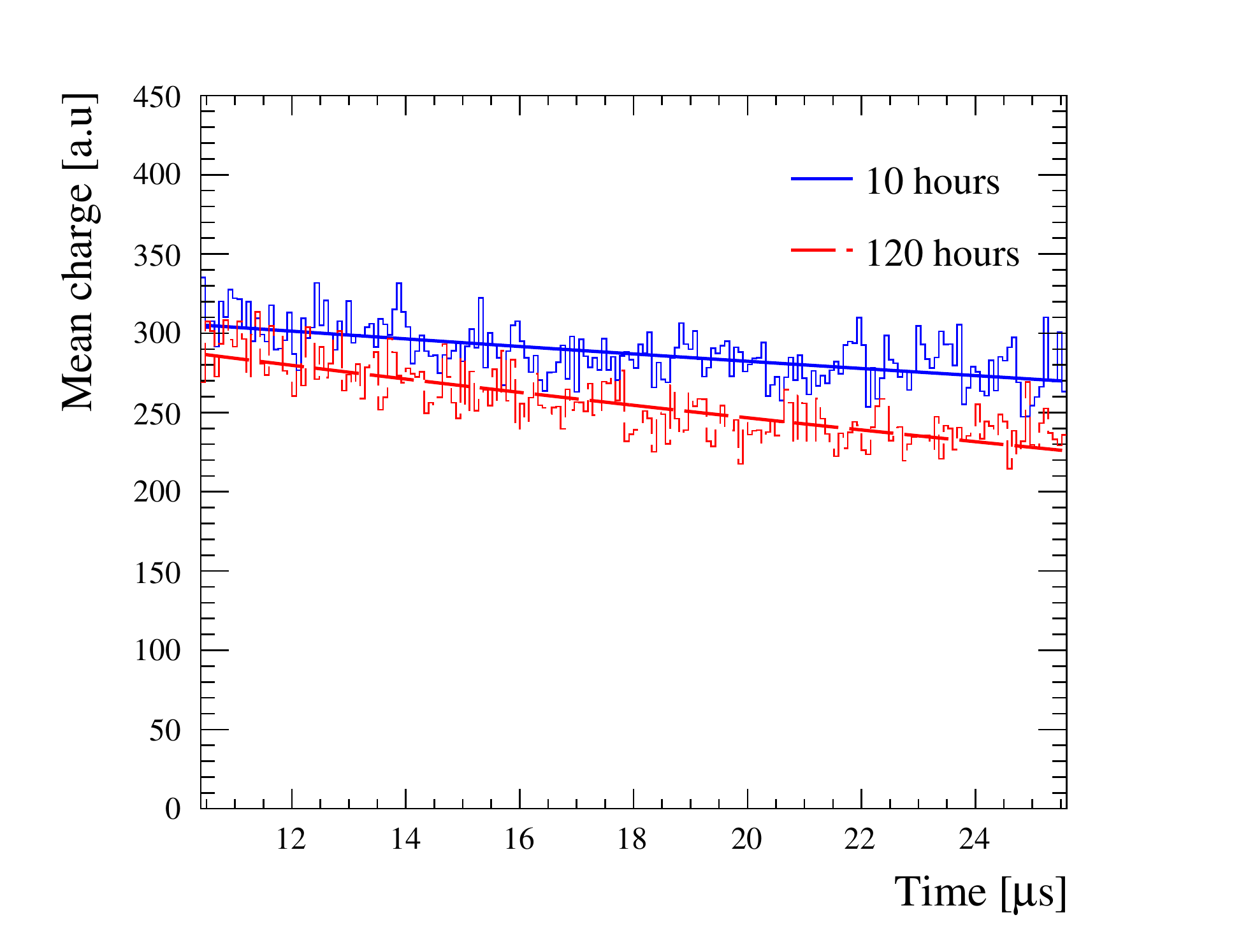}
  \caption{}\label{fig:ele_att_timeA}	

\end{subfigure}
  \begin{subfigure}[b]{0.49\textwidth}
  \centering
  \includegraphics[width=1\textwidth]{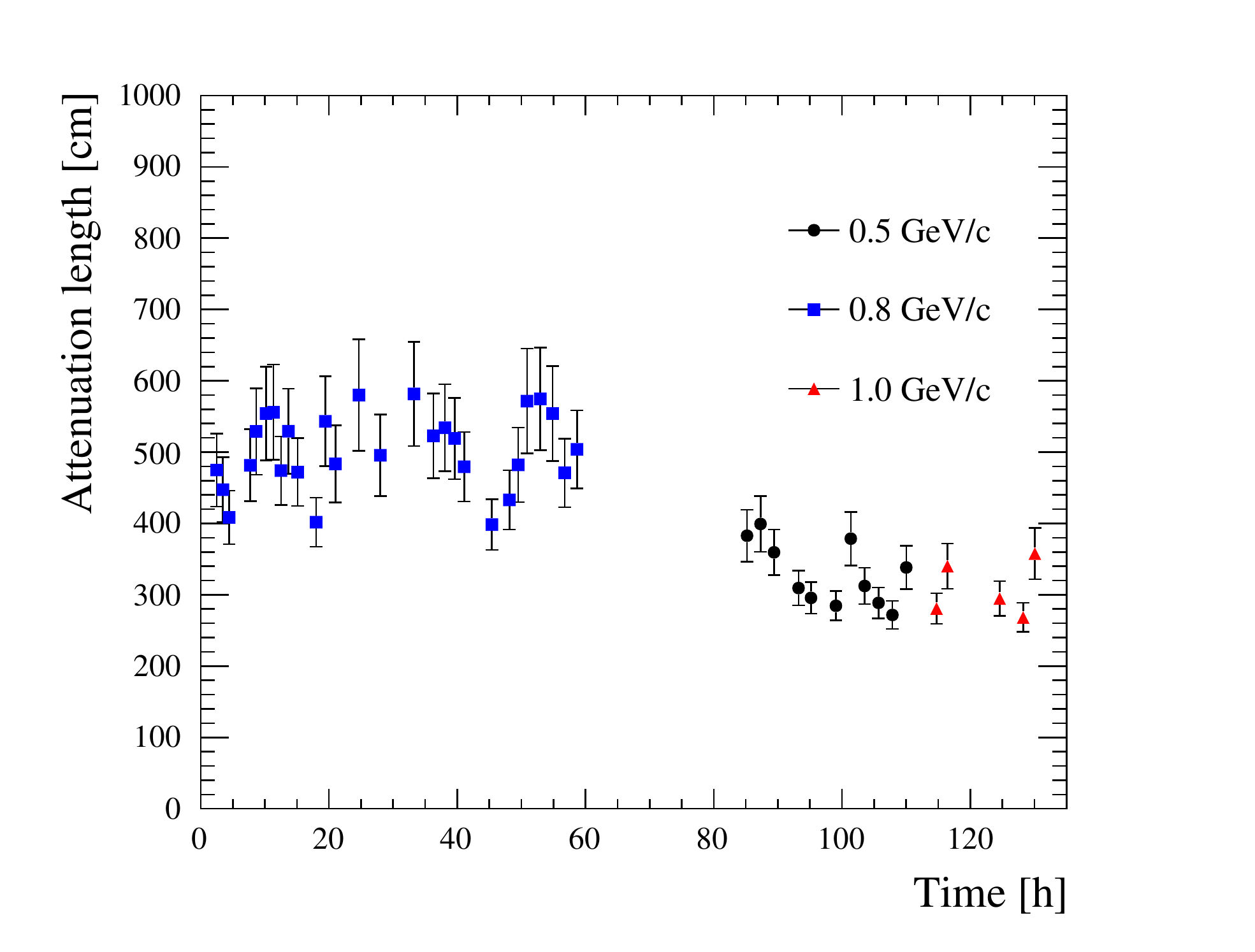}
  \caption{}\label{fig:ele_att_timeB}	
\end{subfigure}%
  \caption{A) Average charge versus drift time for 2 sets of data taken at the beginning and at the end of the test beam period. B) Evolution of the signal attenuation length during the data taking.}
   \label{fig:ele_att}
\end{figure}

\up{The attenuation length of the electrons while drifting in the gas is an indicator of the gas quality. To measure it,} vertical cosmic tracks at different drift distances have been selected. At each distance, the average measured charge has been computed and the resulting distribution fitted with an exponential function as presented in Figure~\ref{fig:ele_att_timeA}. \up{To get the attenuation length into physical units the parameters of the fit in ADC time bins have been converted into a distance by using the drift velocity of each run and the sampling rate.} The obtained attenuation length as a function of the run time is shown in Figure~\ref{fig:ele_att_timeB}.

The precision on the measurement of the attenuation length is limited by the low acceptance for cosmic tracks close to the anode, nonetheless, a reduction of the attenuation length is observed indicating a progressive \up{degradation of the gas quality. The attenuation, however, was reasonably small in all runs being always at least twice the maximum drift length.}

\subsection{Drift velocity}
\label{sec:vel}
\begin{figure} [!ht]
  \centering
\begin{subfigure}[b]{.49\textwidth}
  \centering
  \includegraphics[width=1\linewidth]{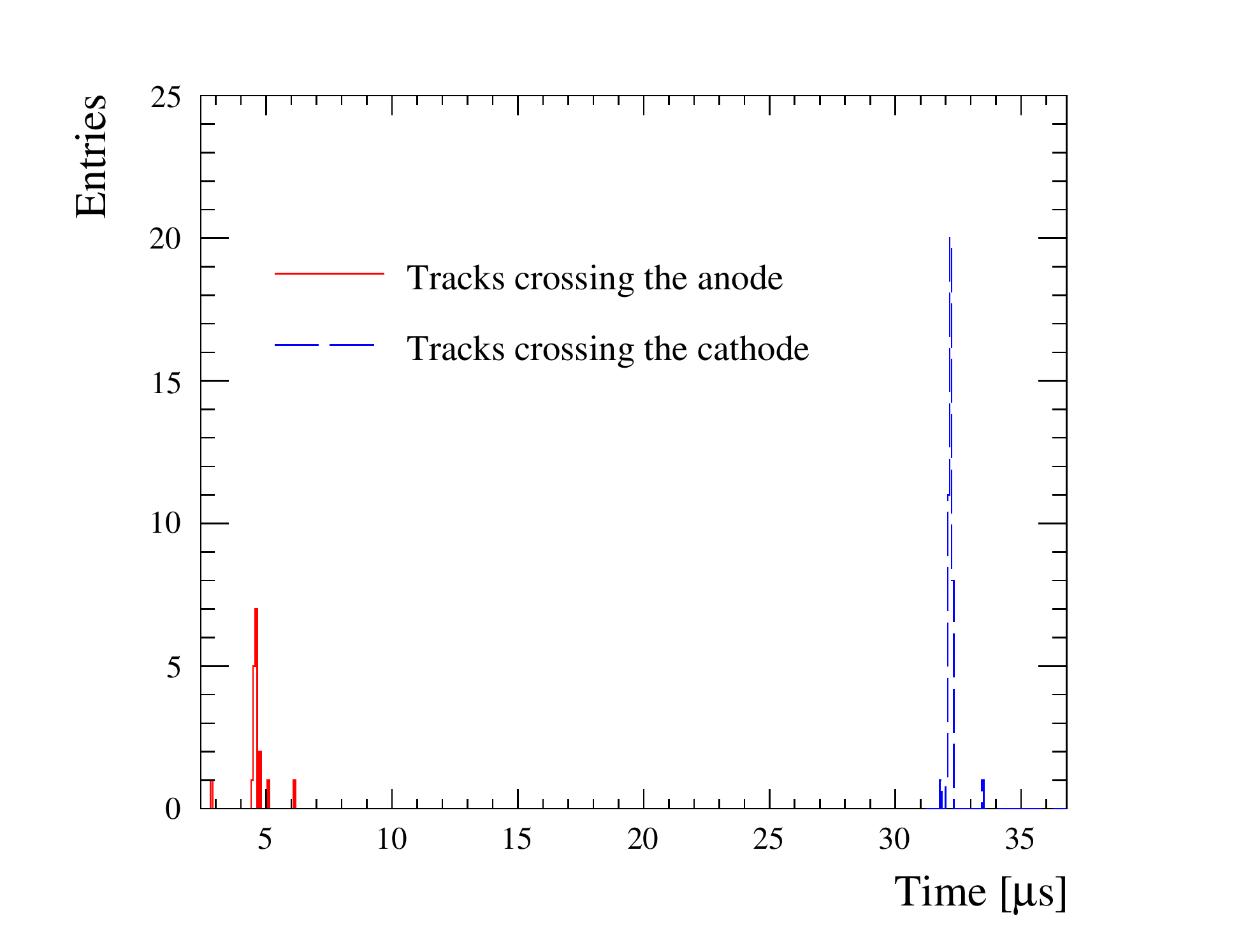}
  	\caption{}\label{fig:drift_vel_time}	
\end{subfigure}
  \begin{subfigure}[b]{0.49\textwidth}
  \centering
  \includegraphics[width=1\textwidth]{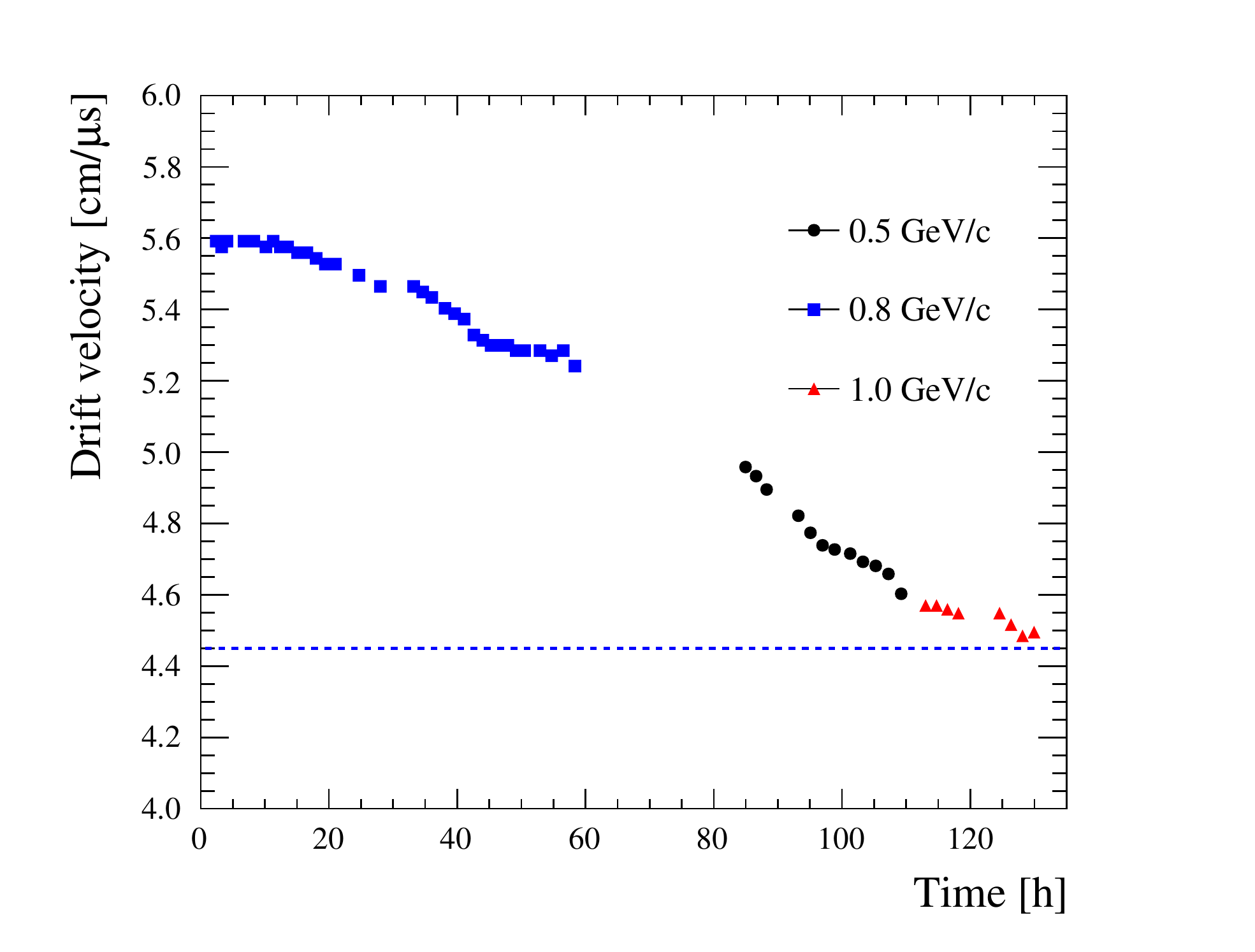}
  \caption{}\label{fig:drift_vel_evolution}	
\end{subfigure}%
\caption{a) Time distribution of tracks crossing the end caps of the chamber for a set of 25k triggers. b) Drift velocity time evolution, the dashed blue line is set at the minimum measurable value. }
  \label{fig:drift_vel}
\end{figure}

The drift velocity can be computed by using the arrival time on the Micromegas of cosmic tracks crossing the cathode 
or the anode, as shown in Figure~\ref{fig:drift_vel_time}. The different amount of tracks at each end was caused by the significantly lower acceptance of the cosmic trigger for tracks close to the anode. 

The evolution of the drift velocity as a function of time is shown in Figure~\ref{fig:drift_vel_evolution}. We observed a smooth reduction of the drift velocity during the data taking, until we reached a point where the end of the chamber could no longer be seen within the acquisition window. Given the electronics sampling rate and the available time bins, the minimal velocity to observe the whole chamber is 4.45~cm/$\mu$s.

\begin{figure} [ht!]
  \centering
  \includegraphics[width=0.5\linewidth]{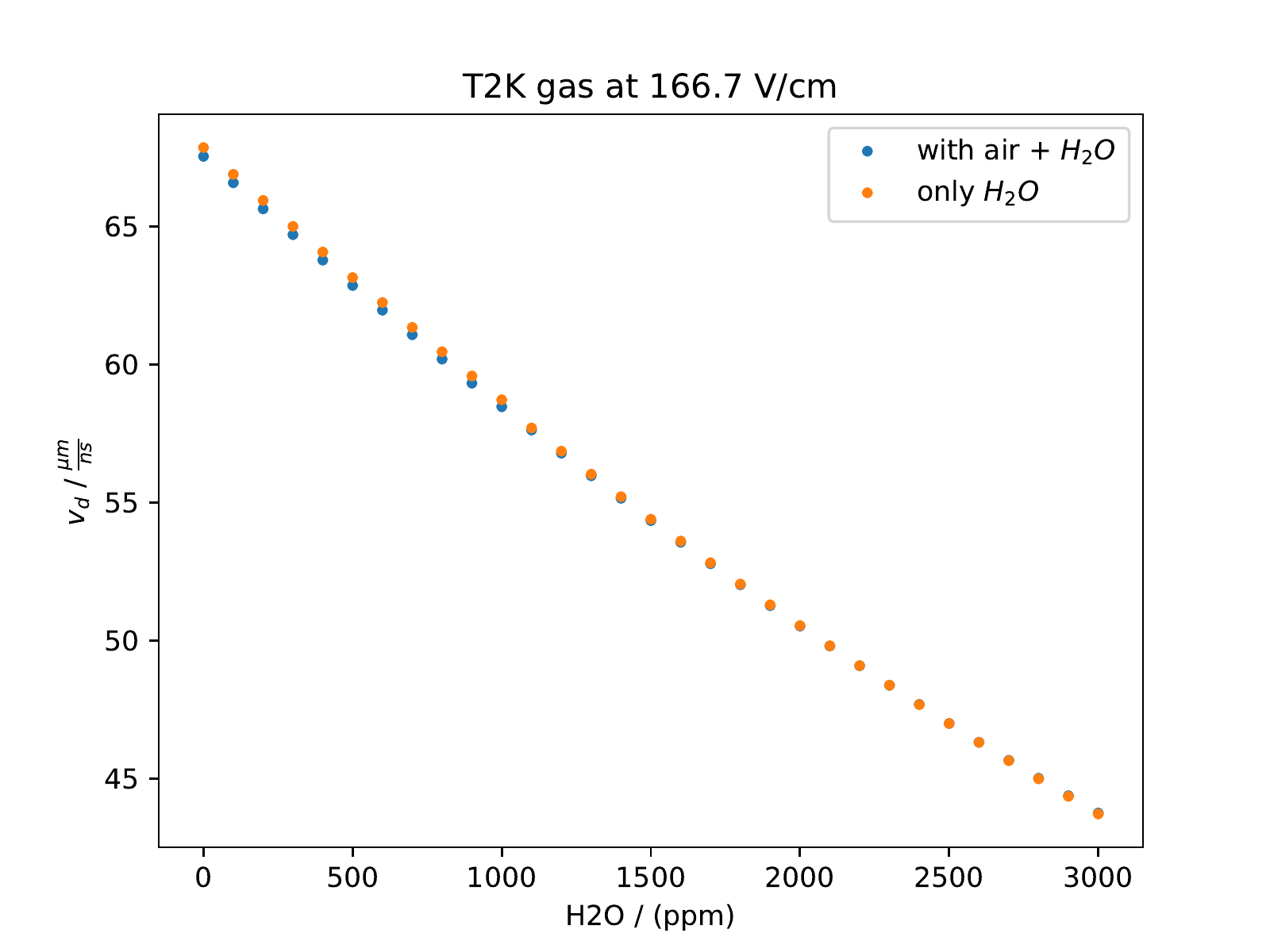}
  \caption{\up{Simulated drift velocity as a function of the water concentration in the TPC using T2K gas and 166.7V/cm electric field.}}\label{fig:gasSim}	
\end{figure}

\up{For the gas mixture that was used under the electric field of 167V/cm, simulations show that the expected drift velocity was 6.8~cm/$\mu$s. However, the maximum drift velocity was observed to be 5.6~cm/$\mu$s.} The observed reduction of the drift velocity is \up{compatible with the presence of $H_2O$ in the gas, as shown in Fig \ref{fig:gasSim}. The increase of the concentration of water with time from 1500~ppm to 3000~ppm explains by itself such a reduction. The mild degradation of the attenuation length in section \ref{sec:att} is also explained by the increase of water in the gas, given that slower drift velocities expose the electrons longer to the attachment effects, thus, decreasing the attenuation length. Given that the TPC was operated in overpressure and that the degradation in the attenuation is not abrupt suggests that the HARP TPC field cage released water in the gas with time. A large humidity on the inner side of the field cage is likely due to the fact that the HARP TPC was stored in air for a long time. The role of humidity was underestimated and the TPC was not dried before bringing it back to operation. The increase of the $H_2O$ ppm inside the gas is explained by the operation procedure during the data taking: initially the gas was flown at 60 litres per hour (L/h), however, the operation was thought to be stable and safe due to over-pressure, and the decision of reducing the flow of gas to 30 L/h (25 L/h) after 30 (35 hours) of the run to save gas would have allowed the concentration of water to double degrading the gas quality with time. } 

%% file: gain.tex

The Micromegas gain and its uniformity are key elements for successful particle identification in the TPC. The gain of the resistive Micromegas module was measured with a $^{55}$Fe source, emitting 5.9 keV gammas, installed at the center of the cathode. The uniformity of the gain was measured with cosmic tracks. 

\subsection{Time evolution of the gain}

\begin{figure}	[!ht]
	\centering
	\begin{subfigure}[b]{0.49\textwidth}
		\centering
		\includegraphics[width=\textwidth]{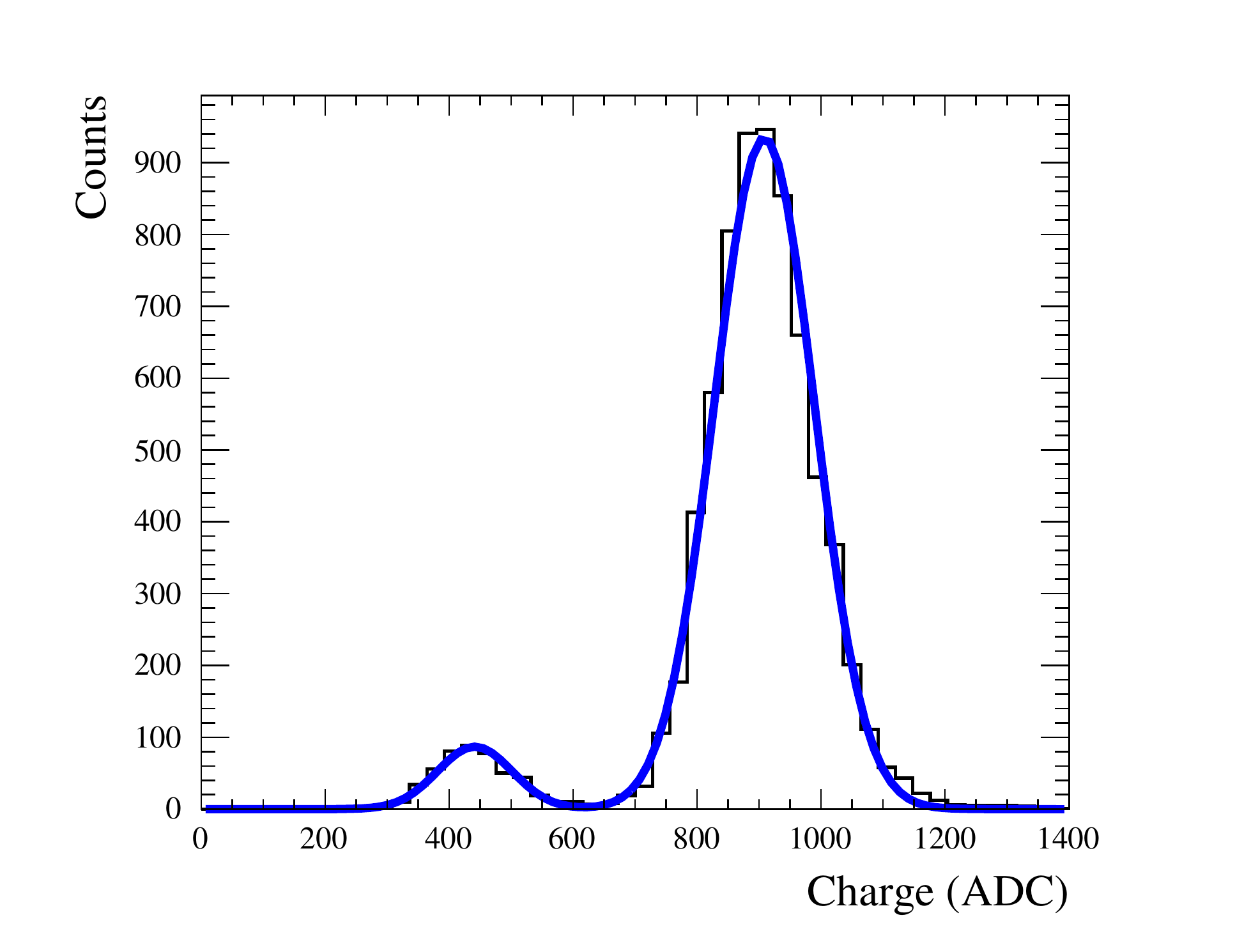}
		\caption{}\label{fig:gain_Fe}		
	\end{subfigure}
	\begin{subfigure}[b]{0.49\textwidth}
		\centering
		\includegraphics[width=\textwidth]{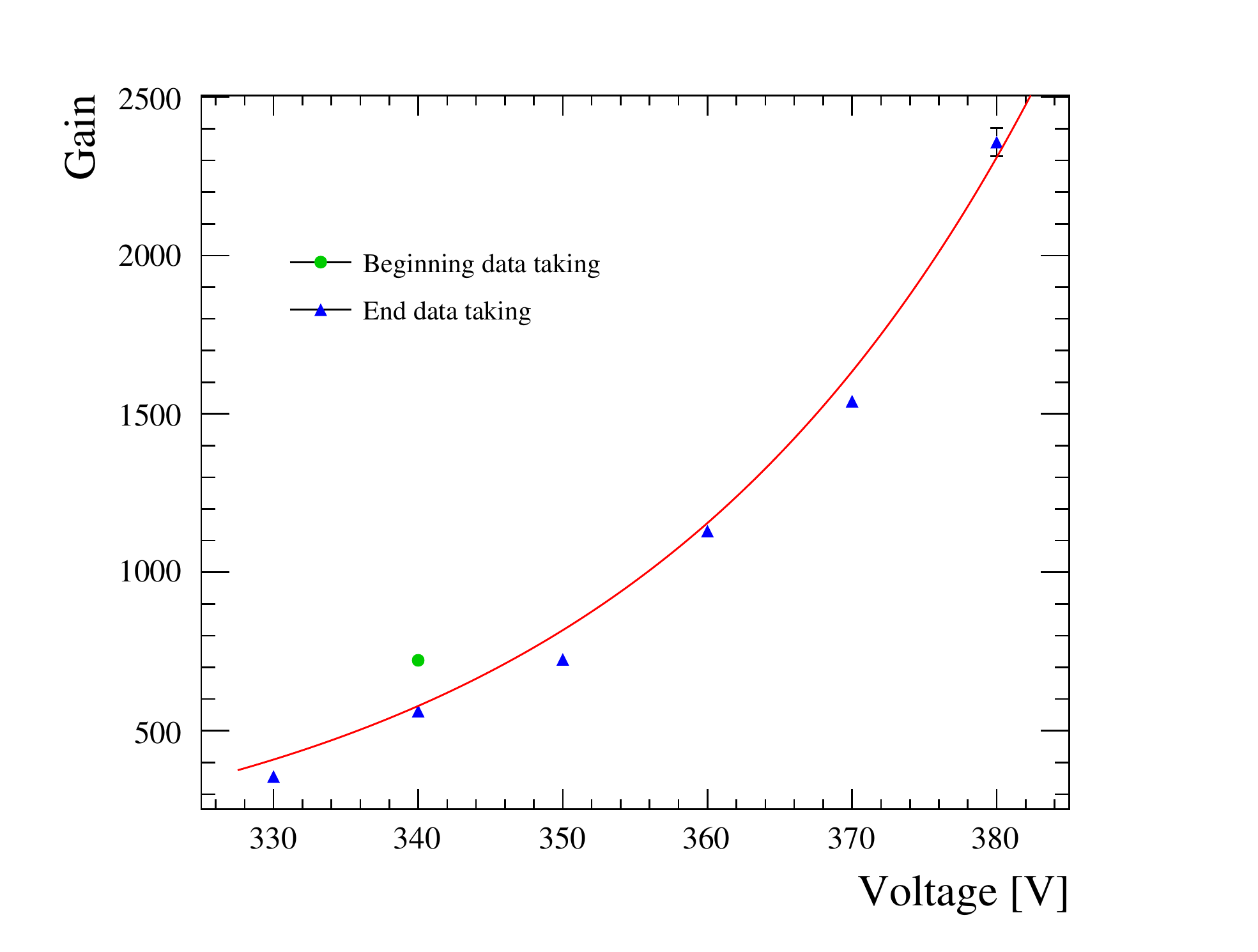}
		\caption{}\label{fig:gain_HV}
	\end{subfigure}
	\\
	\begin{subfigure}[b]{0.49\textwidth}
		\centering
		\includegraphics[width=\textwidth]{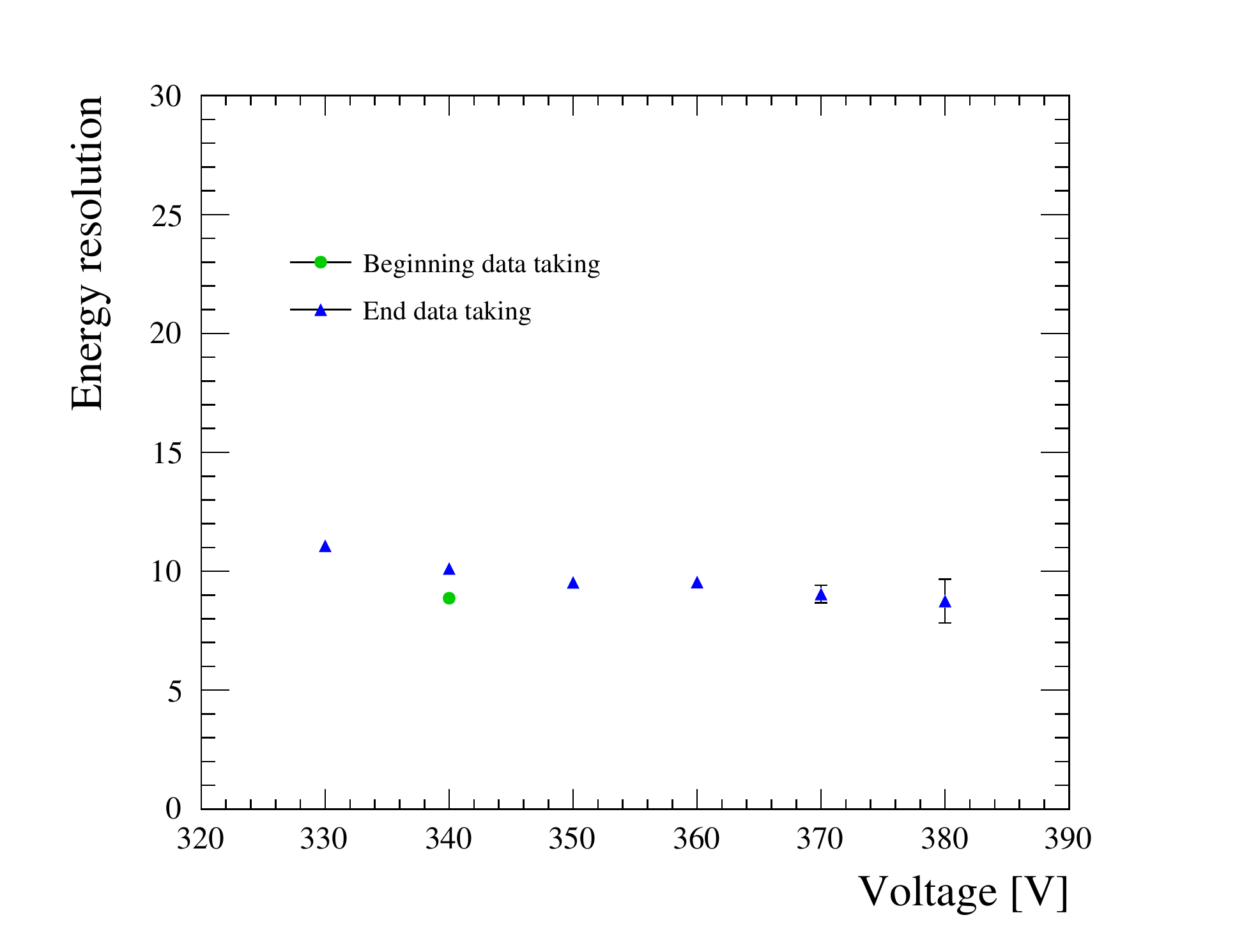}
		\caption{}\label{fig:EnergyRes_HV}		
	\end{subfigure}
	\begin{subfigure}[b]{0.49\textwidth}
		\centering
		\includegraphics[width=\textwidth]{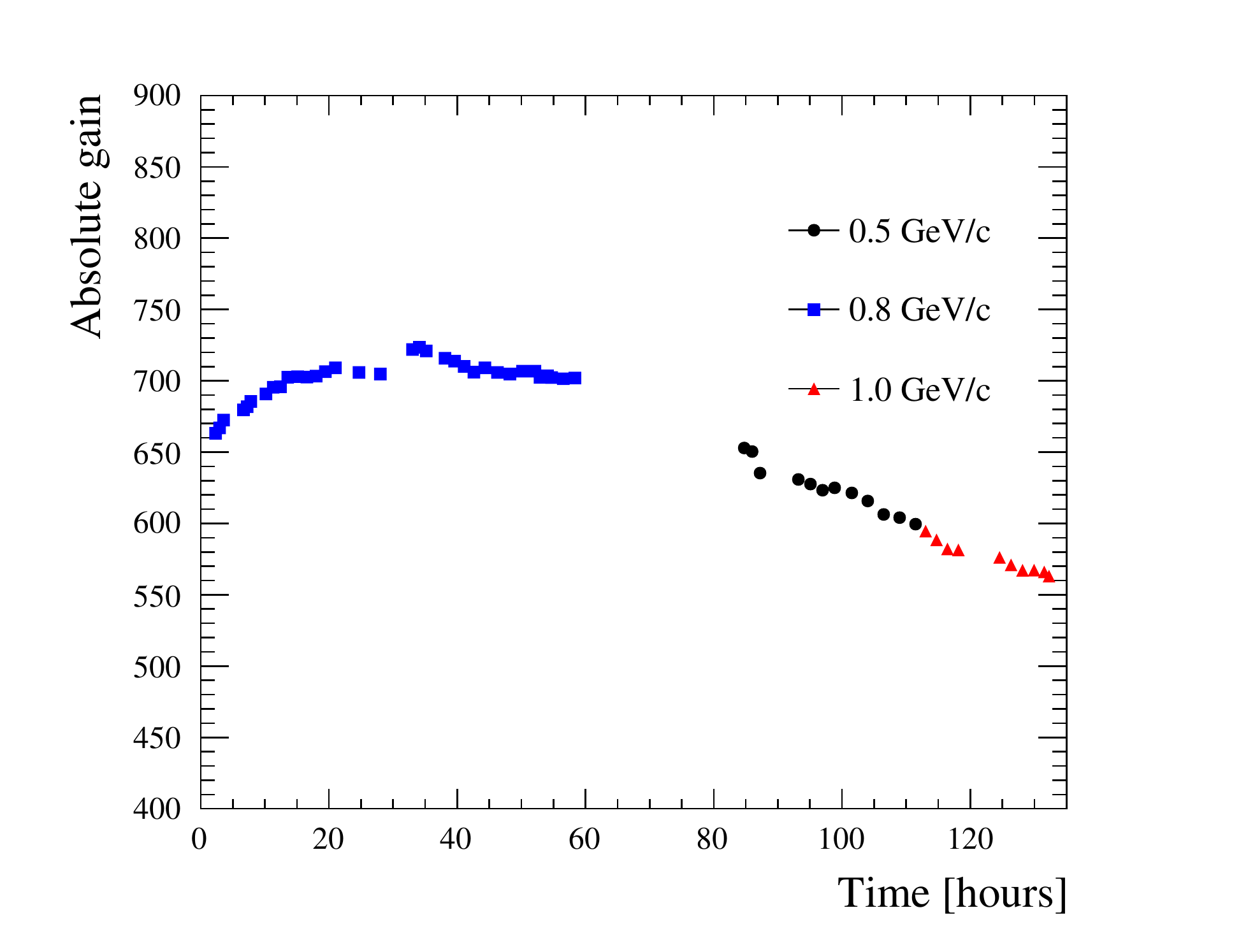}
		\caption{}\label{fig:gain_evolution}
	\end{subfigure}
	\caption{a) X-ray energy spectrum. The energy resolution at 5.9 keV is 8.9$\%$. b) Micromegas gain for different micromesh voltages. c) 5.9 KeV resolution for different micromesh voltages.  d) Time evolution of the gain during the data taking.}\label{fig:gain}
\end{figure}

  A $^{55}$Fe source was placed on a transparent mylar foil window on the cathode to measure the MM gain. \up{Events with an isolated group of fired pads were selected as $^{55}$Fe signals and the total charge of the pads was used to fill a histogram. \secRev{The rate was low enough to have a single gamma deposit for each selected group of pads.}} This radioactive source has a characteristic photon emission at 5.9~keV, as shown in Figure~\ref{fig:gain_Fe}. A resolution of about 8.9\% was obtained, allowing to observe also the 2.9~keV escape line in argon. 

This value is obtained with a source at a distance of $\sim1.5~m$ from the Micromegas module and cannot be directly compared with the results for bulk Micromegas shown in~\cite{Abgrall:2010hi}, obtained with -350 V on the Micromegas and with the source placed close to the read-out plane and hence not affected by distortions or by electron attachment. 

\secRev{In order to measure the gain the ratio between the number of primary electrons and the number of measured electrons is computed. The first is estimated dividing the known $^{55}$Fe gamma deposit by the mean ionization energy for the was mixture used in the TPC. The later is computed converting the ADC mean value in Figure~\ref{fig:gain_Fe} to electrons using the electronics specifications in section \ref{sec:DataSamples}.}
The dependence of the gain and of the resolution on the Micromegas voltage, ranging from -330~V to -380~V is shown in Figure~\ref{fig:gain_HV} and Figure~\ref{fig:EnergyRes_HV}. The gain grows exponentially with the voltage with values for the absolute gain ranging from 500 to 2500 and the resolution improves. 
The voltage scan, however, was performed at the end of the test beam, when the gas quality was lower, affecting both the gain and the energy resolution. To have a reference, the data from Figure~\ref{fig:gain_Fe} taken at -340~V and at the beginning of the data taking is also shown. 

The evolution of the absolute gain, for data taken with a Micromegas voltage of -340~V  is presented in Figure~\ref{fig:gain_evolution}. A decrease of the gain of \secRev{$\sim20\%$} during the data taking was observed. 
\secRev{ This reduction is compatible with the gas quality degradation explained in the section \ref{sec:vel}. The increase of H$_2$O in the gas slows down the accelerating electrons during the avalanche process reducing the gain. In addition, variations on the room temperature and atmospheric pressure, which were not strictly controlled, could also have contributed to small gain fluctuations.}

\subsection{Gain uniformity}
The iron source cannot be used to study the uniformity of the gain, since it illuminates only the central pads of the module.

\begin{figure} [!ht]
    \centering
        \includegraphics[width=0.8\textwidth]{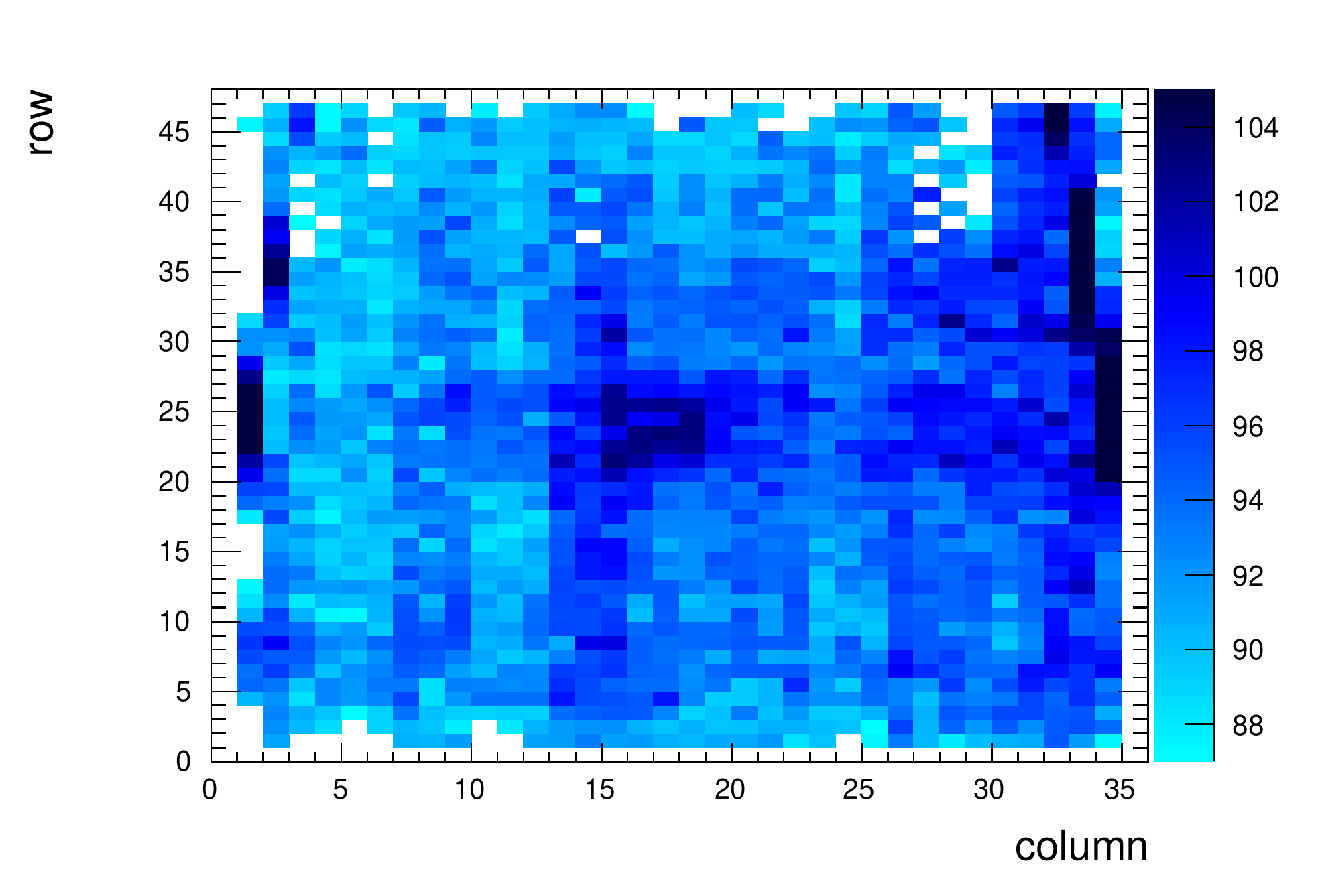}
   
    \caption{Truncated \up{charge} mean for each pad of the Micromegas module}
   \label{fig:gain_uniformity2}
\end{figure}

The uniformity of the gain, however, can be assessed by selecting vertical cosmic tracks, crossing the whole module, and looking at the average charge on each pad. As it will be also described in Sect.~\ref{sec:dedx}, a better estimator of the gain is the truncated average of the charge per pad, obtained by retaining only a certain fraction of measurements with lower charge. For the study presented in this section, a truncation fraction of 0.7 was used to compute the truncated mean.

\begin{figure}	[!ht]
	\centering
	\begin{subfigure}[b]{0.48\textwidth}
		\centering
		\includegraphics[width=\textwidth]{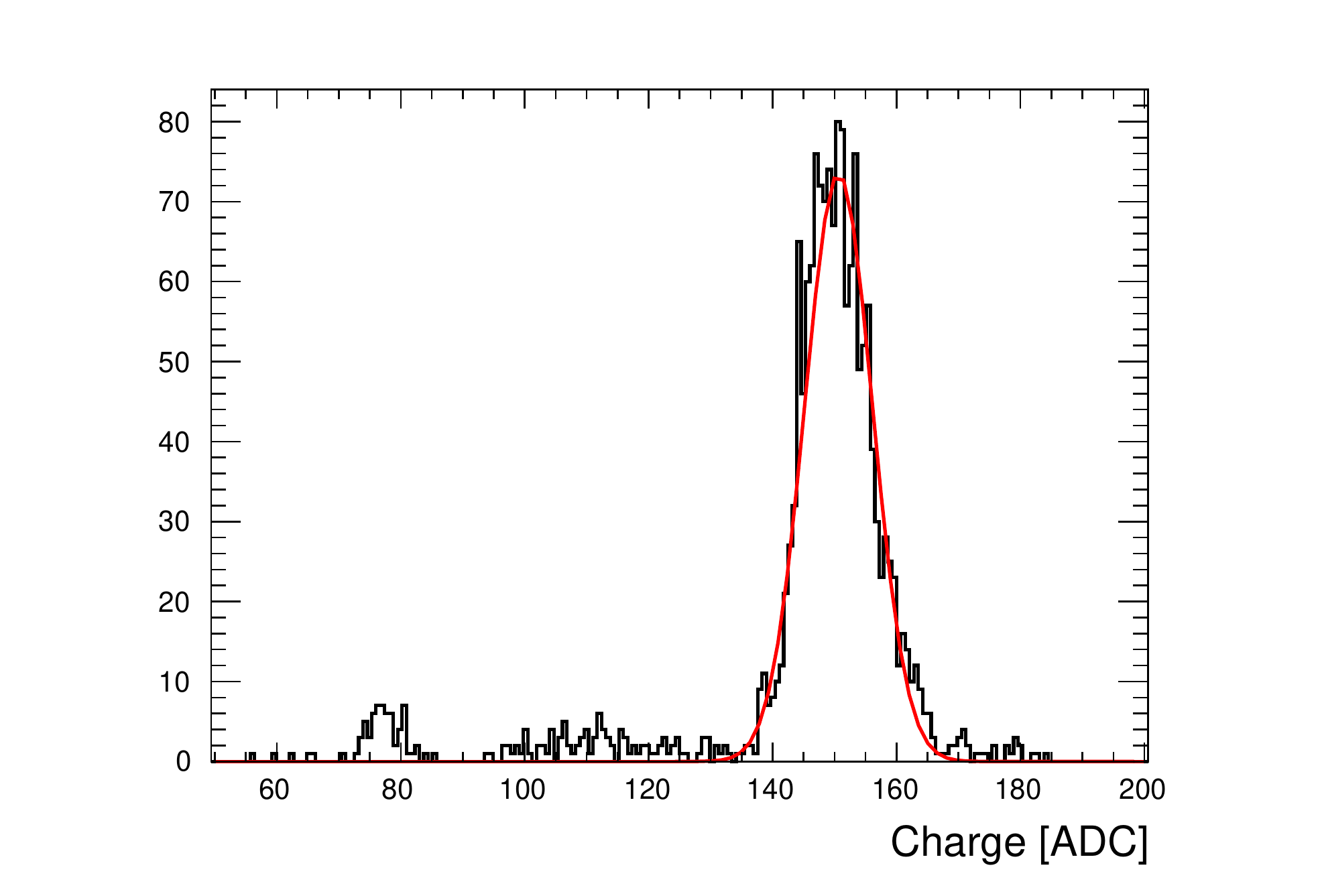}
		\caption{}\label{fig:gain_uniformityA}		
	\end{subfigure}
	\quad
	\begin{subfigure}[b]{0.48\textwidth}
		\centering
		\includegraphics[width=\textwidth]{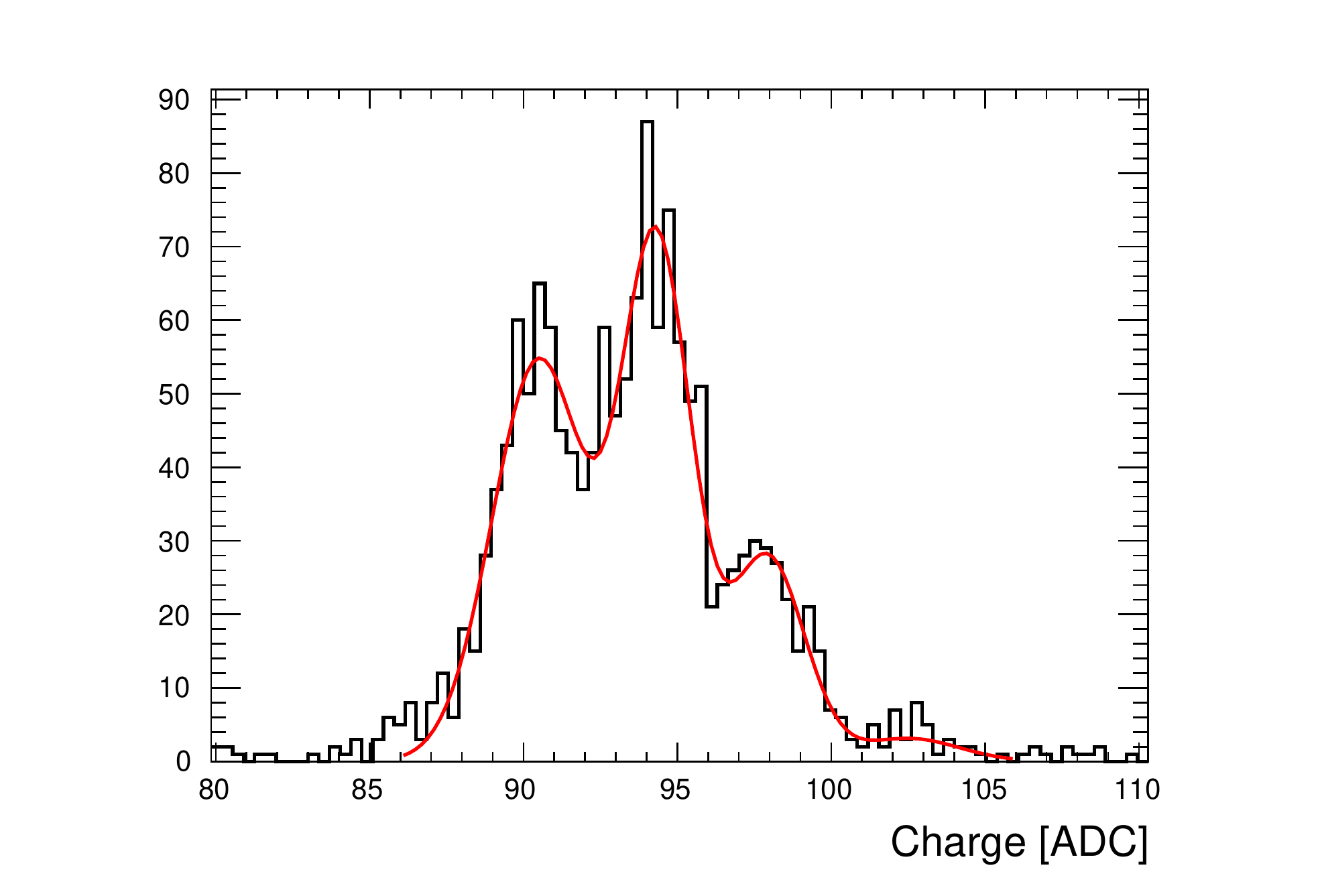}
		\caption{}\label{fig:gain_uniformityB}
	\end{subfigure}
	\caption{Distribution of the mean (left) and truncated mean (right) of the \up{collected charge} for the Micromegas pads obtained with cosmics.}\label{fig:gain_uniformity}
\end{figure}

The distribution of the mean and of the truncated mean for all the pads is shown in Figure~\ref{fig:gain_uniformity}. Except for the pads at the edge of the Micromegas, that \up{collect roughly half of the charge with respect to the other pads due to the smaller number of surrounding pads contributing to the spreading on the measured pad}, the gain uniformity is better than 3\%. By taking the truncated mean (Figure~\ref{fig:gain_uniformityB}), it is also possible to recognize different regions on the module with slightly different gains, as shown in Figure~\ref{fig:gain_uniformity2}. This could be due to non-uniformities of the resistive plane or to electronics routing and it shows that the intrinsic Micromegas gain uniformity is at the 1\% level.

%% file: spread.tex

\begin{figure}[!ht]
 \begin{center}
  \includegraphics[width=0.48\linewidth]{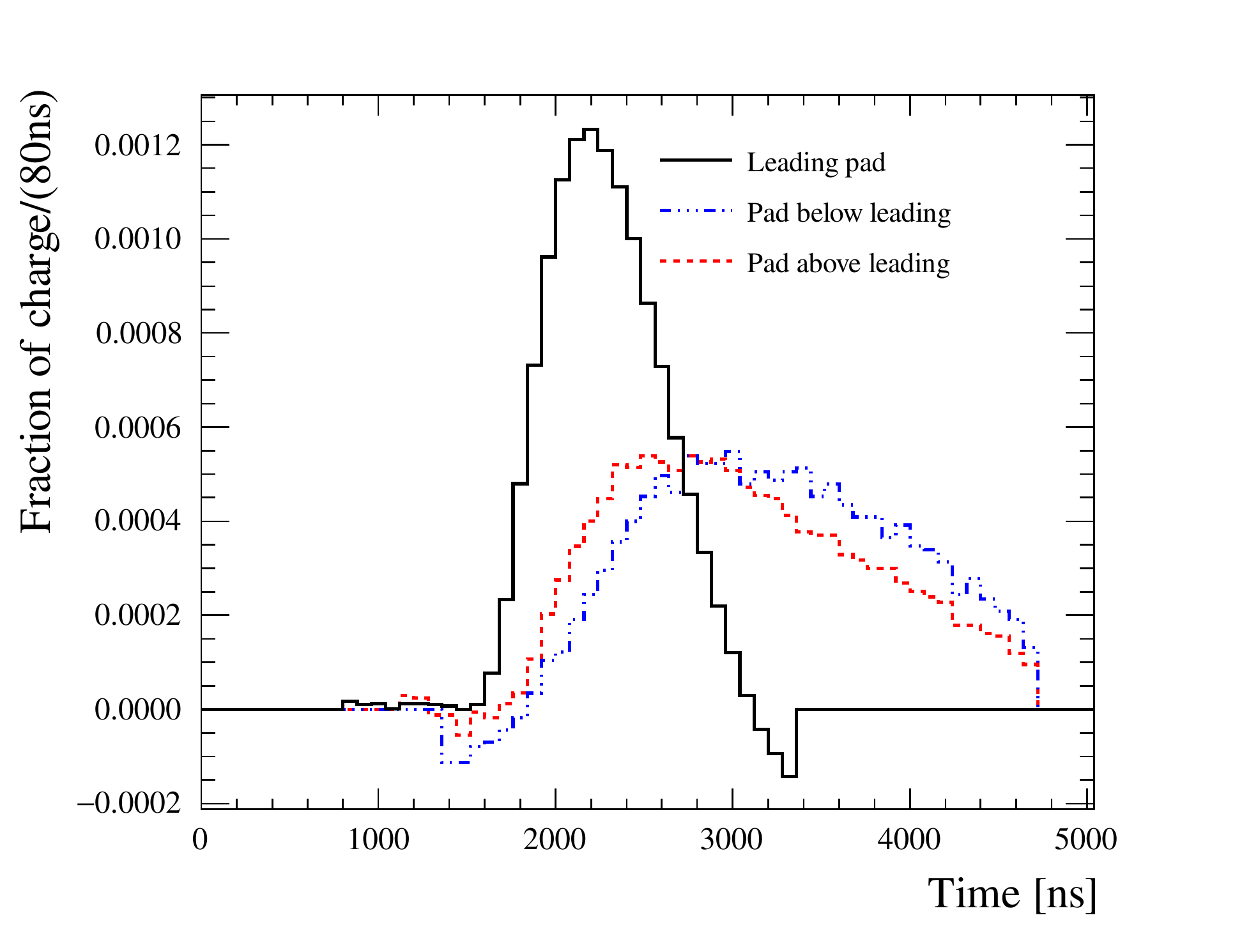}
\end{center}
\caption{Signal waveforms in the three leading pads 600~ns shaping time. The total area of each waveform is normalized to one. The value of each time bin is 80~ns.
 \label{fig:waveforms}}
\end{figure}
\begin{figure} [!ht]
 \begin{center}
  \includegraphics[width=0.49\linewidth]{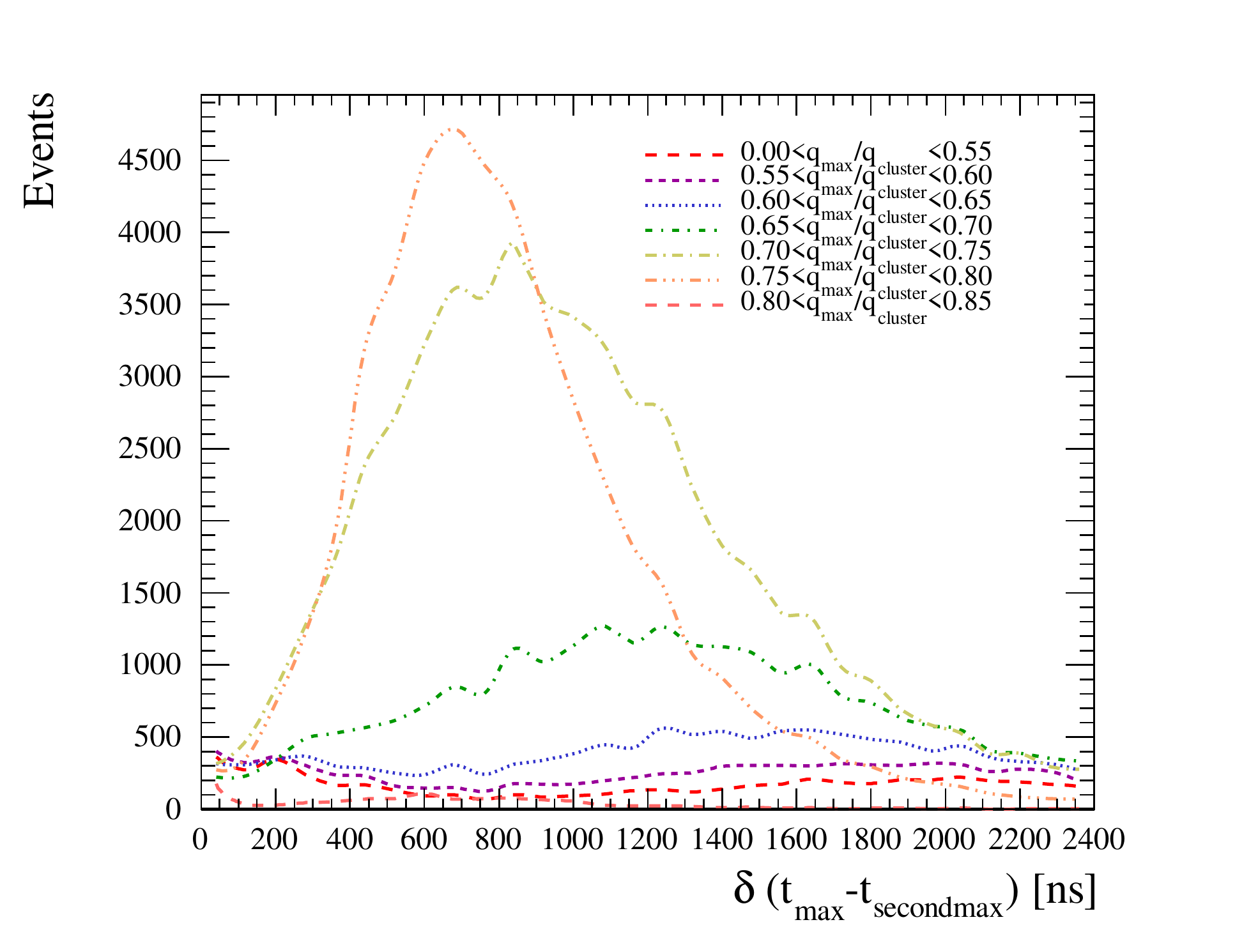}
  \includegraphics[width=0.49\linewidth]{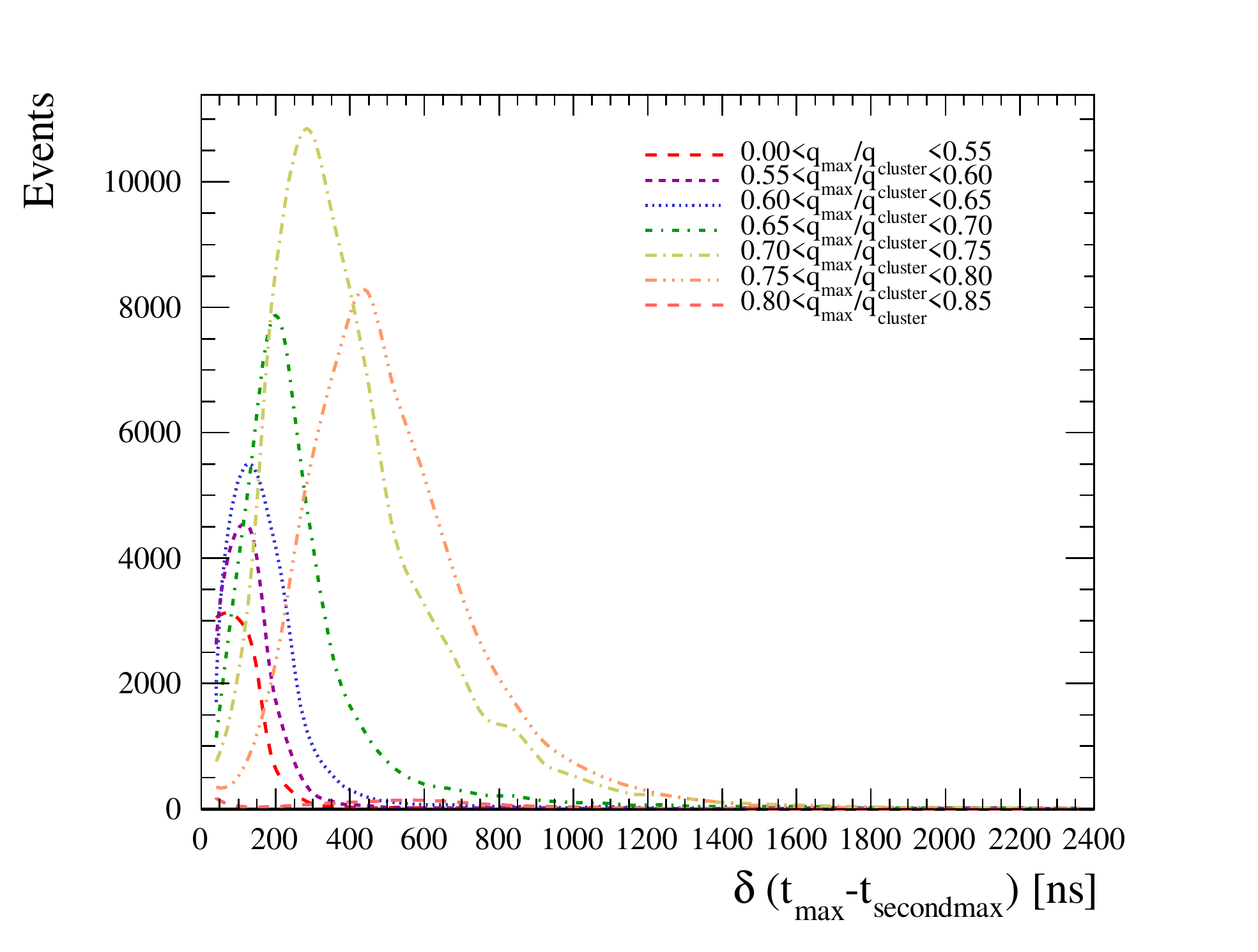}
  \includegraphics[width=0.49\linewidth]{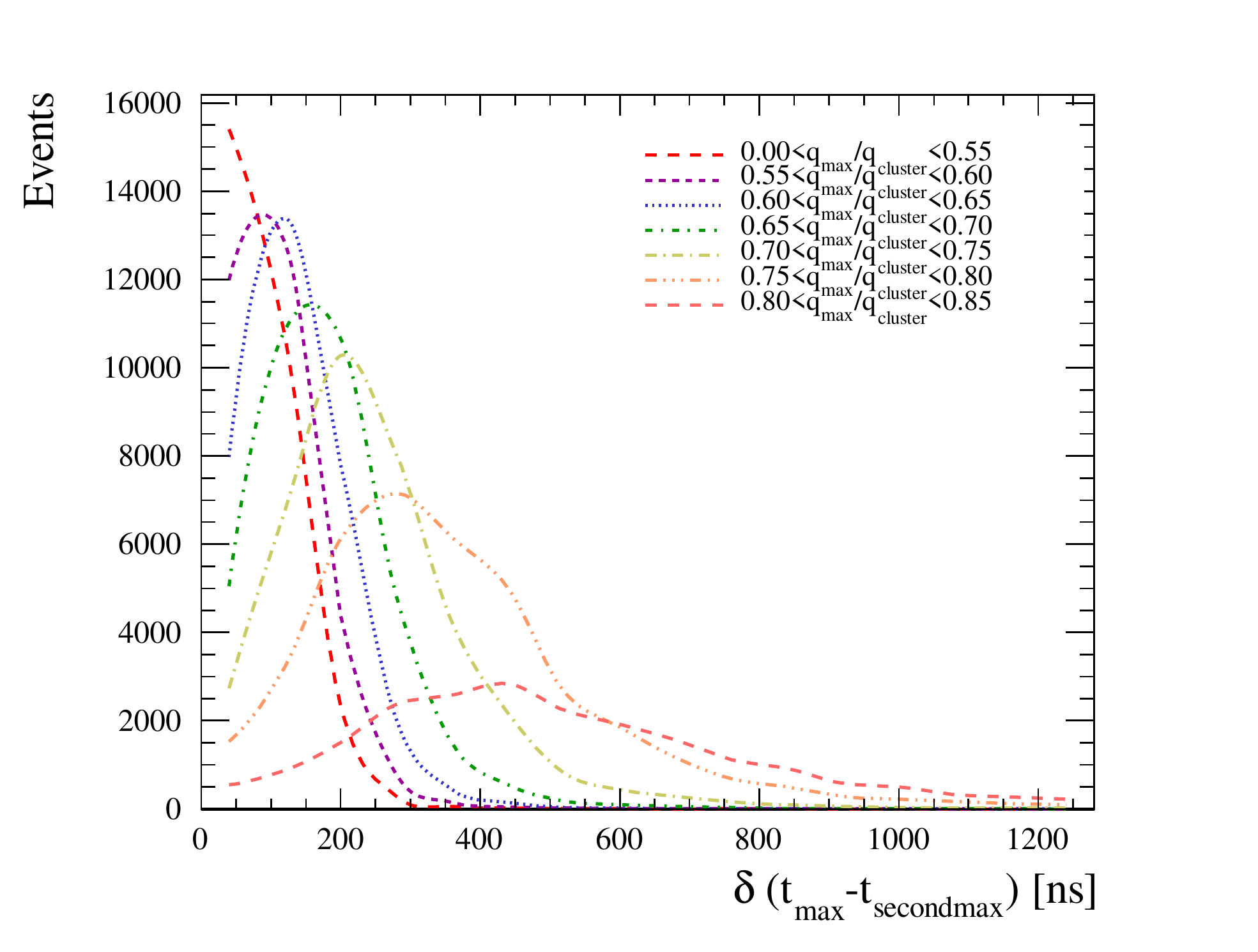}
 \includegraphics[width=0.49\linewidth]{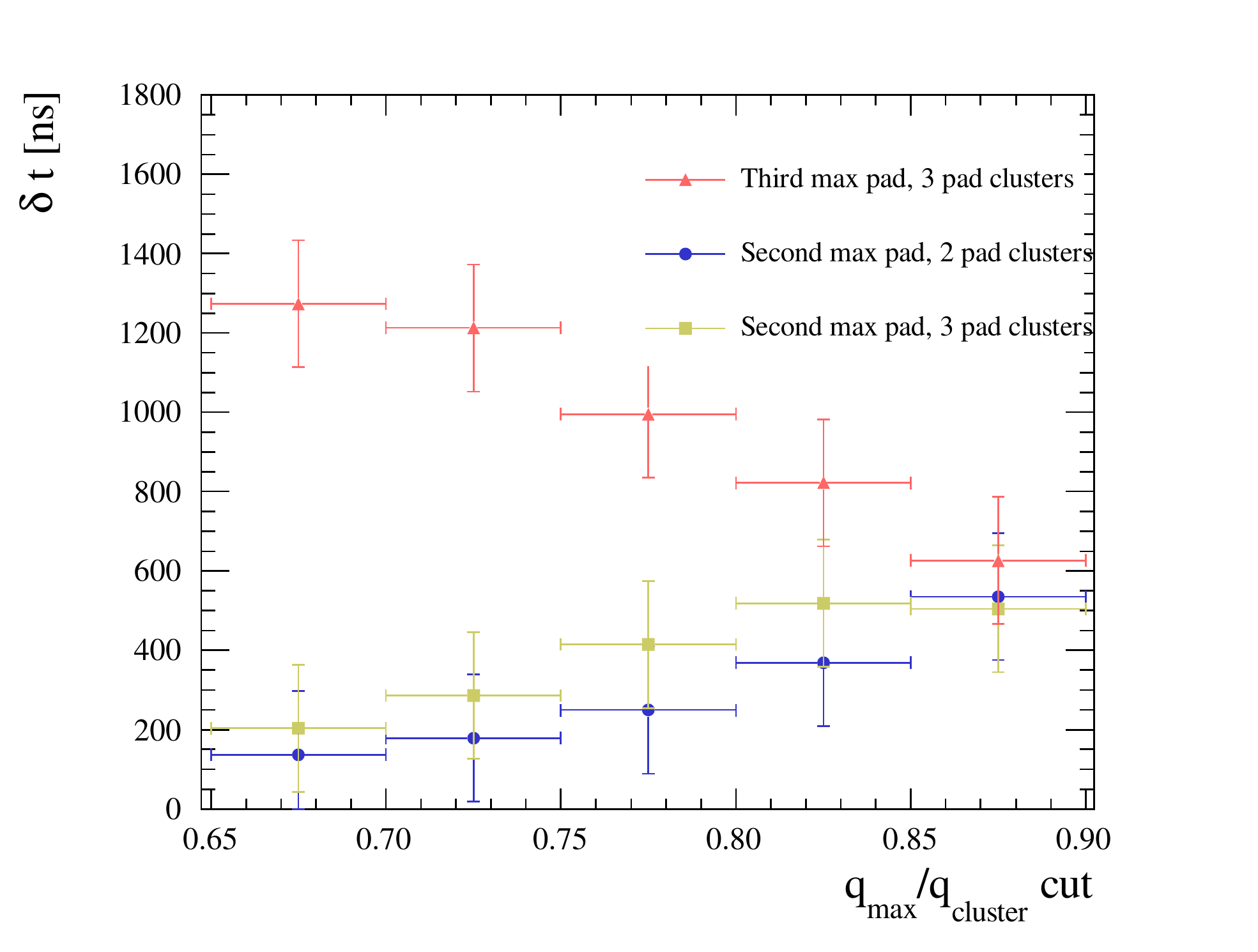}
 \end{center}
\caption{Distribution of peak-time differences between pads, for different cuts on the fraction of charge in the leading pad. Top left (right): difference between the leading and the second (third) pad in 3-pads clusters. Bottom left: difference between the leading and the second pad for 2-pads clusters. Bottom right: peak of the time difference distributions as a function of the cut value. The time is measured with a granularity of 80~ns. The error bars corresponds to two time-bins.
 \label{fig:chargespreadintime}}
\end{figure}

As described in Sec.~\ref{sec:Micromegas}, the resistive Micromegas technology produces a spreading of the collected charge into neighboring pads. This is a novelty with respect to the ND280 TPCs. The charge spreading can be clearly observed in the signal waveforms of adjacent pads, as shown in Figure~\ref{fig:waveforms}. The charge spreading phenomenon which drives the waveform shape is described in~\cite{Dixit:2006ge}. 
\begin{figure} [!ht]
 \begin{center}
  \includegraphics[width=0.49\linewidth]{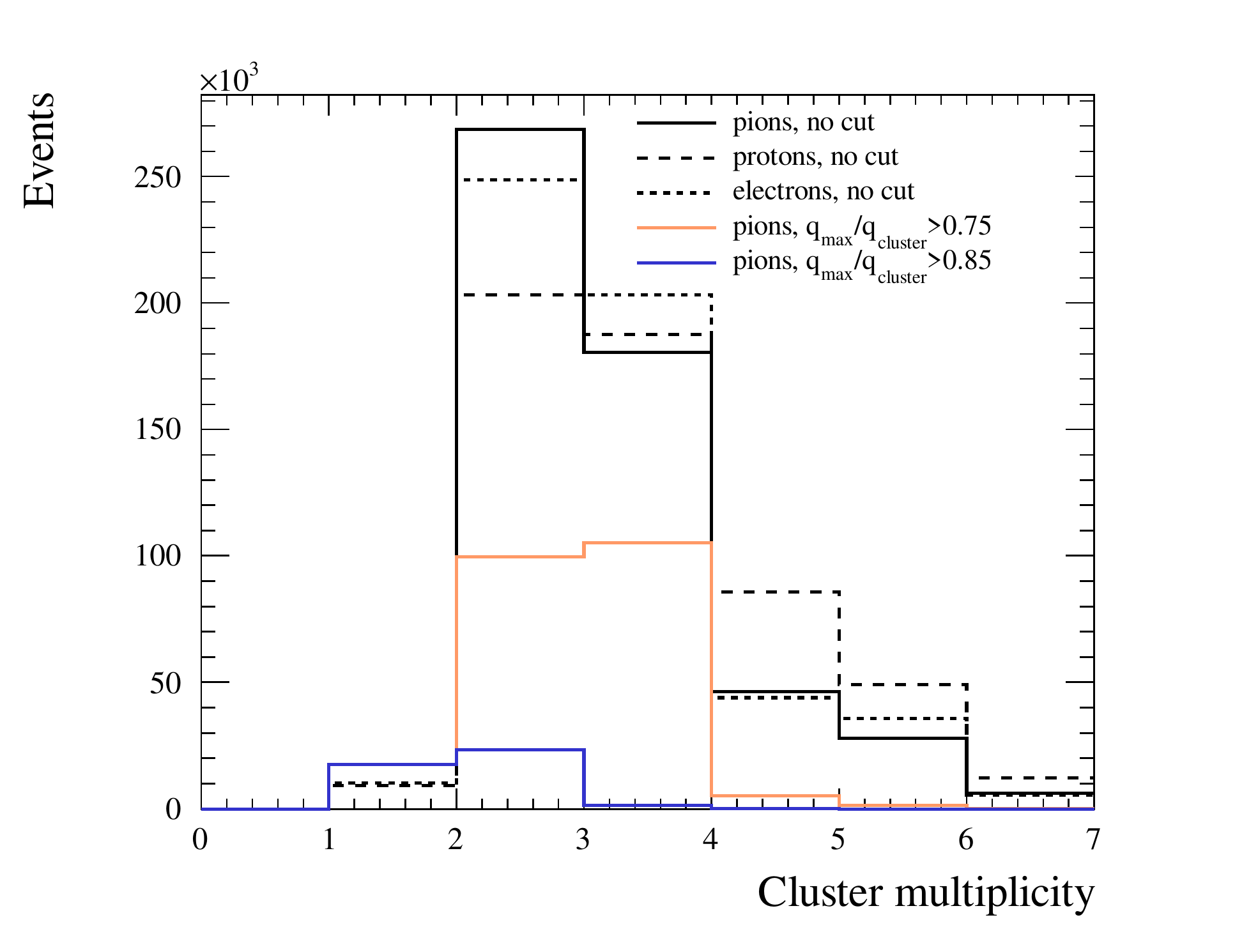}
    \includegraphics[width=0.49\linewidth]{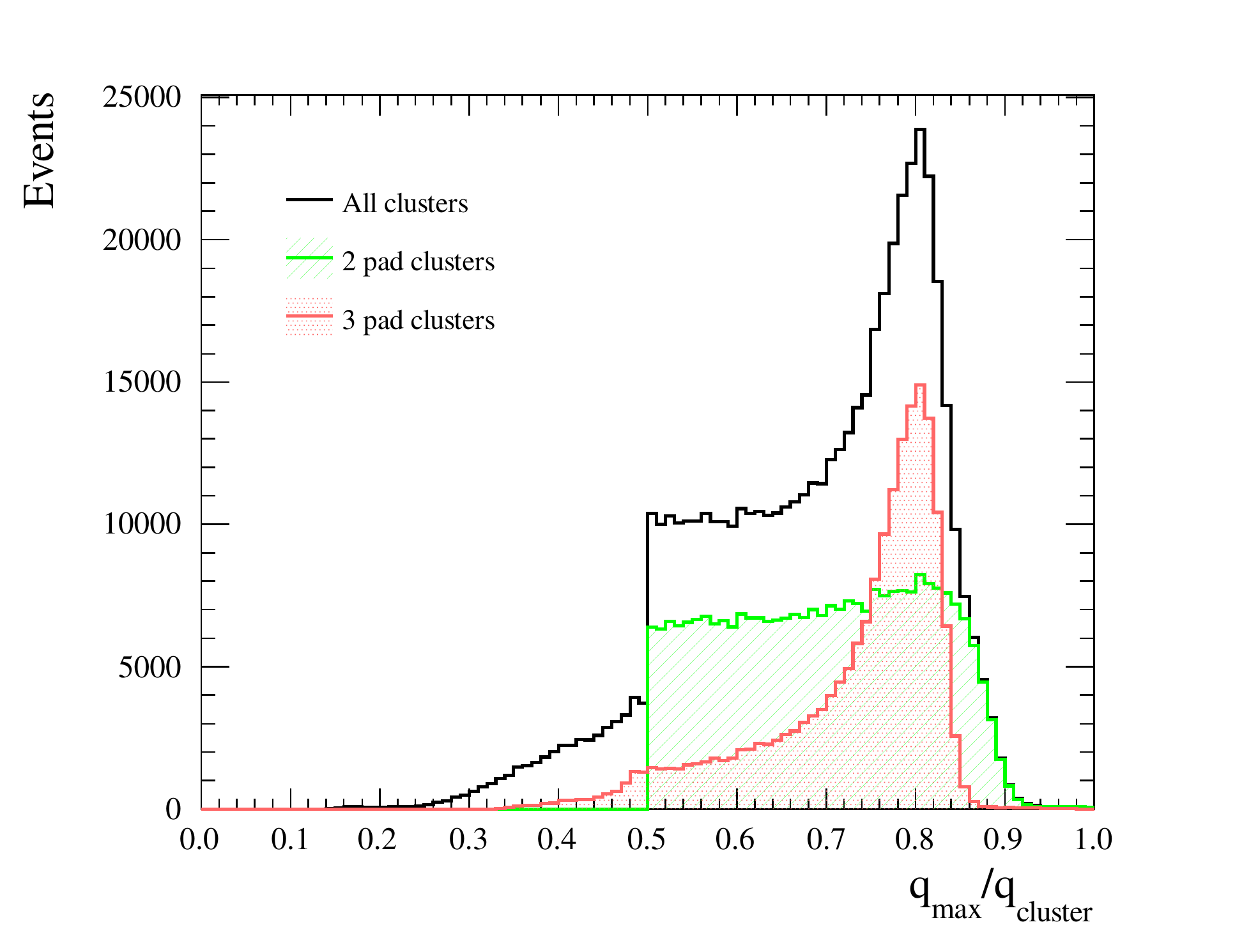}
\end{center}
\caption{\up{Pad} multiplicity \up{in the cluster} (left) and fraction of the cluster charge which is collected in the pad with largest signal (right). The histograms for pion, electron and proton clusters are normalized to the same area. 
 \label{fig:chargespreadinspace}}
\end{figure}
The pads adjacent to the one collecting most of the charge contain a signal smaller in amplitude and longer in time. In order to characterize the size and the timing of the signal spread, we considered nearly horizontal tracks selected with pion trigger, as described in~Sec.\ref{sec:setup}. 
A cluster is defined as the ensemble of the pads on the same column (row in case of vertical tracks) and \up{a pad multiplicity is a number of pads per cluster}. In Figure~\ref{fig:chargespreadinspace} the pad multiplicity per cluster and the fraction of charge in the pad with largest signal ($q_{max}/q_{cluster}$) are shown. Most of the clusters are formed by 2 or 3 pads and the pad with largest signal contains typically 80\% of the total collected charge. 
Figure~\ref{fig:chargespreadintime} shows the difference in peak time between the leading pad and the other pads. The delay of the charge spreading can be up to few $\mu$s for columns with signal above the threshold in 3 pads. 

In Figure~\ref{fig:chargespreadintime} the time difference between the leading and the second (third) pad is also shown as a function of the fraction of signal in the leading pad. Such fraction is a proxy for the position of the track and can be exploited to extract an estimation of the velocity of the charge spreading. A large fraction corresponds to a charge deposition in the middle of the leading pad. Decreasing the fraction corresponds to move closer to (away from) the second (third) pad. 
In the limit of large charge fraction in the leading pad the difference in time peak between the second and the third pad should converge to the same value, which corresponds to the time needed by the charge to spread along half a pad. From these consideration, an  {\itshape effective} charge spreading velocity of about 0.6 cm/$\mu$s can be extracted.


%% file: dedx.tex

The TPCs are pivotal to perform particle identification in ND280. The PID method is based on the measurement of the energy loss by a charged particle in the gas. In the existing ND280 TPCs, the deposited energy resolution for a minimum ionizing particle crossing two Micromegas modules is $\sim8\%$. In this section the performances of the resistive Micromegas module are evaluated for different particle types. 

The method used to estimate the energy loss in the TPC is the truncated mean method: the clusters of each track are ordered as a function of the charge, and only a fraction of them with the lowest charge is used to compute the mean deposited energy. As shown in Sect.~\ref{sec:gain}, clusters on the edge of the Micromegas \up{collect less charge} and are removed from the computation of the truncated mean. 

\begin{figure} [!ht]
  \centering
\begin{subfigure}[b]{.49\textwidth}
  \centering
  \includegraphics[width=1\linewidth]{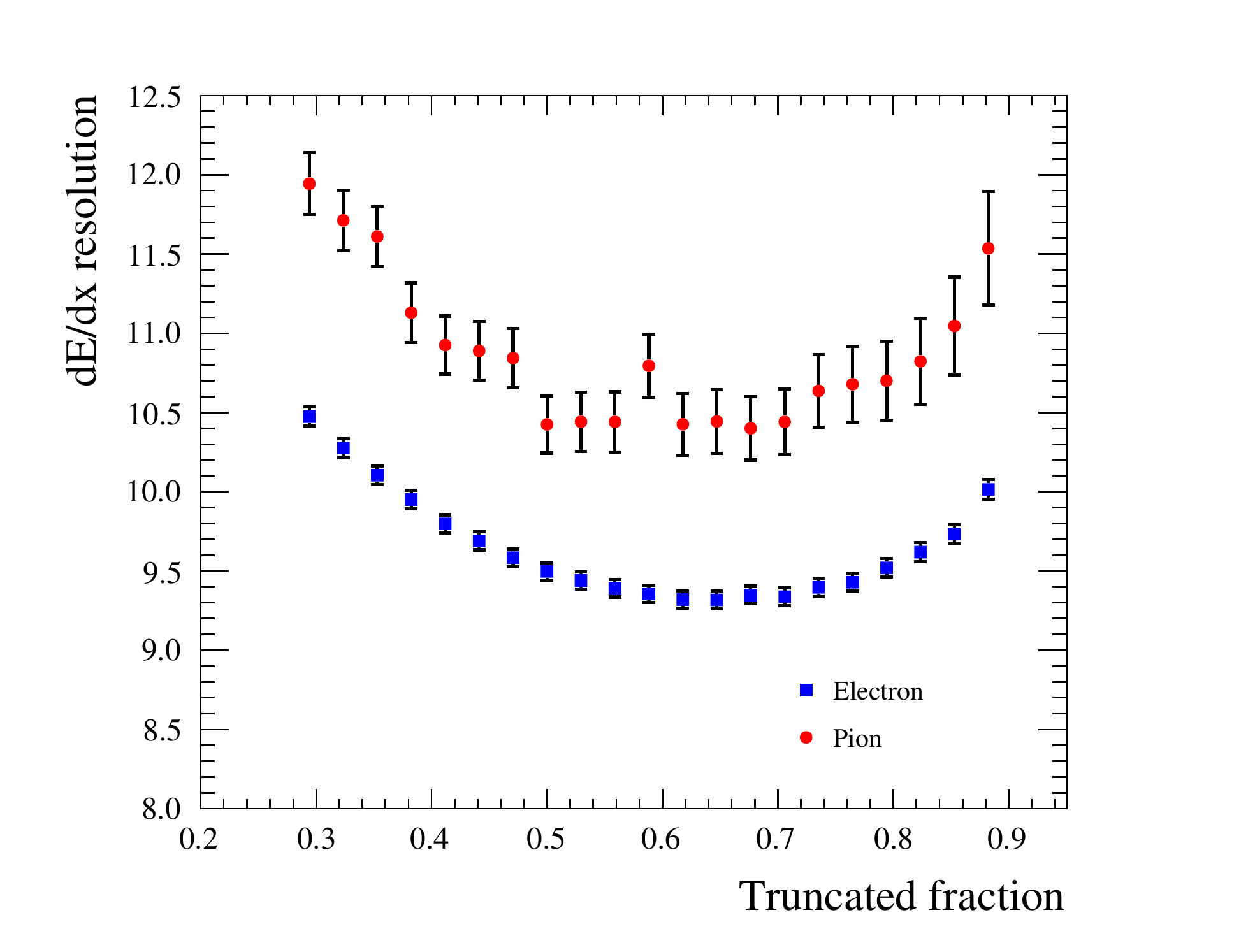}
  \caption{}\label{fig:dedx_truncation_trunc}
\end{subfigure}
  \begin{subfigure}[b]{0.49\textwidth}
  \centering
  \includegraphics[width=1\textwidth]{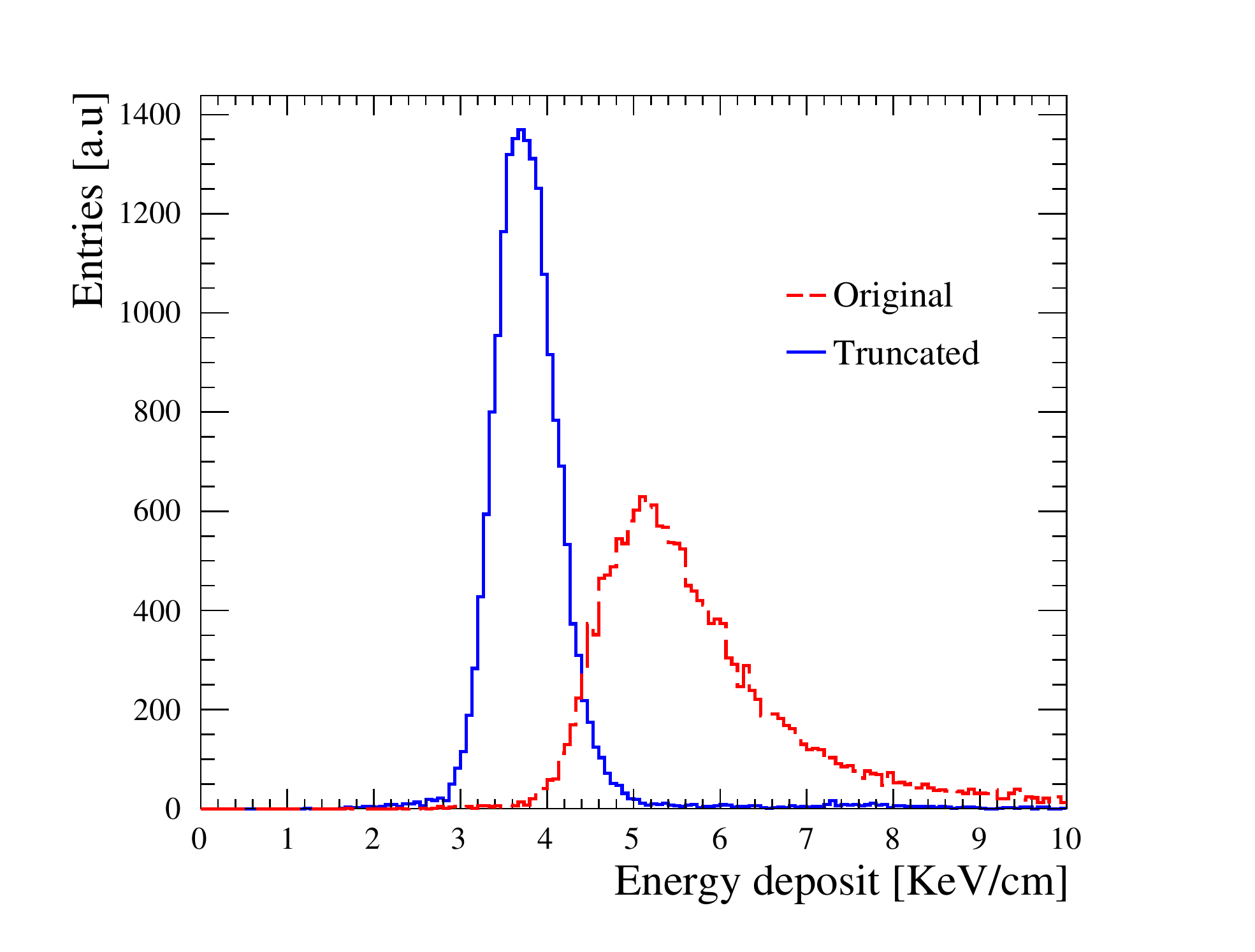}
  \caption{}\label{fig:dedx_truncation_dist}
\end{subfigure}%
  \caption{a) Resolution on the energy deposit per unit of
  track length for different truncation values. b) Comparison on the energy deposited per unit length on the same beam sample of positrons at 0.8~GeV/c with and without applying the truncation mean method.}
   \label{fig:dedx_truncation}
\end{figure}

The impact of the truncated mean method on the deposited energy resolution is shown in Figure~\ref{fig:dedx_truncation_trunc}. The best resolution is obtained for values of truncation fraction between 50\% and 70\%. For the results presented in this section, a truncation factor of 62.5\% was used, corresponding to retaining 21 measurements for tracks with 34 clusters.



\begin{figure}[!ht]
  \centering
  \begin{subfigure}{.49\textwidth}
  \centering
  \includegraphics[width=1\linewidth]{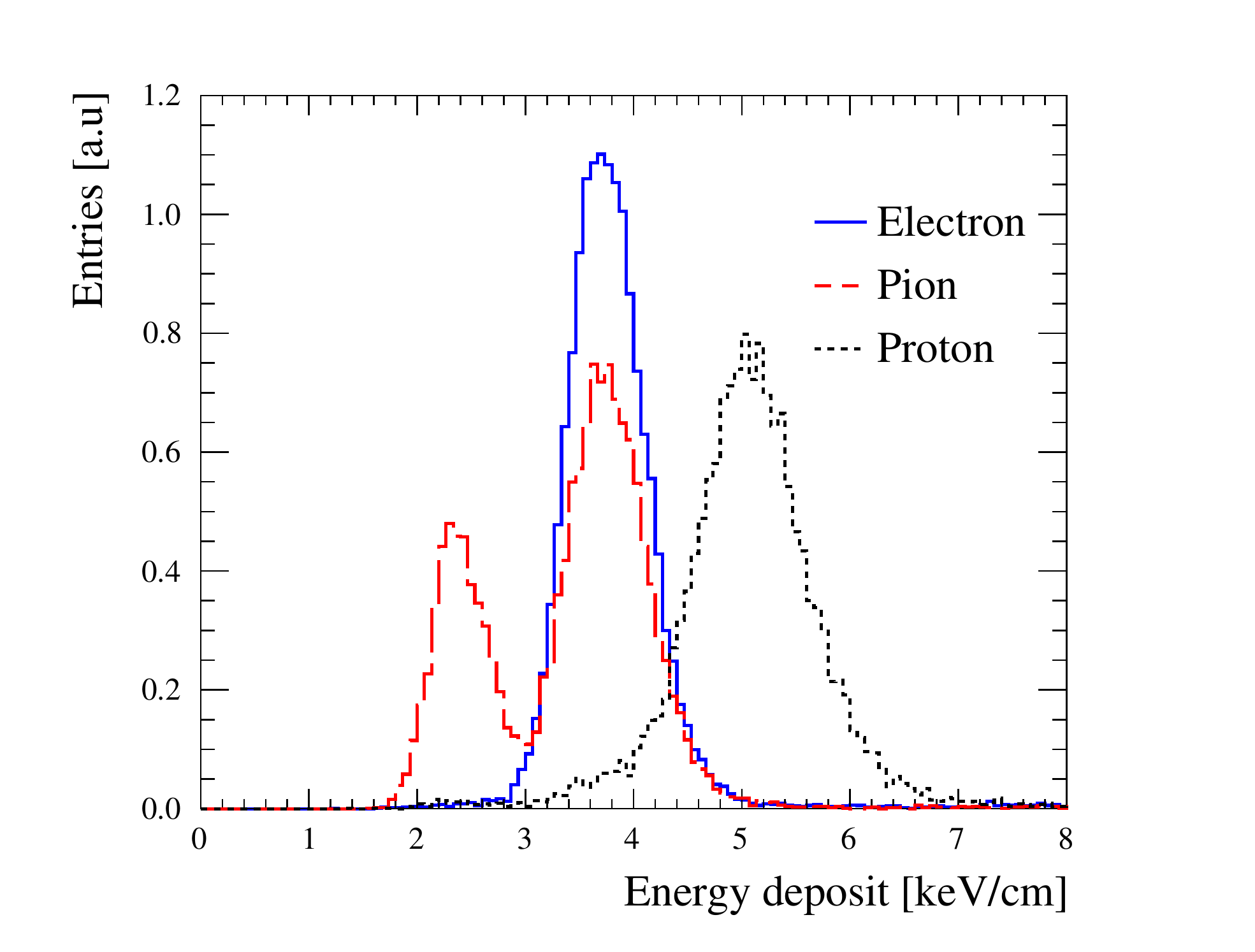}
  \caption{}\label{fig:dedx_spectrum_all}
\end{subfigure}
  \begin{subfigure}{0.49\textwidth}
  \centering
  \includegraphics[width=1\textwidth]{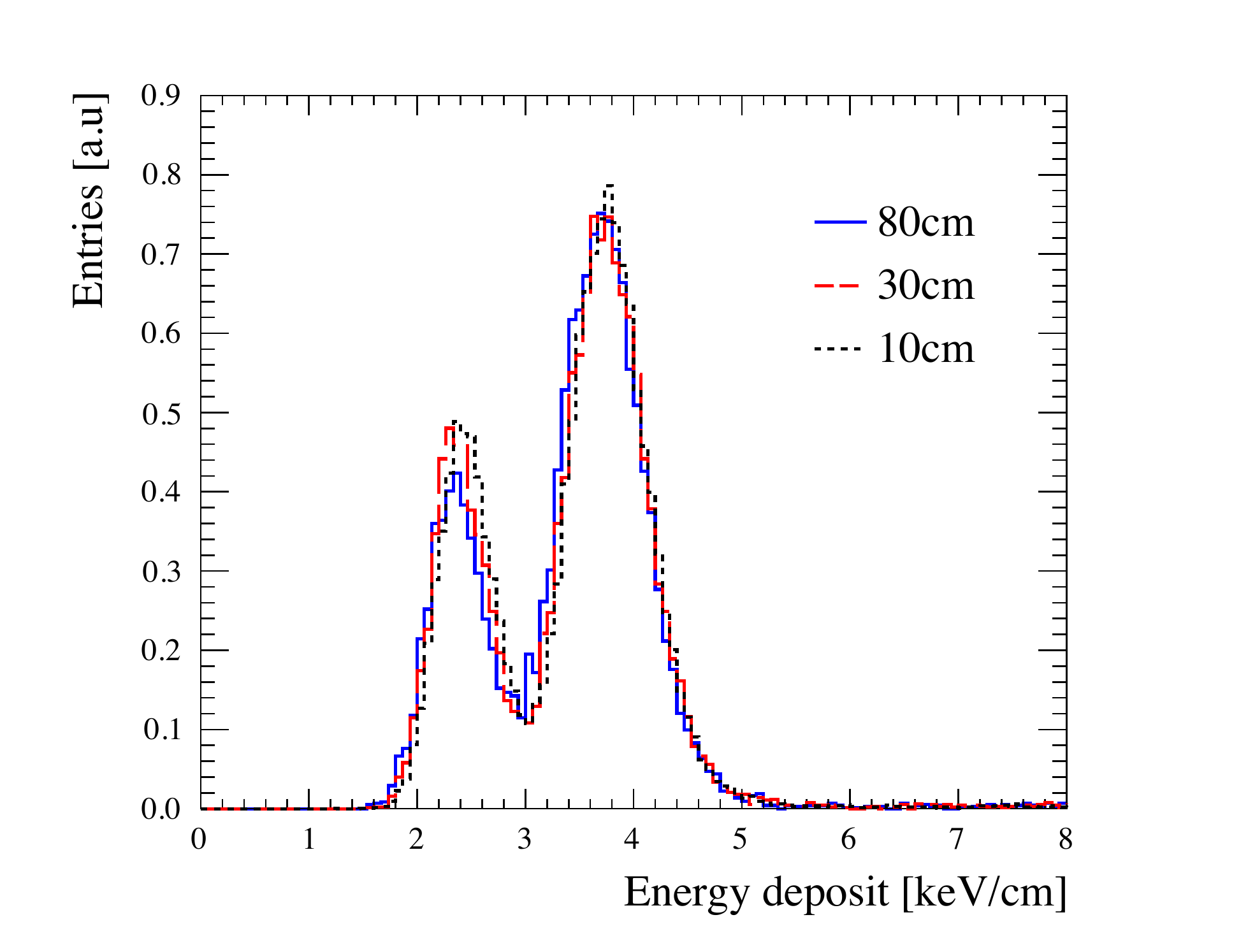}
  \label{fig:dedx_spectrum}
    \caption{}\label{fig:dedx_spectrum_dist}
\end{subfigure}%
  \caption{a) Energy deposited for three different triggers. The particles crossed the chamber at 30 cm from the Micromegas module. b) Energy deposited using pion trigger at 10, 30, and 80~cm drift distance.}
  \centering
    \includegraphics[width=0.7\textwidth]{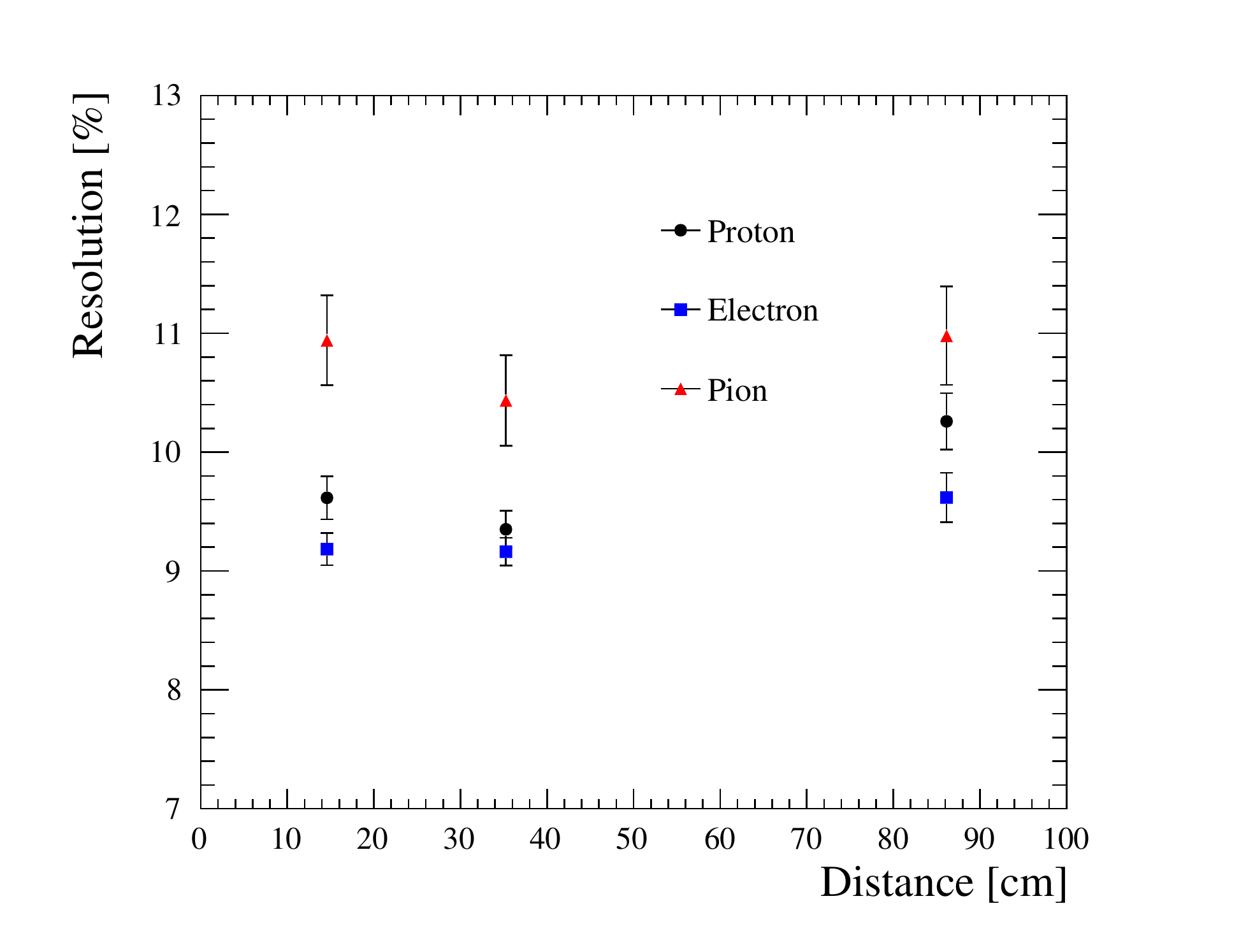}
  \caption{Deposited energy resolution for 0.8~GeV/c positrons, pions, and protons at different drift distances. The total drift distance of the new ND280 horizontal TPCs will be 90~cm.}
  \label{fig:dedx_resolution}
\end{figure}

The deposited energy distribution for different triggers is presented in Figure~\ref{fig:dedx_spectrum_all}. The double peaked spectrum of the distribution with pion trigger indicates a low purity on the selected tracks since a large fraction of them were positrons. 
Figure~\ref{fig:dedx_spectrum_dist} shows the deposited energy for pion triggers taken with different drift distances. After applying corrections to account for the different gain and attenuation lengths (see Sect.~\ref{sec:gain} and \ref{sec:gas}) the deposited energy spectrum is not affected by the drift distance, as it is also shown in Figure~\ref{fig:dedx_resolution}. 
Deposited energy resolutions ranging from 9.0$\%$ to 11.2$\%$ were obtained. These measurements cover drift distances up to 85~cm, close to the maximal drift distance of the TPCs in the upgraded ND280 (90~cm). 
In the horizontal TPCs for the ND280 upgrade, tracks moving perpendicular to the beam direction are expected to cross two Micromegas modules. A larger number of measurements will then be available for each track and this will improve the deposited energy resolution. 
\begin{figure} [!ht]
  \centering
    \includegraphics[width=0.8\textwidth]{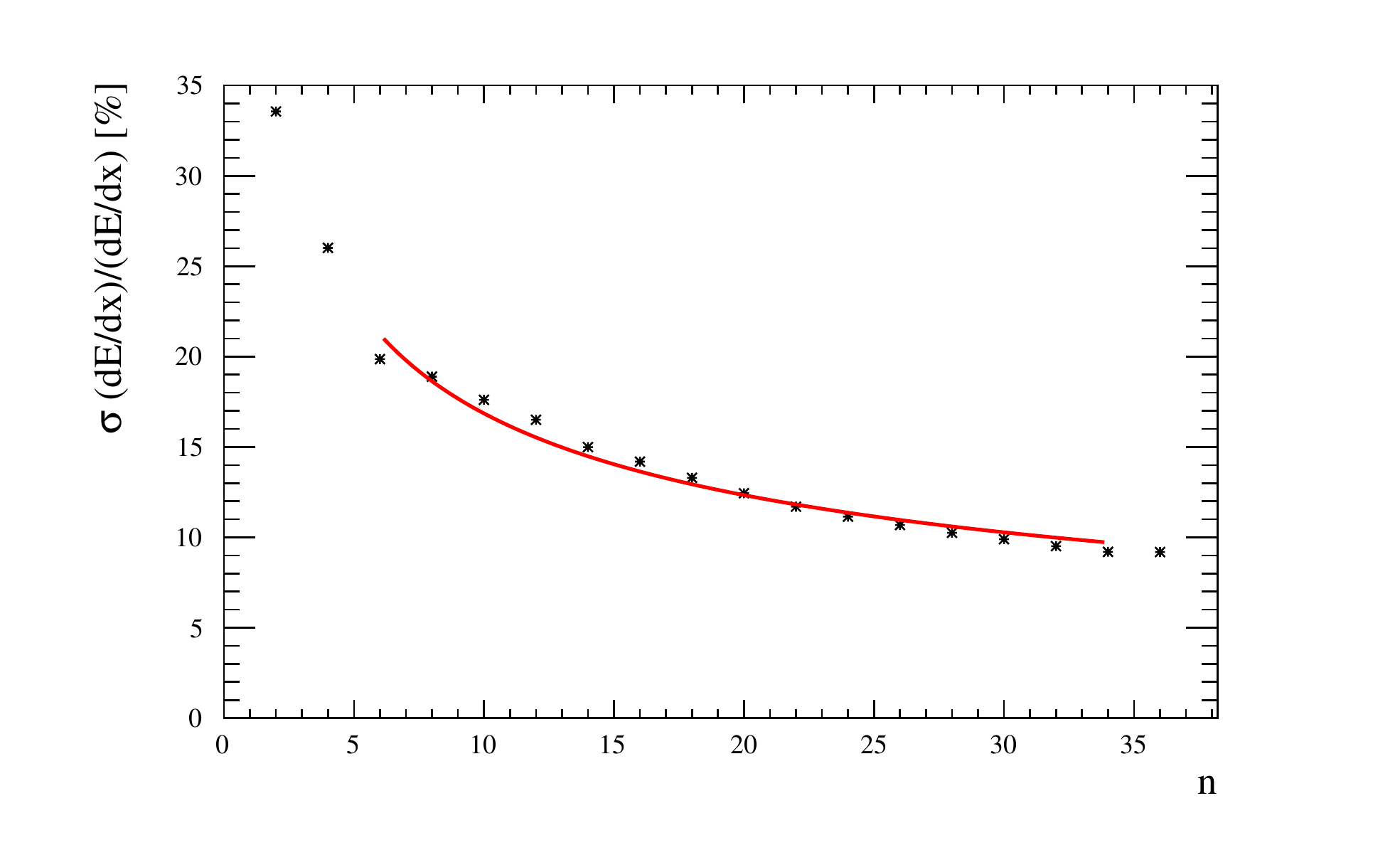}
  \caption{Deposited energy resolution versus number of clusters for positrons at 30 cm of drift distance. }
  \label{fig:ABfits}
\end{figure}
To extrapolate the expected deposited energy resolution to the whole TPC, we measured the deposited energy resolution as a function of the number of clusters used to compute the truncated mean. The distribution for positrons is shown in Figure~\ref{fig:ABfits}. The curve is fitted with 
\begin{equation}
    \centering
	\frac{\sigma (dE/dx)}{(dE/dx)} = A\:n^{B}
	\label{eq:fit_ab}
\end{equation}

Where $n$ is the number of clusters. The fitted values of A and B are shown in tab.~\ref{table:fit_ab} for different drift distances and confirm that, with two Micromegas modules a resolution of $\sim7\%$ is expected. 


\begin{table} [!ht]
	\centering 
	\begin{tabular}{ c | c | c | c | c }
		\hline
		trigger & momentum [GeV/c] & drift distance [cm] & A & B \\ 
		\hline
		 & 0.8 & 80 & $51.9\pm2.9$ & $-0.46\pm0.02$ \\
		positron & 0.8 & 30 & $47.5\pm2.6$ & $-0.45\pm0.02$ \\
		& 0.8 & 10 & $46.2\pm2.5$ & $-0.44\pm0.02$ \\
		\hline
	\end{tabular}
	\caption{ Values of \up{A and B} for different triggers.}
	\label{table:fit_ab}
\end{table}



%% file: spatial.tex


As mentioned above, the main advantage of the TPC with the resistive anode is improved transverse spatial resolution while keeping the same pad size~\cite{Dixit:2003qg}. We performed the analysis using collected data and estimated the accuracy of the transverse position for different particle types at various drift distances.

\subsection{Pad Response Function method}
Due to charge spreading discussed in Sect.~\ref{sec:spread}, the simple weighted mean of the pad position (Center Of Charge, CoC) will not give an accurate position \up{because of the large pad size with respect to the charge spreading region}. A better estimator of the position is obtained by using the Pad Response Function (PRF), defined as 
\begin{equation}
Q_{pad}/Q_{cluster} =
PRF\left(x_{track} - x_{pad}\right)
\label{eq:prf_def}
\end{equation}
where  $x_{pad}$ is the pad center, $Q_{pad}$ is the charge on the pad and $Q_{cluster}$ is the total charge on the cluster containing the pad.

To parametrize the PRF we used the ratio of two symmetric \text{$4^{th}$} order polynoms proposed in~\cite{Boudjemline:2006hf}:
\begin{equation}
PRF(x, \Gamma, \Delta, a, b)=\frac{1+a_2x^2+a_4x^4}{1+b_2x^2+b_4x^4}
\label{eq:GL}
\end{equation}

The coefficients $a_2$ and $a_4$, and $b_2$ and $b_4$ can be expressed in terms of the full width half maximum $\Gamma$, the base width $\Delta$ of the PRF, and two scale parameters $a$ and $b$.




To infer the position of the track, we minimize the $\chi^2$ between $Q_{pad}/Q_{cluster}$ and $PRF(x_{track} - x_{pad})$:
\begin{equation}
\chi^2=\sum_{pads}\frac{Q_{pad}/Q_{cluster} - PRF\left(x_{track} - x_{pad}\right)}{\sqrt{Q_{pad}}/Q_{cluster}}
\end{equation}

\begin{figure}[H]
    \begin{minipage}[h]{0.49\linewidth}
        \center{\includegraphics[width=\linewidth]{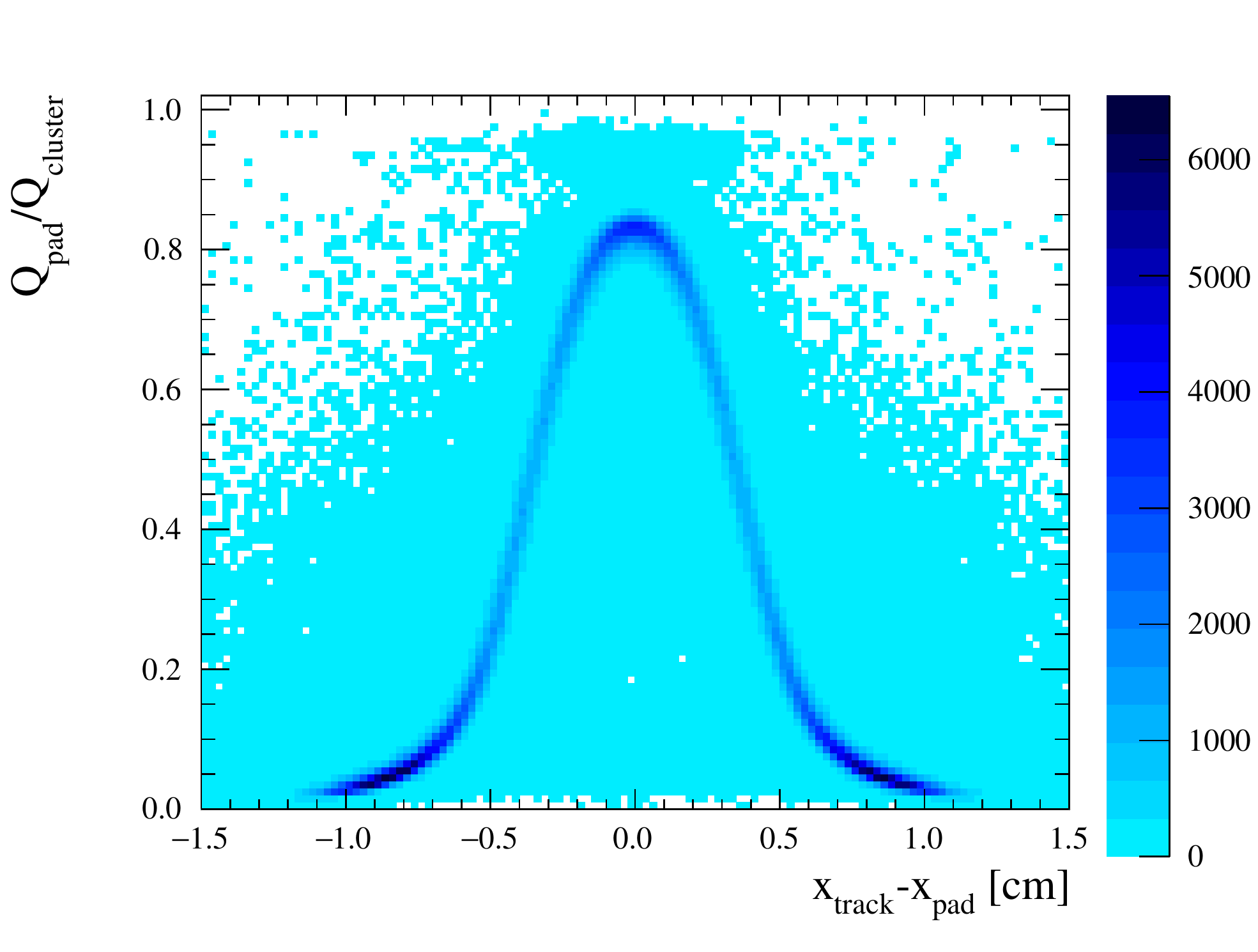} \\ a)}
    \end{minipage}
    \hfill
    \begin{minipage}[h]{0.49\linewidth}
        \center{\includegraphics[width=\linewidth]{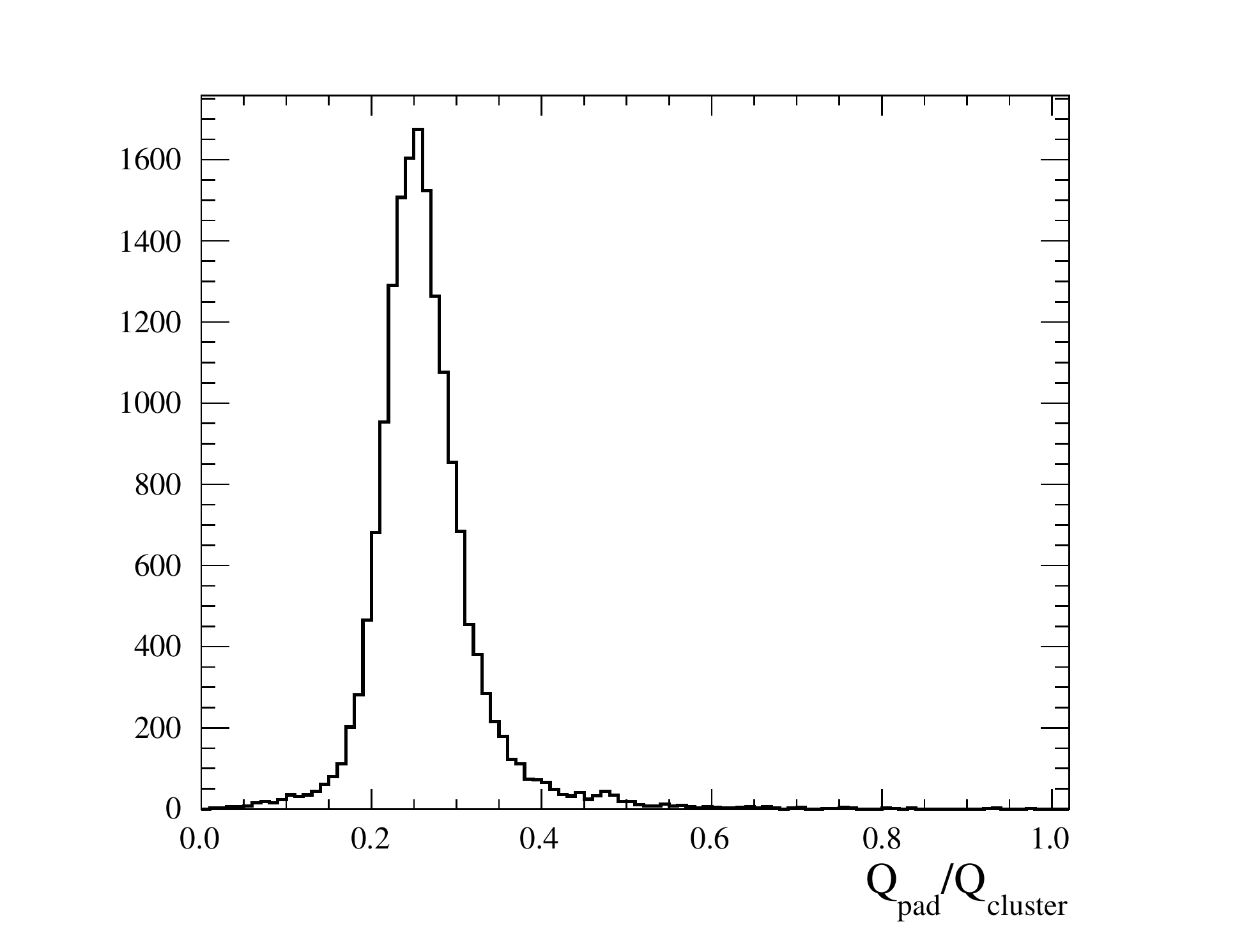} \\ b)}
    \end{minipage}
    \hfill
    \centering
    \begin{minipage}[h]{0.49\linewidth}
        \center{\includegraphics[width=\linewidth]{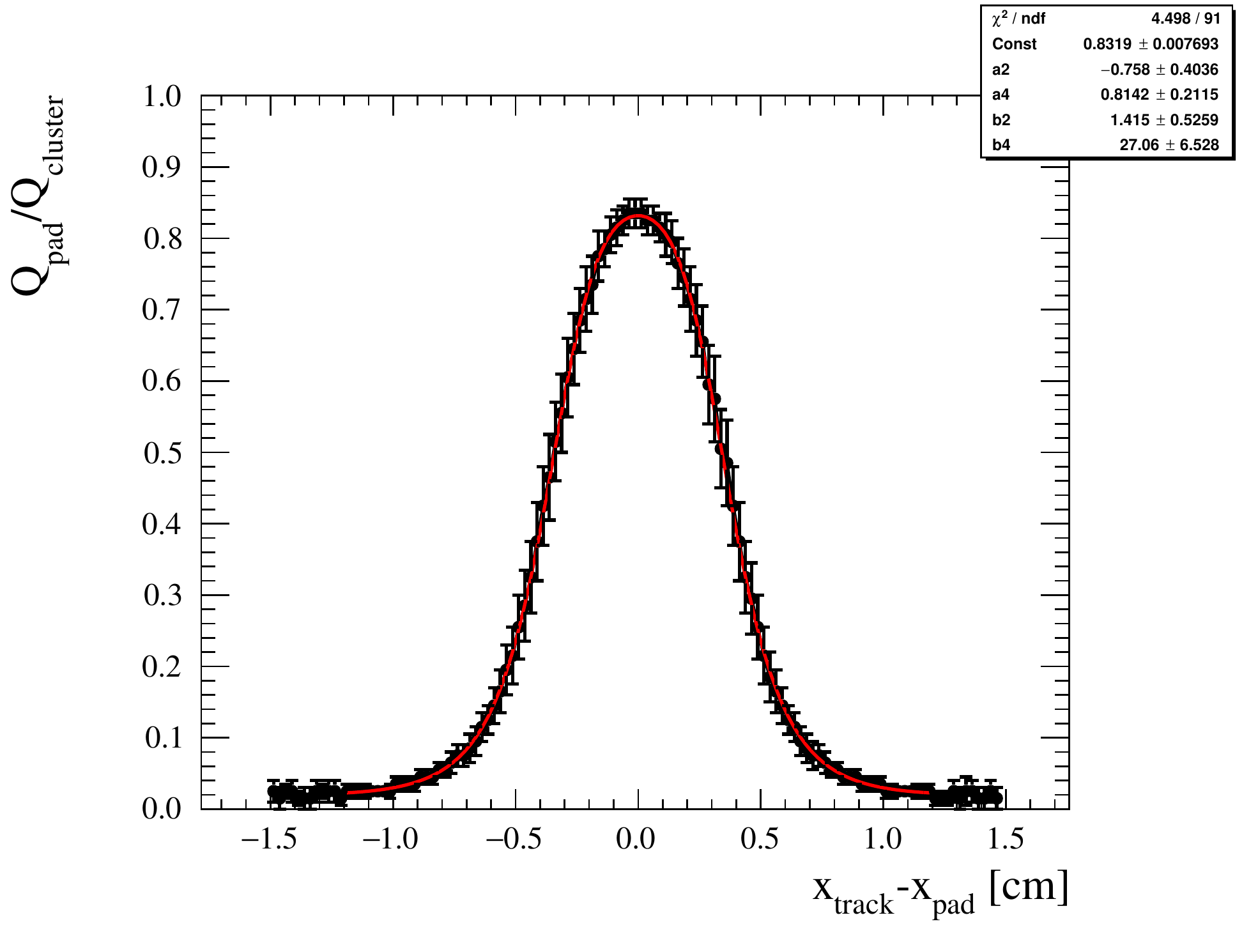} \\ c)}
    \end{minipage}
    \caption{PRF from the pion 1 GeV/c sample with 30 cm drift distance with 360 V at MM: a) 2D histogram, b) example of the 1D histogram for $x_{track} - x_{pad} = -0.5$, c) final PRF plot with errors and fit with analytical function.}
    \label{fig:PRF}
\end{figure}

The analysis was performed in several iterations. In the first iteration the CoC method is used to extract the track position. Then, based on the CoC, the track is fitted with a straight line. The PRF scatter plot is filled  and the distribution of $Q_{pad}/Q_{cluster}$ is obtained for each bin in $x_{track}-x_{pad}$. The peak and the full width half maximum are then taken as estimator of $Q_{pad}/Q_{cluster}$ and its uncertainty. This procedure is shown in Figure~\ref{fig:PRF}.



With this method the PRF function is estimated with the CoC method as a prior. Since this prior could be far from the $x$ position, we repeated the procedure for several iterations. At each iteration, the analytical formula from the previous step is used to estimate the track position and then a new fit is performed. We observed the improvement of the fit in the first few iterations, while the fit quality does not improve further after about 5 iterations.

\begin{figure}[H]
    \begin{minipage}[h]{0.49\linewidth}
        \center{\includegraphics[width=\linewidth]{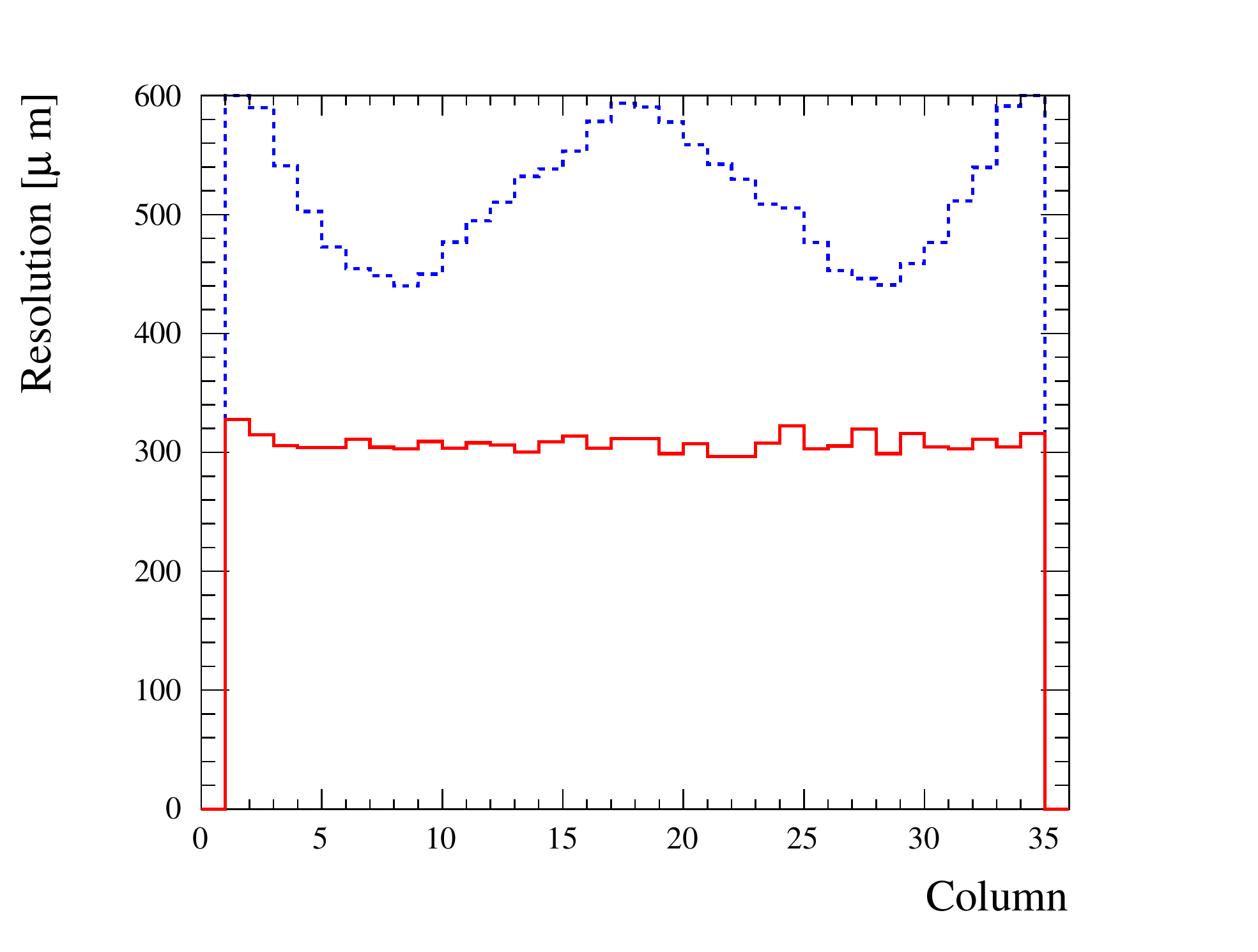} \\ a)}
    \end{minipage}
    \hfill
    \begin{minipage}[h]{0.49\linewidth}
        \center{\includegraphics[width=\linewidth]{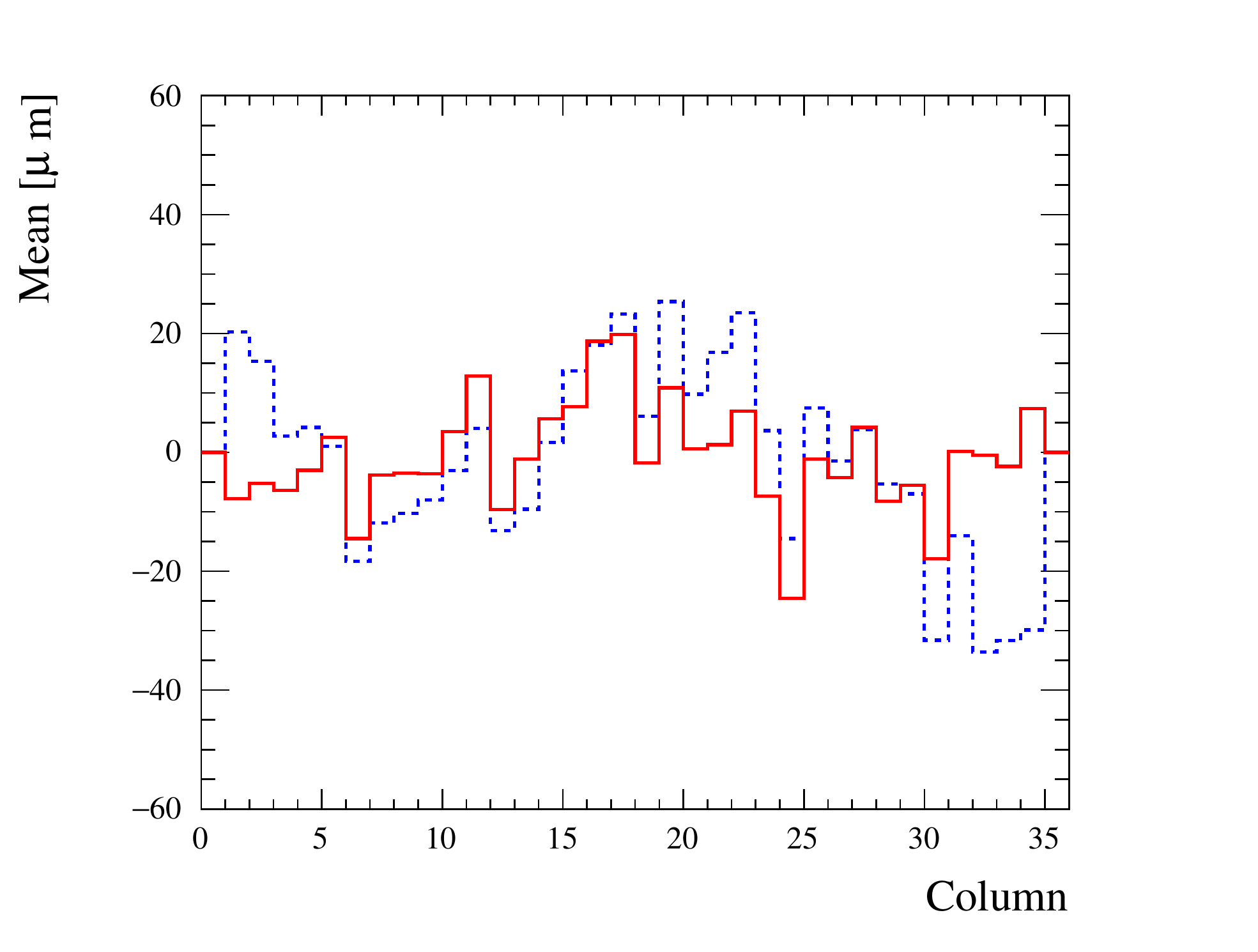} \\ b)}
    \end{minipage}
    \caption{Comparison between PRF method (red solid line) and CoC method (blue dashed line). a) sigma of the residual as a function of the column number, b) mean of the residual as a function of the column number.}
    \label{fig:residual}
\end{figure}

The spatial resolution is defined with the residuals $x_{track} - x_{fit}$ for each cluster. The obtained distribution is fit with a Gaussian and the $\sigma$ is considered as the final accuracy. 
\up{The improvement} obtained with the PRF method with respect to the CoC method are shown in Figure~\ref{fig:residual}: a spatial resolution at the level of 300 $\mu m$ is observed for horizontal tracks over the whole Micromegas.

\begin{figure}[H]
\centering
   \includegraphics[width=0.5\linewidth]{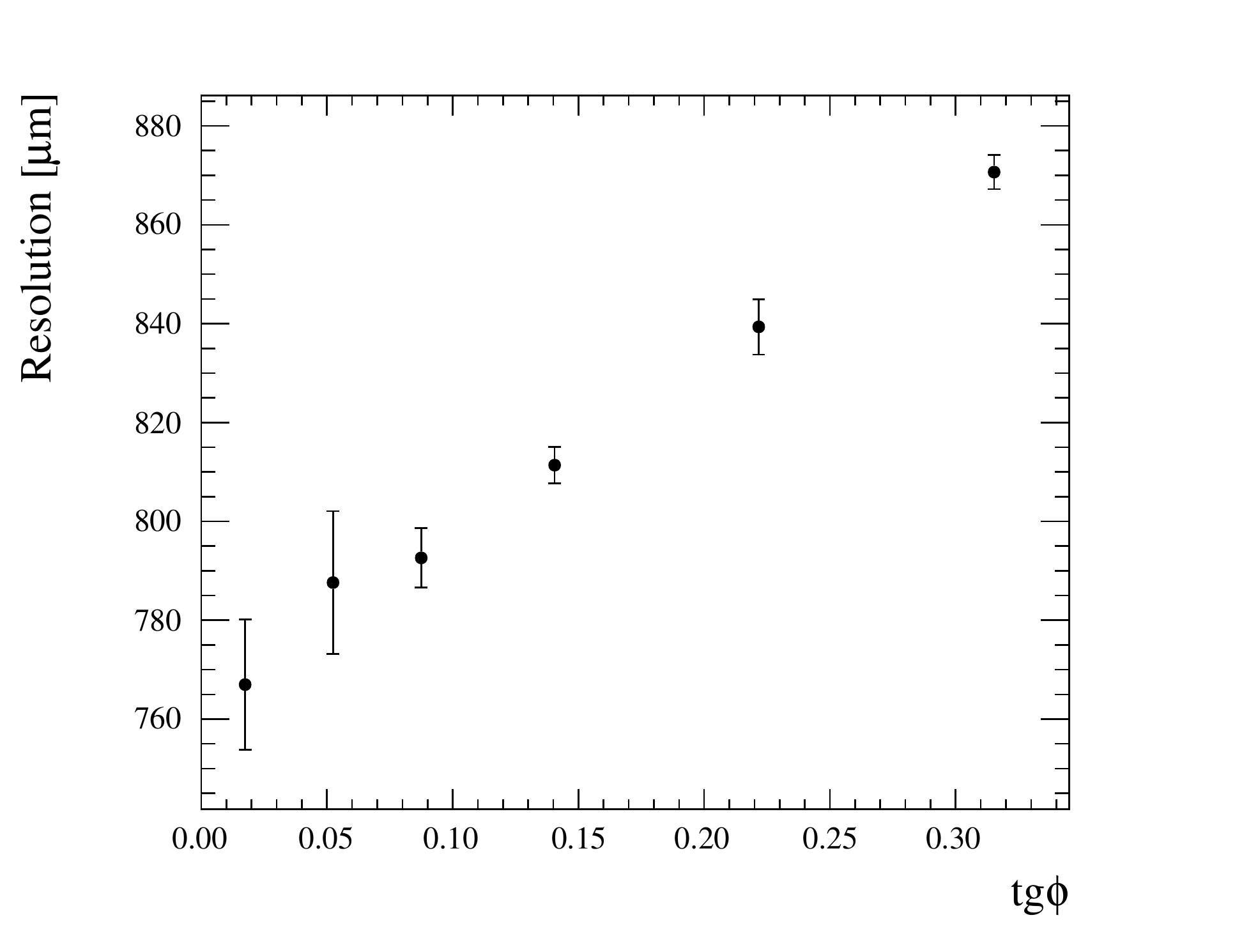} 
    \caption{Spatial resolution as function of the tangent of
the angle in the detector plane for cosmics.  The CoC method is used to compute the spatial resolution}
    \label{fig:spatialcosmics}
\end{figure}

The spatial resolution performances are expected to depend on the angle of the track with respect to the Micromegas pads. This is due to the fact that ionization fluctuations 
increase the variance of charge sharing for tracks at an angle to pad boundaries. 
Such behavior cannot be tested with test beam data, since mostly horizontal tracks were produced, and we investigated it using cosmics. The dependence of the spatial resolution on the angle in the detector plane, using the CoC method, is shown in Figure~\ref{fig:spatialcosmics}. Due to the limited geometrical acceptance of the cosmic trigger described in Sect.~\ref{sec:setup}, only tracks with drift distances comprised between 50 and 100~cm and with angles between 0 and 20 degrees  could be used for this study.  
The difference with respect to beam tracks, for which the resolution with the CoC method is 590 $\mu m$ at 80~cm, is due to the smaller pad length in the vertical direction (0.7~cm) with respect to the horizontal direction (1.0~cm).

\subsection{Dependence of spatial resolution on the drift distance}
\begin{figure}[H]
    \includegraphics[width=\linewidth]{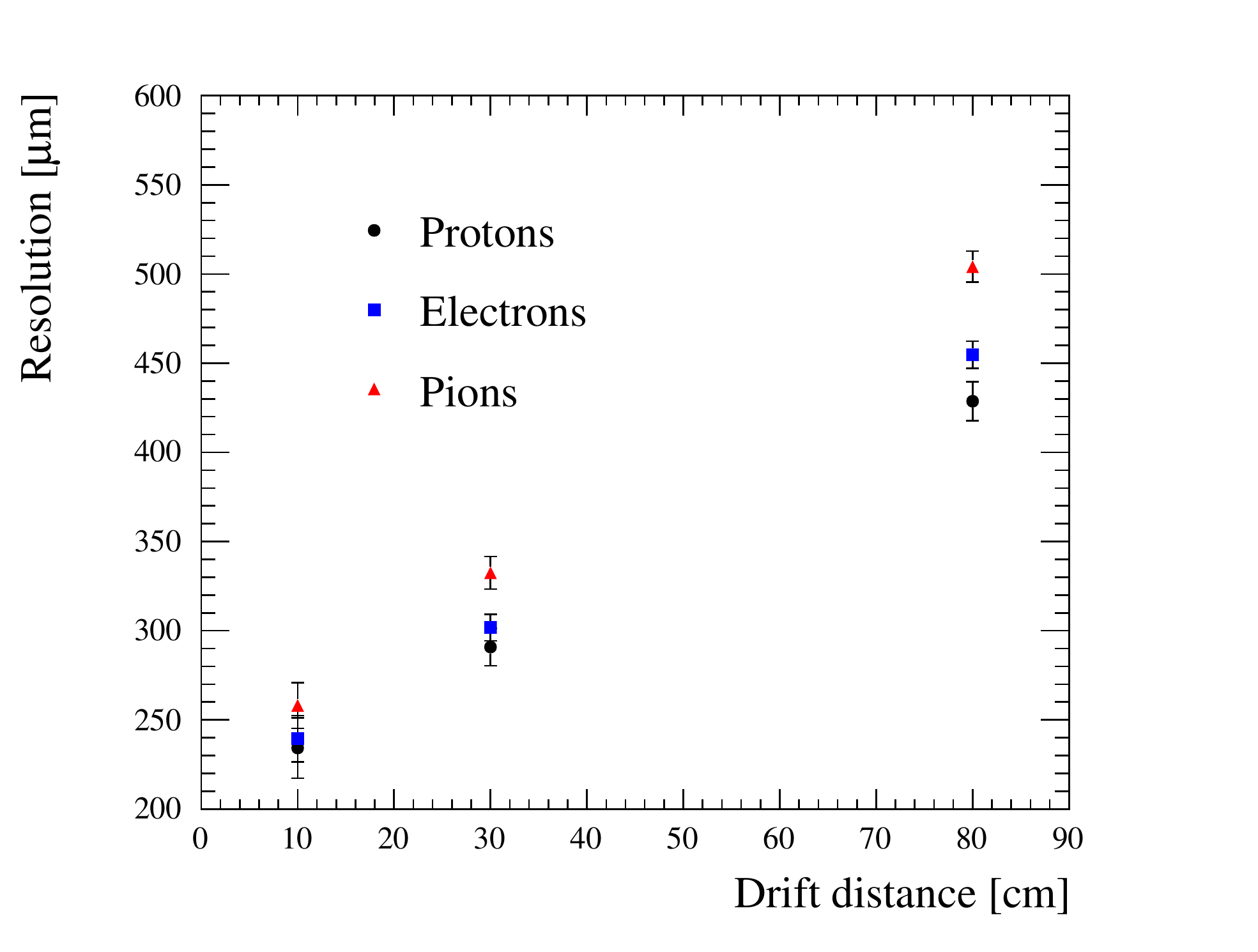} 
    \caption{Spatial resolution for different particles at different drift distances. For each sample the average value on the 36 columns was taken while the RMS is shown as the error bar for each point.}
    \label{fig:spatial_res}
\end{figure}

The spatial resolution is expected to be affected by the drift distance, since tracks with larger drift distances will be more affected by transverse and longitudinal diffusion.
The average spatial resolution for pions, positrons and protons at different drift distances is shown in Figure~\ref{fig:spatial_res} and, as expected, the spatial resolution is worst for long drift distances. The dependence of the resolution on the particle type can be explained with the different energy loss through the ionisation. Higher energy deposition causes larger charge spreading resulting in higher \up{pad} multiplicity. As a result with a higher number of pads per cluster the track position could be reconstructed more precisely.

\subsection{Dependence of spatial resolution on the Micromegas voltage}
Finally we studied the spatial resolution for different Micromegas voltages. These data were taken only with 1 GeV/c pion trigger at drift distance of 30~cm.

\begin{figure}[H]
    \center{\includegraphics[width=0.6\linewidth]{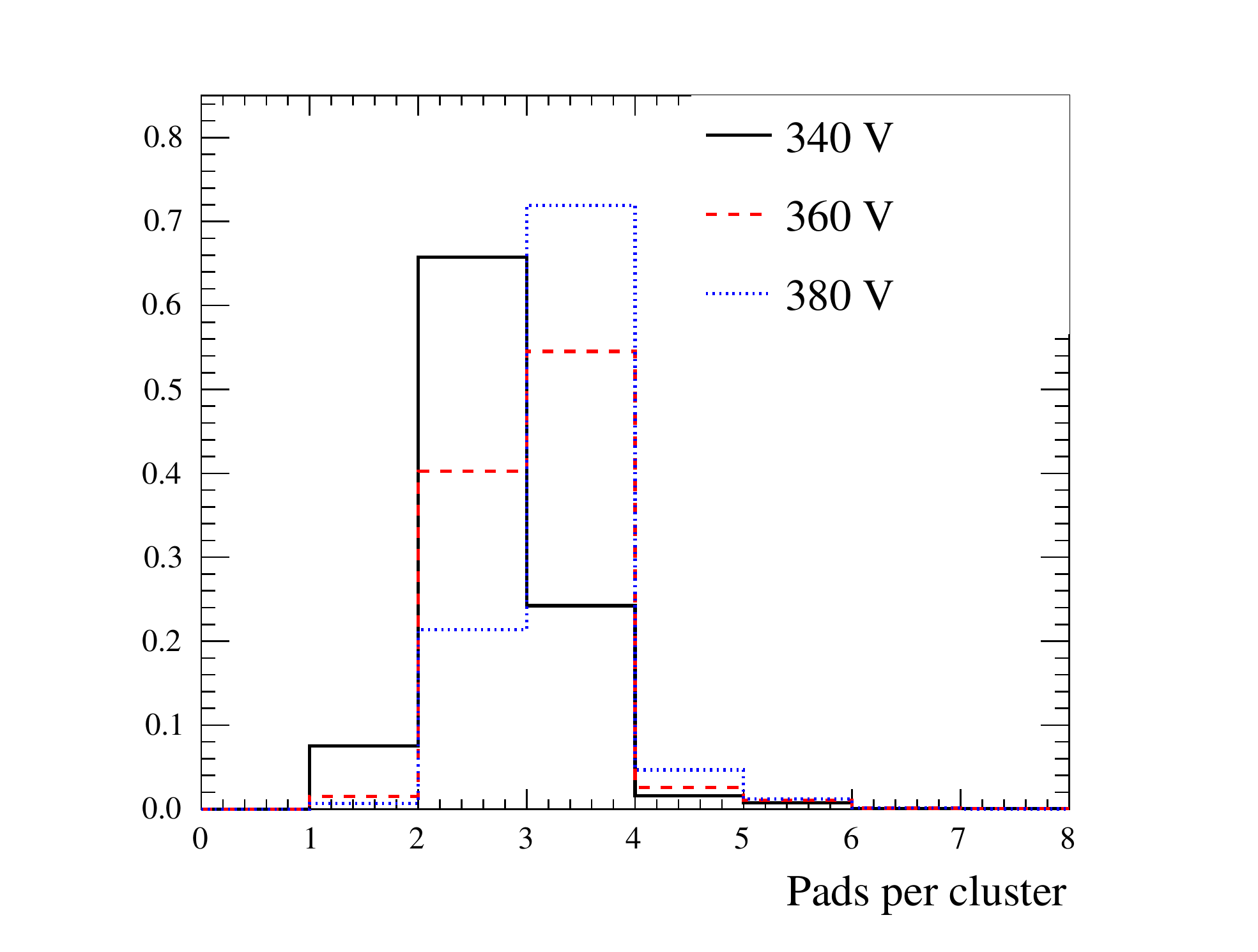}}
    \caption{Pad multiplicity  in the cluster for different Micromegas voltages for 1~GeV/c pion sample with 30 cm drift distance.}
    \label{fig:HV_mult}
\end{figure}

\begin{figure}[H]
    \begin{minipage}[h]{0.49\linewidth}
        \center{\includegraphics[width=\linewidth]{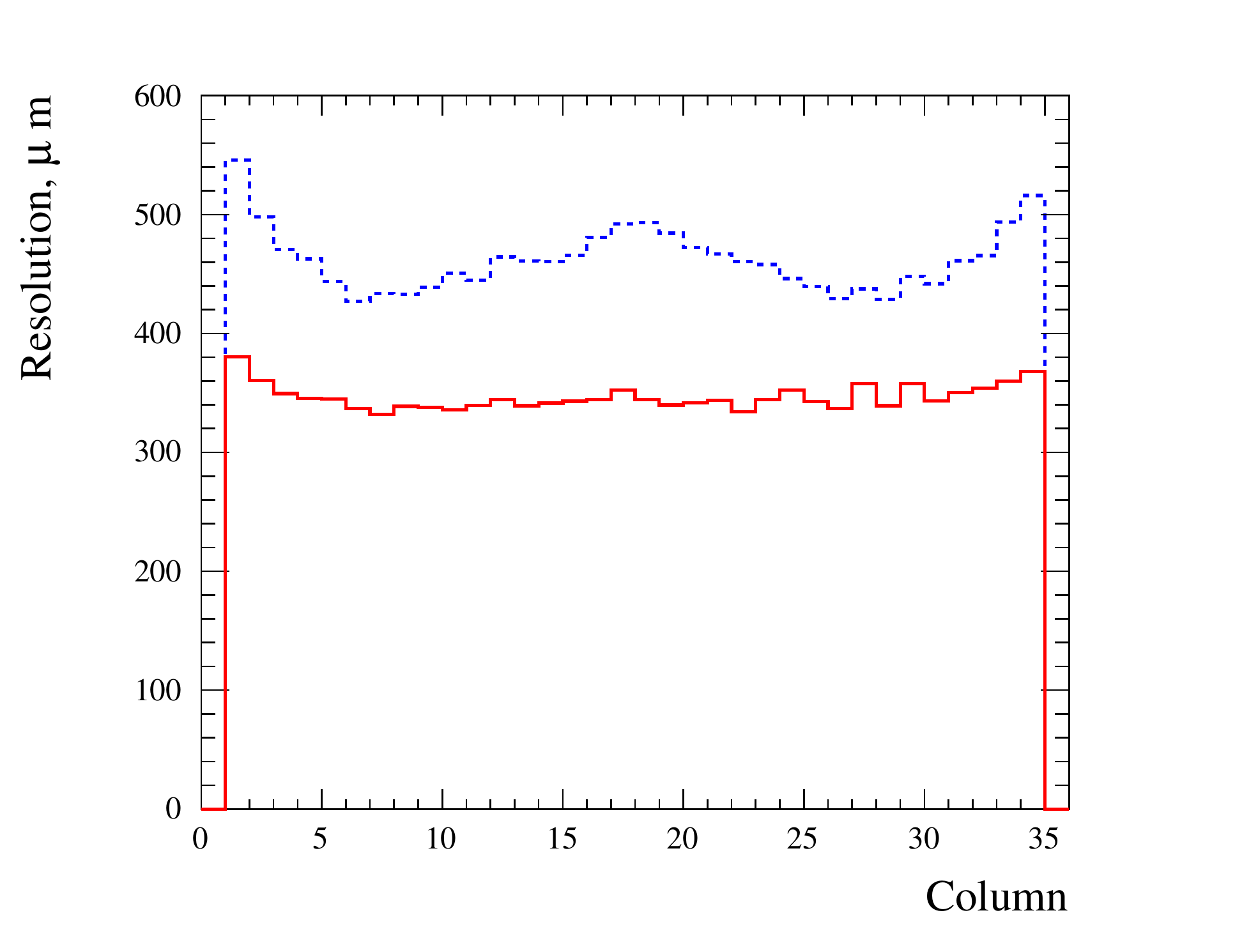} \\ 340 V}
    \end{minipage}
    \hfill
    \begin{minipage}[h]{0.49\linewidth}
        \center{\includegraphics[width=\linewidth]{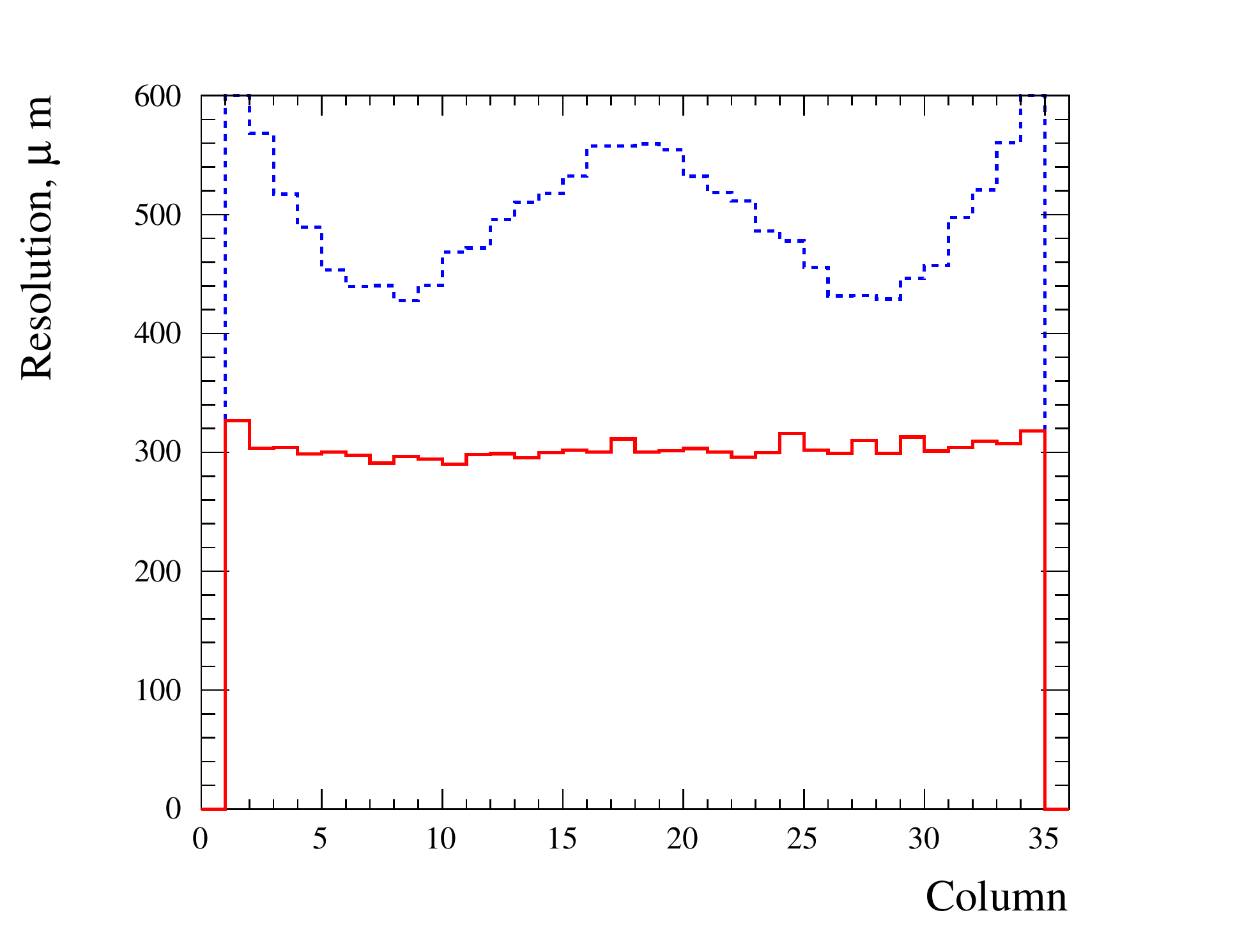} \\ 360 V}
    \end{minipage}
    \vfill
    \begin{minipage}[h]{0.49\linewidth}
        \center{\includegraphics[width=\linewidth]{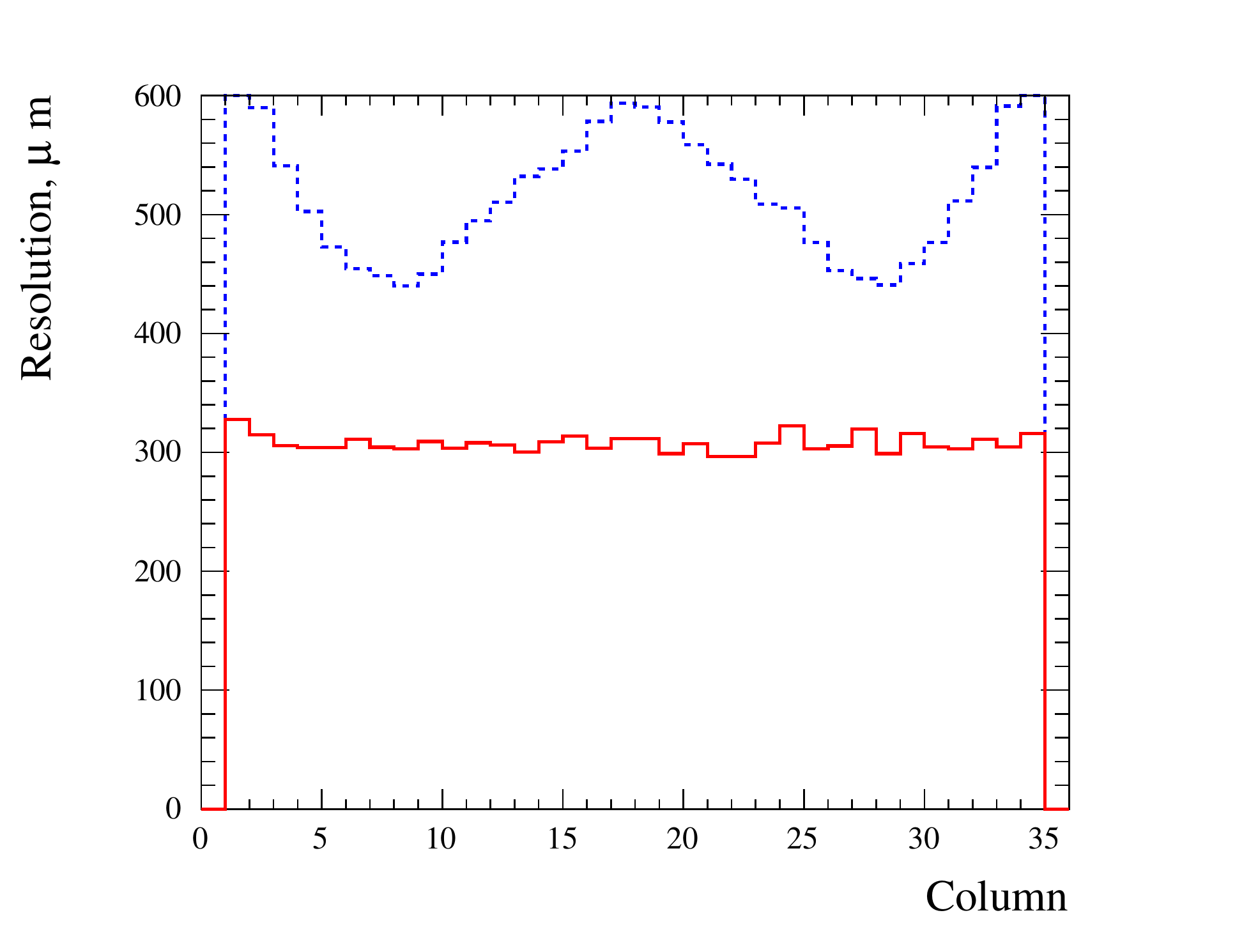} \\ 380 V}
    \end{minipage}
    \hfill
    \begin{minipage}[h]{0.49\linewidth}
        \center{\includegraphics[width=\linewidth]{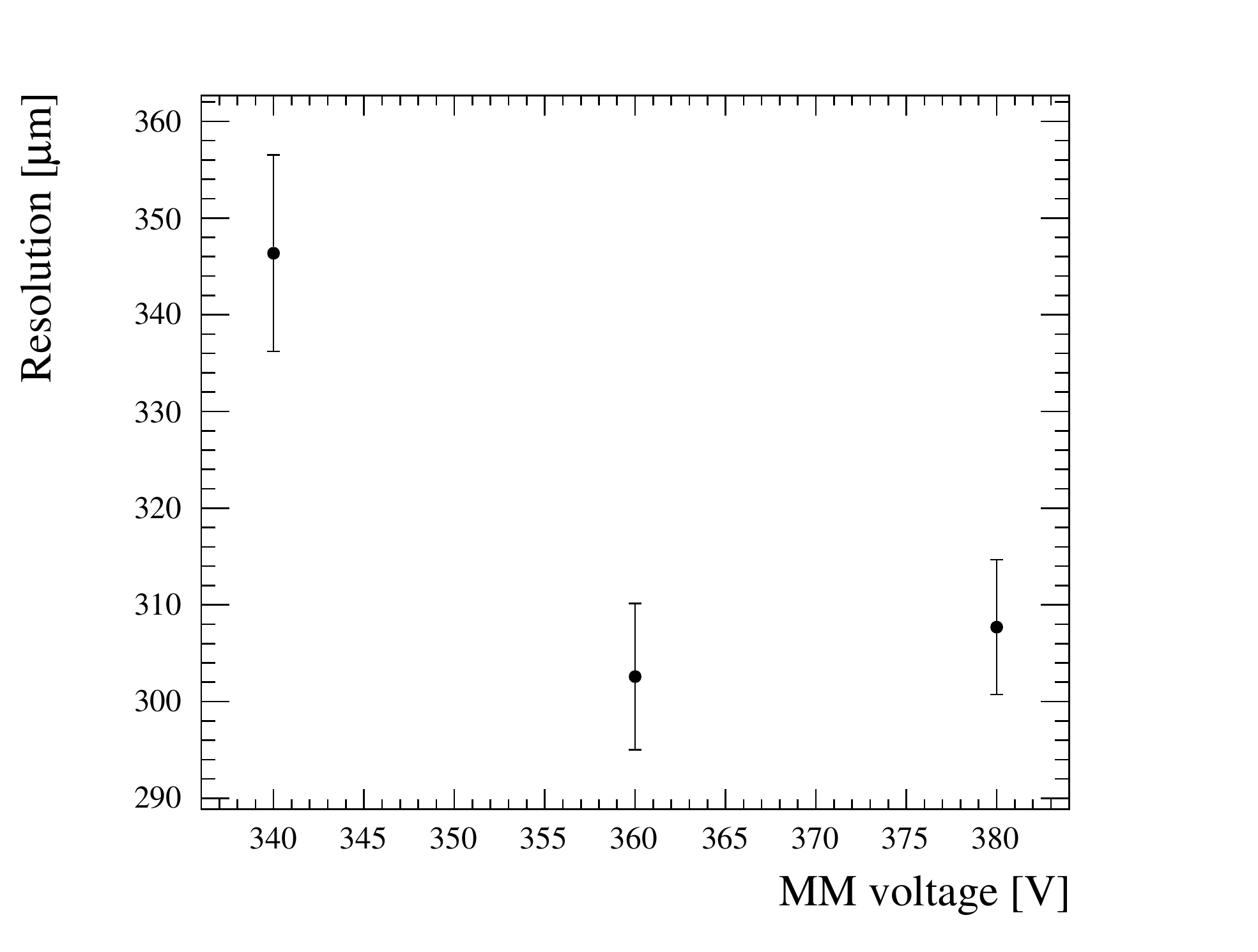} \\ Summary}
    \end{minipage}
    \caption{Spatial resolution versus the voltage at the Micromegas for 1~GeV/c pion sample with 30 cm drift distance. Blue dashed line corresponds to the CoC method, red line corresponds to the PRF method. The summary is shown with PRF method.}
    \label{fig:HV_spatial}
\end{figure}

The higher voltage provides larger amplification of the signal. That is helpful for detection of the small charges at the edge of charge spreading. This is shown in Figure~\ref{fig:HV_mult} where the number of pads in each cluster is shown for different voltages.
The spatial resolution for different voltages is shown in Figure~\ref{fig:HV_spatial}. The spatial resolution improvements was reached with the PRF method, while the CoC method shows the result degradation.

%% file: conclusions.tex
In this paper we showed the performances of a resistive Micromegas module  installed in the HARP field cage and exposed to a beam of charged particles at CERN, with momenta ranging from 500 MeV/c to 2 GeV/c. The Micromegas had an excellent uniformity of the gain and the data were used to characterize the properties of the charge spreading process of the collected charge. 

The performances of the Micromegas in terms of spatial resolution and deposited energy resolution have been evaluated and found to be satisfactory for the TPCs that will be built for the upgrade of the T2K Near Detector. 

Thanks to the resistive anode a spatial resolution of 300~$\mu m$ for horizontal electrons at 30~cm drift distance was measured while for the dE/dx we observed a resolution of 9\% for positrons of 800~MeV/c by using 34 measurements of the charge.

%% file: resistiveMM.bbl
\begin{thebibliography}{10}
\expandafter\ifx\csname url\endcsname\relax
  \def\url#1{\texttt{#1}}\fi
\expandafter\ifx\csname urlprefix\endcsname\relax\def\urlprefix{URL }\fi
\expandafter\ifx\csname href\endcsname\relax
  \def\href#1#2{#2} \def\path#1{#1}\fi

\bibitem{Abe:2011ks}
K.~Abe, et~al., {The T2K Experiment}, Nucl. Instrum. Meth. A659 (2011)
  106--135.
\newblock \href {http://arxiv.org/abs/1106.1238} {\path{arXiv:1106.1238}},
  \href {http://dx.doi.org/10.1016/j.nima.2011.06.067}
  {\path{doi:10.1016/j.nima.2011.06.067}}.

\bibitem{Abe:2011sj}
K.~Abe, et~al., {Indication of Electron Neutrino Appearance from an
  Accelerator-produced Off-axis Muon Neutrino Beam}, Phys. Rev. Lett. 107
  (2011) 041801.
\newblock \href {http://arxiv.org/abs/1106.2822} {\path{arXiv:1106.2822}},
  \href {http://dx.doi.org/10.1103/PhysRevLett.107.041801}
  {\path{doi:10.1103/PhysRevLett.107.041801}}.

\bibitem{Abe:2013hdq}
K.~Abe, et~al., {Observation of Electron Neutrino Appearance in a Muon Neutrino
  Beam}, Phys. Rev. Lett. 112 (2014) 061802.
\newblock \href {http://arxiv.org/abs/1311.4750} {\path{arXiv:1311.4750}},
  \href {http://dx.doi.org/10.1103/PhysRevLett.112.061802}
  {\path{doi:10.1103/PhysRevLett.112.061802}}.

\bibitem{Abe:2018wpn}
K.~Abe, et~al., {Search for CP Violation in Neutrino and Antineutrino
  Oscillations by the T2K Experiment with $2.2\times10^{21}$ Protons on
  Target}, Phys. Rev. Lett. 121~(17) (2018) 171802.
\newblock \href {http://arxiv.org/abs/1807.07891} {\path{arXiv:1807.07891}},
  \href {http://dx.doi.org/10.1103/PhysRevLett.121.171802}
  {\path{doi:10.1103/PhysRevLett.121.171802}}.

\bibitem{Assylbekov:2011sh}
S.~Assylbekov, et~al., {The T2K ND280 Off-Axis Pi-Zero Detector}, Nucl.
  Instrum. Meth. A686 (2012) 48--63.
\newblock \href {http://arxiv.org/abs/1111.5030} {\path{arXiv:1111.5030}},
  \href {http://dx.doi.org/10.1016/j.nima.2012.05.028}
  {\path{doi:10.1016/j.nima.2012.05.028}}.

\bibitem{Amaudruz:2012agx}
P.~A. Amaudruz, et~al., {The T2K Fine-Grained Detectors}, Nucl. Instrum. Meth.
  A696 (2012) 1--31.
\newblock \href {http://arxiv.org/abs/1204.3666} {\path{arXiv:1204.3666}},
  \href {http://dx.doi.org/10.1016/j.nima.2012.08.020}
  {\path{doi:10.1016/j.nima.2012.08.020}}.

\bibitem{Abgrall:2010hi}
N.~Abgrall, et~al., {Time Projection Chambers for the T2K Near Detectors},
  Nucl. Instrum. Meth. A637 (2011) 25--46.
\newblock \href {http://arxiv.org/abs/1012.0865} {\path{arXiv:1012.0865}},
  \href {http://dx.doi.org/10.1016/j.nima.2011.02.036}
  {\path{doi:10.1016/j.nima.2011.02.036}}.

\bibitem{Allan:2013ofa}
D.~Allan, et~al., {The Electromagnetic Calorimeter for the T2K Near Detector
  ND280}, JINST 8 (2013) P10019.
\newblock \href {http://arxiv.org/abs/1308.3445} {\path{arXiv:1308.3445}},
  \href {http://dx.doi.org/10.1088/1748-0221/8/10/P10019}
  {\path{doi:10.1088/1748-0221/8/10/P10019}}.

\bibitem{Aoki:2012mf}
S.~Aoki, et~al., {The T2K Side Muon Range Detector (SMRD)}, Nucl. Instrum.
  Meth. A698 (2013) 135--146.
\newblock \href {http://arxiv.org/abs/1206.3553} {\path{arXiv:1206.3553}},
  \href {http://dx.doi.org/10.1016/j.nima.2012.10.001}
  {\path{doi:10.1016/j.nima.2012.10.001}}.

\bibitem{Friend:2017oav}
M.~Friend, {J-PARC accelerator and neutrino beamline upgrade programme}, J.
  Phys. Conf. Ser. 888~(1) (2017) 012042.
\newblock \href {http://dx.doi.org/10.1088/1742-6596/888/1/012042}
  {\path{doi:10.1088/1742-6596/888/1/012042}}.

\bibitem{Abe:2019fux}
K.~Abe, et~al., {J-PARC Neutrino Beamline Upgrade Technical Design Report}\href
  {http://arxiv.org/abs/1908.05141} {\path{arXiv:1908.05141}}.

\bibitem{Abe:2019whr}
K.~Abe, et~al., {T2K ND280 Upgrade - Technical Design Report}\href
  {http://arxiv.org/abs/1901.03750} {\path{arXiv:1901.03750}}.

\bibitem{Kawamoto2013}
T.~Kawamoto, S.~Vlachos, L.~Levinson, C.~Amelung, G.~Mikenberg, L.~Pontecorvo,
  D.~Lellouch, J.~Dubbert, C.~Dallapiccola, R.~Richter, P.~Iengo,
  \href{http://cds.cern.ch/record/1552862/}{{New Small Wheel Technical Design
  Report}}, CERN-LHCC-2013-006.
\newline\urlprefix\url{http://cds.cern.ch/record/1552862/}

\bibitem{Giomataris:2004aa}
I.~Giomataris, R.~De~Oliveira, S.~Andriamonje, S.~Aune, G.~Charpak, P.~Colas,
  A.~Giganon, P.~Rebourgeard, P.~Salin, {Micromegas in a bulk}, Nucl. Instrum.
  Meth. A560 (2006) 405--408.
\newblock \href {http://arxiv.org/abs/physics/0501003}
  {\path{arXiv:physics/0501003}}, \href
  {http://dx.doi.org/10.1016/j.nima.2005.12.222}
  {\path{doi:10.1016/j.nima.2005.12.222}}.

\bibitem{Dixit:2003qg}
M.~S. Dixit, J.~Dubeau, J.~P. Martin, K.~Sachs, {Position sensing from charge
  dispersion in micropattern gas detectors with a resistive anode}, Nucl.
  Instrum. Meth. A518 (2004) 721--727.
\newblock \href {http://arxiv.org/abs/physics/0307152}
  {\path{arXiv:physics/0307152}}, \href
  {http://dx.doi.org/10.1016/j.nima.2003.09.051}
  {\path{doi:10.1016/j.nima.2003.09.051}}.

\bibitem{Colas:2010zz}
P.~Colas, {First test results from a Micromegas large TPC prototype}, Nucl.
  Instrum. Meth. A623 (2010) 100--101.
\newblock \href {http://dx.doi.org/10.1016/j.nima.2010.02.161}
  {\path{doi:10.1016/j.nima.2010.02.161}}.

\bibitem{Attie:2011zz}
D.~Attie, {Beam tests of Micromegas LC-TPC large prototype}, JINST 6 (2011)
  C01007.
\newblock \href {http://dx.doi.org/10.1088/1748-0221/6/01/C01007}
  {\path{doi:10.1088/1748-0221/6/01/C01007}}.

\bibitem{Baron:2008zza}
P.~Baron, D.~Calvet, E.~Delagnes, X.~de~la Broise, A.~Delbart, F.~Druillole,
  E.~Monmarthe, E.~Mazzucato, F.~Pierre, M.~Zito, {AFTER, an ASIC for the
  readout of the large T2K time projection chambers}, IEEE Trans. Nucl. Sci. 55
  (2008) 1744--1752.
\newblock \href {http://dx.doi.org/10.1109/TNS.2008.924067}
  {\path{doi:10.1109/TNS.2008.924067}}.

\bibitem{prior2003harp}
G.~Prior, H.~collaboration, et~al., The harp time projection chamber, Nuclear
  Physics B-Proceedings Supplements 125 (2003) 37--42.

\bibitem{Ester96adensity-based}
M.~Ester, H.-P. Kriegel, J.~Sander, X.~Xu, A density-based algorithm for
  discovering clusters in large spatial databases with noise, in: Proceedings
  of the Second International Conference on Knowledge Discovery and Data
  Mining, AAAI Press, 1996, pp. 226--231.

\bibitem{Dixit:2006ge}
M.~S. Dixit, A.~Rankin, {Simulating the charge dispersion phenomena in micro
  pattern gas detectors with a resistive anode}, Nucl. Instrum. Meth. A566
  (2006) 281--285.
\newblock \href {http://arxiv.org/abs/physics/0605121}
  {\path{arXiv:physics/0605121}}, \href
  {http://dx.doi.org/10.1016/j.nima.2006.06.050}
  {\path{doi:10.1016/j.nima.2006.06.050}}.

\bibitem{Boudjemline:2006hf}
K.~Boudjemline, M.~S. Dixit, J.~P. Martin, K.~Sachs, {Spatial resolution of a
  GEM readout TPC using the charge dispersion signal}, Nucl. Instrum. Meth.
  A574 (2007) 22--27.
\newblock \href {http://arxiv.org/abs/physics/0610232}
  {\path{arXiv:physics/0610232}}, \href
  {http://dx.doi.org/10.1016/j.nima.2007.01.017}
  {\path{doi:10.1016/j.nima.2007.01.017}}.

\end{thebibliography}
